\documentclass[11pt]{article}

\usepackage{amsmath}
\usepackage{amsthm}
\usepackage{amsfonts}
\usepackage{amssymb}
\usepackage{latexsym} 
\usepackage{epsfig}
\usepackage{graphicx}

\usepackage{calc}  
\usepackage[matrix,tips,graph,curve,web]{xy}

\makeatletter

\def\@maketitle{%
  \newpage
  \null
  \let \footnote \thanks
    {\normalfont\sffamily\bfseries\Large\noindent\@title \par}%
    \vskip 1em%
    {\normalfont\sffamily 
        \noindent
        \@author
        \par}
  \par
  \vskip 4em}
\def\@seccntformat#1{\csname the#1\endcsname{.\ }}
\renewcommand\section{\@startsection {section}{1}{\z@}%
                                   {-3.0ex \@plus -1ex \@minus -.2ex}%
                                   {1.5ex \@plus.2ex}%
                                   {\normalfont\large\bfseries}}
\renewcommand\subsection{\@startsection{subsection}{2}{\z@}%
                                     {-2.75ex\@plus -1ex \@minus -.2ex}%
                                     {1.5ex \@plus .2ex}%
                                   {\normalfont\large}}
\def\fnum@figure{\normalfont\footnotesize\figurename~\thefigure}

\renewcommand\tableofcontents{%
    \section*{\contentsname
        \@mkboth{%
           \MakeUppercase\contentsname}{\MakeUppercase\contentsname}}%
    \@starttoc{toc}%
    }
\renewcommand*\l@part[2]{%
  \ifnum \c@tocdepth >-2\relax
    \addpenalty\@secpenalty
    \addvspace{2.25em \@plus\p@}%
    \begingroup
      \setlength\@tempdima{3em}%
      \parindent \z@ \rightskip \@pnumwidth
      \parfillskip -\@pnumwidth
      {\leavevmode
       \large \bfseries #1\hfil \hb@xt@\@pnumwidth{\hss #2}}\par
       \nobreak
       \if@compatibility
         \global\@nobreaktrue
         \everypar{\global\@nobreakfalse\everypar{}}%
      \fi
    \endgroup
  \fi}
\renewcommand*\l@section[2]{%
  \ifnum \c@tocdepth >\z@
    \addpenalty\@secpenalty
    \addvspace{1.0em \@plus\p@}%
    \setlength\@tempdima{1.5em}%
    \begingroup
      \parindent \z@ \rightskip \@pnumwidth
      \parfillskip -\@pnumwidth
      \leavevmode \sffamily\bfseries
      \advance\leftskip\@tempdima
      \hskip -\leftskip
      #1\nobreak\hfil \nobreak\hb@xt@\@pnumwidth{\hss #2}\par
    \endgroup
  \fi}
\renewcommand*\l@subsection{\sffamily\@dottedtocline{2}{1.5em}{2.3em}}
\renewcommand*\l@subsubsection{\@dottedtocline{3}{3.8em}{3.2em}}
\renewcommand*\l@paragraph{\@dottedtocline{4}{7.0em}{4.1em}}
\renewcommand*\l@subparagraph{\@dottedtocline{5}{10em}{5em}}
\makeatother


\theoremstyle{plain}
\newtheorem{theorem}{Theorem}[section]
\newtheorem{corollary}[theorem]{Corollary}
\newtheorem{lemma}[theorem]{Lemma}
\newtheorem{proposition}[theorem]{Proposition}
\newtheorem{conjecture}[theorem]{Conjecture}

\theoremstyle{definition}
\newtheorem{definition}[theorem]{Definition}

\newtheorem{example}[theorem]{Example}

  {\begin{list}{}%
    {%
    \settowidth{\labelwidth}{#1}%
    \setlength{\itemindent}{0pt}%
    \setlength{\labelsep}{1em}%
    \setlength{\leftmargin}{\labelwidth+\parindent+\labelsep}%
    \setlength{\itemsep}{0pt}%
    \setlength{\parsep}{.6ex}}}%
  {\end{list}}

  {\begin{list}{}%
    {%
    \settowidth{\labelwidth}{#1}%
    \setlength{\itemindent}{0pt}%
    \setlength{\labelsep}{1em}%
    \setlength{\leftmargin}{\labelwidth+\labelsep}%
    \setlength{\itemsep}{0pt}%
    \setlength{\parsep}{.6ex}}}%
  {\end{list}}

\makeatletter
\newenvironment{subequations*}{
  \begingroup 
  \let\protect\@nx
  \edef\@tempa{\def\@nx\theparentequation{\theequation}}%
  \@xp\endgroup\@tempa
  \setcounter{parentequation}{\value{equation}}%
  \setcounter{equation}{0}%
  \def\theequation{\theparentequation\alph{equation}}%
  \ignorespaces
}{%
  \setcounter{equation}{\value{parentequation}}%
  \global\@ignoretrue
}

\newcommand{\sgn}{\rm sgn\,}

\renewcommand\det{{\rm det\,}}

\def\d/{/\mspace{-6.0mu}/}

\usepackage{helvet}
\usepackage{graphicx,amssymb,amsmath,amsfonts,amsthm,color,url, placeins, enumitem}
\usepackage{wrapfig}
\usepackage{tabstackengine}
\usepackage{multirow,ctable} 
\usepackage{makecell}

\renewcommand\section{\@startsection{section}{1}{\z@}%
                                   {-3.0ex \@plus -1ex \@minus -.2ex}%
                                   {1.5ex \@plus.2ex}%
                                   {\normalfont\sffamily\large\bfseries}}
\renewcommand\subsection{\@startsection{subsection}{2}{\z@}%
                                     {-2.75ex\@plus -1ex \@minus -.2ex}%
                                     {1.5ex \@plus .2ex}%
                                   {\normalfont\sffamily\large}}
\renewcommand\subsubsection{\@startsection{subsubsection}{3}{\z@}%
                                     {-2.75ex\@plus -1ex \@minus -.2ex}%
                                     {1.5ex \@plus .2ex}%
                                   {\normalfont\sffamily\large}}

\newcommand{\od}{\stackrel{\mbox {\tiny {def}}}{=}}

\def\RR{\mathbb{R}}

\def\d{\mathrm{d}}

\def\RR{\mathbb{R}}

\newtheorem{rules}{Rule}

\def\S{{\mathcal{S}}}

\def\RR{\mathbb{R}}

\def\det{\operatorname{det}}
\def\S{{\mathcal{S}}}

\def\max{\mathrm{max}}

\def\supp{\operatorname{supp}}

\def\od{\stackrel{\mathrm{def}}{=}}

\def\supp{\operatorname{supp}}

\def\sgn{\operatorname{sgn}}
\def\FP{\operatorname{FP}}

\def\idx{\operatorname{idx}}

\usepackage{color}
\definecolor{cherry}{rgb}{0.9,.1,.2}

\begin{document}

\noindent {\large \bf Sequential attractors in combinatorial threshold-linear networks}\\
\noindent Caitlyn Parmelee, Juliana Londono Alvarez, Carina Curto*, Katherine Morrison* \\
{\footnotesize * equal contribution}\\

\section*{Abstract}
Sequences of neural activity arise in many brain areas, including cortex, hippocampus, and central pattern generator circuits that underlie rhythmic behaviors like locomotion. While network architectures supporting sequence generation vary considerably, a common feature is an abundance of inhibition. 
In this work, we focus on architectures that support sequential activity in recurrently connected networks with inhibition-dominated dynamics. Specifically, we study emergent sequences in a special family of threshold-linear networks, called combinatorial threshold-linear networks (CTLNs), whose connectivity matrices are defined from directed graphs.  Such networks naturally give rise to an abundance of sequences whose dynamics are tightly connected to the underlying graph. We find that architectures based on generalizations of cycle graphs produce limit cycle attractors that can be activated to generate transient or persistent (repeating) sequences. Each architecture type gives rise to an infinite family of graphs that can be built from arbitrary component subgraphs. Moreover, we prove a number of {\it graph rules} for the corresponding CTLNs in each family. The graph rules allow us to strongly constrain, and in some cases fully determine, the fixed points of the network in terms of the fixed points of the component subnetworks. Finally, we also show how the structure of certain architectures gives insight into the sequential dynamics of the corresponding attractor. 

\tableofcontents

\section{Introduction}
Sequences of neural activity arise in many brain areas, including cortex \cite{Luczak-PNAS, Yuste-cortex-sequences, Yuste-CPG}, hippocampus \cite{Stark-PNAS, Eva-science, JNeuro}, and central pattern generator circuits that underlie rhythmic behaviors like locomotion \cite{Marder-CPG, CPG-review}. Moreover, fast sequences during sharp wave ripple events in hippocampus are believed to be critical for memory processing and cortico-hippocampal communication \cite{Colgin-hipp-rhythms, Zugaro-NN, Ripple-learning}.  Such sequences are examples of emergent or \emph{internally-generated activity}: that is, neural activity that is shaped primarily by the structure of a recurrent network rather than inherited from a changing external input.  A fundamental question is to understand how a network's connectivity shapes neural activity, and what types of network architectures underlie emergent sequences.

Inhibition has long been viewed as a key component of sequence generation in CPGs. It also plays an important role in generating rhythmic and sequential activity in cortex and hippocampus \cite{Yuste-CPG, Marder-CPG, Rotstein-inhibition, Carandini-inhibition, Yuste-inhibition, Karnani-inhibition, Inhibition-hippocampus}. Roughly speaking, inhibition creates competition among neurons, resulting in a tendency for neurons to take turns reaching peak activity levels and thus to fire in sequence. In particular, inhibition-dominated networks exhibit emergent sequences even in the absence of an obvious chain-like architecture, such as a synfire chain \cite{Kopell2, Bazhenov-hipp-ripples}.  In this work, we analyze a variety of network architectures that give rise to sequential neural activity in a simple  nonlinear model of recurrent networks with inhibition-dominated dynamics.  

\paragraph{Mathematical setup\\}
We study sequential dynamics in a family of threshold-linear networks (TLNs).
The firing rates $x_1(t),\ldots,x_n(t)$ of $n$ recurrently-connected neurons evolve in time according to the standard TLN equations:
\begin{equation}\label{eq:dynamics}
\dfrac{dx_i}{dt} = -x_i + \left[\sum_{j=1}^n W_{ij}x_j+b_i \right]_+, \quad i = 1,\ldots,n,
\vspace{-.04in}
\end{equation}
where $[\cdot]_+ = \max\{0,\cdot\}$ is the threshold nonlinearity.   A given TLN is specified by the choice of a connection strength matrix $W$ and a vector of external inputs $b \in \RR^n$. TLNs have been widely used in computational neuroscience as a framework for modeling recurrent neural networks, including associative memory networks \cite{AppendixE,Seung-Nature,HahnSeungSlotine, XieHahnSeung, flex-memory, net-encoding, pattern-completion, Fitzgerald2020}.

In order to investigate how network architectures support sequential dynamics, we consider the special family of \emph{combinatorial threshold-linear networks} (CTLNs). These are inhibition-dominated TLNs where the matrix $W = W(G,\varepsilon,\delta)$ is determined by a simple\footnote{A graph is \emph{simple} if it does not have self-loops or multiple edges (in the same direction) between a pair of nodes.}
 directed graph $G$, as follows:
\begin{equation} \label{eq:binary-synapse}
W_{ij} = \left\{\begin{array}{ll} \phantom{-}0 & \text{ if } i = j, \\ -1 + \varepsilon & \text{ if } j \rightarrow i \text{ in } G,\\ -1 -\delta & \text{ if } j \not\rightarrow i \text{ in } G. \end{array}\right. \quad \quad \quad \quad
\end{equation}
Note that $ j \rightarrow i$ indicates the presence of an edge from $j$ to $i$ in the graph $G$, while $j \not\rightarrow i$ indicates the absence of such an edge. Additionally, CTLNs typically have a constant external input $b_i=\theta$ in order to ensure the dynamics are internally generated and not inherited from a changing or spatially heterogeneous input.  We require the three parameters to satisfy $\theta>0$, $\delta >0$, and $0 < \varepsilon < \frac{\delta}{\delta+1}$; when these conditions are met, we say that the parameters are within the \emph{legal range}.\footnote{The upper bound on $\varepsilon$ is motivated by a theorem in \cite{CTLN-preprint}. It ensures that subgraphs consisting of a single directed edge $i \to j$ are not allowed to support stable fixed points.} 
Note that the upper bound on $\varepsilon$ implies $\varepsilon < 1$, and so the $W$ matrix is always effectively inhibitory.

One of the most striking features of CTLNs is the strong connection between dynamic attractors and unstable fixed points \cite{book-chapter, rule-of-thumb}.  A fixed point $x^*$ of a TLN is a solution that satisfies $dx_i/dt|_{x=x^*} = 0$ for each $i \in [n]$. The  {\it support} of a fixed point is the subset of active neurons, $\supp{x} = \{i \mid x_i>0\}$.
For a given network, there can be at most one fixed point per support.  Thus, we can label all the fixed points of a network by their support, $\sigma = \supp{x^*} \subseteq [n],$ where $[n] \od \{1, \ldots, n\}$.  We denote this collection of supports by 
\vspace{-.05in}
\[\FP(G) = \FP(G, \varepsilon, \delta)\od \{\sigma \subseteq [n] ~|~  \sigma \text{ is a fixed point support of } W(G,\varepsilon,\delta) \}.\]
  In prior work, a series of {\it graph rules} were proven that can be used to determine fixed points of a CTLN by analyzing the structure of the graph $G$ \cite{fp-paper, stable-fp-paper}.  These rules are all independent of the choice of parameters 
$\varepsilon, \delta,$ and $\theta$.

\paragraph{Sequences from limit cycles\\}
Limit cycles are dynamic attractors corresponding to periodic solutions. A {\it sequential} limit cycle produces a repeating sequence of neural activations. Limit cycles thus provide a basic mechanism for generating sequences in the context of attractor neural networks. 

\begin{figure}[!ht]
\begin{center}
\includegraphics[width=.85\textwidth]{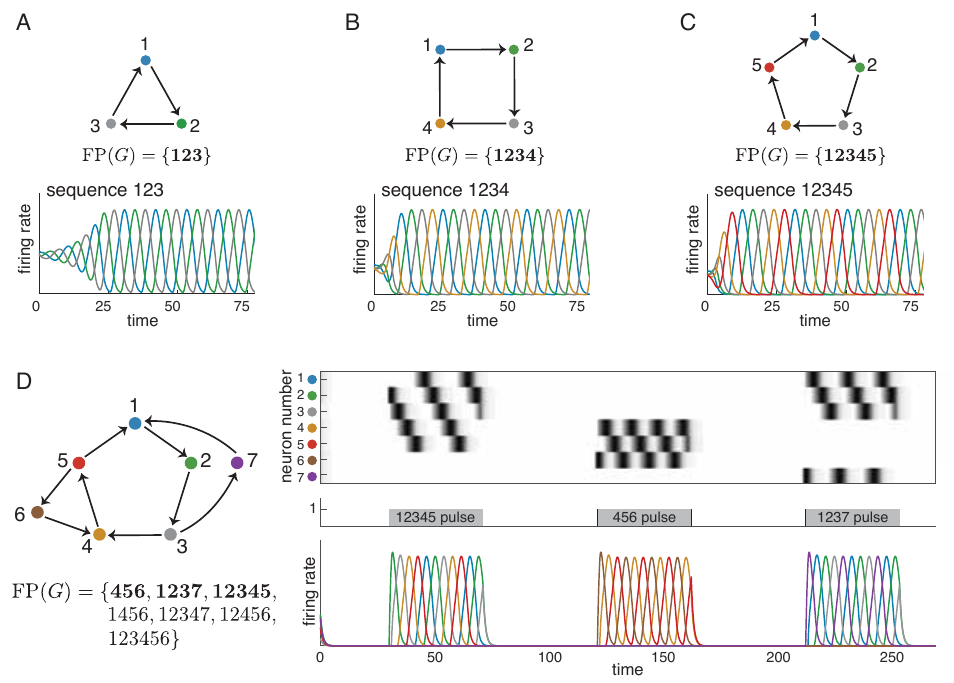}
\end{center}
\caption{Sequential attractors from cycle graphs. (A-C) CTLNs corresponding to a $3$-cycle, a $4$-cycle, and a $5$-cycle each produce a limit cycle where the neurons reach their peak activations in the expected sequence. Colored curves correspond to solutions $x_i(t)$ for matching node $i$ in the graph. 
(D) Attractors corresponding to the embedded $3$-cycle, $4$-cycle, and $5$-cycle of the network are transiently activated to produce sequences matching those of the isolated cycle networks in A-C.  For each network in A-D, $\FP(G)$ is shown, with the minimal fixed points bolded.  To simplify notation for $\FP(G)$, we denote a subset $\{i_1, \ldots, i_k\}$ by $i_1\cdots i_k$.  For example, $12345$ denotes the set $\{1,2,3,4,5\}$. Unless otherwise noted (as in Section~\ref{sec:applications}), all simulations have CTLN parameters $\varepsilon=0.25, \delta = 0.5,$ and $\theta =1$.
}
\label{fig:sequences-setup}
\end{figure}

It is easy to see computationally that a CTLN corresponding to a cyclic graph produces a sequential attractor. Figure~\ref{fig:sequences-setup}A-C shows limit cycles corresponding to the graph $G$ being a $3$-cycle (panel A), a $4$-cycle (panel B), or a $5$-cycle (panel C). In each case, the solution exhibits a sequence of peak activations that matches the order of neurons in the cycle of the graph. Note that although all connections are effectively inhibitory, the activity appears to follow the edges in the graph. A rigorous proof for the existence of these limit cycles was given in \cite{Horacio-paper}.

To obtain shorter sequences, these attractors may be transiently activated by an external drive that is time dependent. Figure~\ref{fig:sequences-setup}D shows the solution for a CTLN with a graph on seven neurons (left). Here we have chosen $\theta = 0$ as a baseline, with step function pulses of $\theta_i = 1$ for different subsets of neurons. A single simulation is shown, with localized pulses activating the $5$-cycle, the $3$-cycle, and finally the $4$-cycle. Although these cycles overlap, each pulse activates a sequence involving only the neurons in the stimulated subnetwork. Depending on the duration of the pulse, the sequence may play only once or repeat two or more times. 

Notice that the minimal fixed points of the network in Figure~\ref{fig:sequences-setup}D reflect the subsets of neurons active in the attractors.  In related work, we have seen a close correspondence between certain minimal fixed points, called \emph{core motifs}, and the attractors of a network \cite{rule-of-thumb}.  Thus, $\FP(G)$ is often predictive of limit cycles and other dynamic attractors of a network.

The above mechanism for sequence generation differs from that observed in synfire chains \cite{synfire-chain1,synfire-chain2,synfire-chain3} where neural activity flows through a feedforward network, transiently activating neurons in sequence.  In Section~\ref{sec:dir-chains}, we provide a generalization of synfire chain structure, known as \emph{directional chains}, that allow for some local recurrence while still yielding sequences from their transient activity. But the primary focus of this work is on architectures that support sequential attractors, such as limit cycles, with transient sequences emerging from transient activation of these networks. 

Graphs that are cycles were the most obvious candidate to produce sequential attractors. But not all CTLN attractors are limit cycles, and not all limit cycles generate sequences. What other architectures can support sequential attractors? This is the main question we address in this paper. We investigate four architectures that generalize the cyclic structure of graphs that are cycles. These are: cyclic unions, directional cycles, simply-embedded partitions, and simply-embedded directional cycles. A common feature of all these architectures is that the neurons of the network are partitioned into components $\tau_1,\ldots,\tau_N$, organized in a cyclic manner, whose disjoint union equals the full set of neurons $[n] \od \{1,\ldots,n\}$. The induced subgraphs $G|_{\tau_i}$ are called component subgraphs.  We will prove a series of theorems about these architectures connecting the fixed points of a graph $G$ to the fixed points of the component subgraphs $G|_{\tau_i}$.  As shown in \cite{rule-of-thumb}, there is a striking correspondence between certain unstable fixed points of a network and its dynamic attractors.  Our theorems about the fixed points thus provide valuable insight into the dynamics associated to these network architectures.

\begin{figure}[!h]
\begin{center}
\includegraphics[width=.85\textwidth]{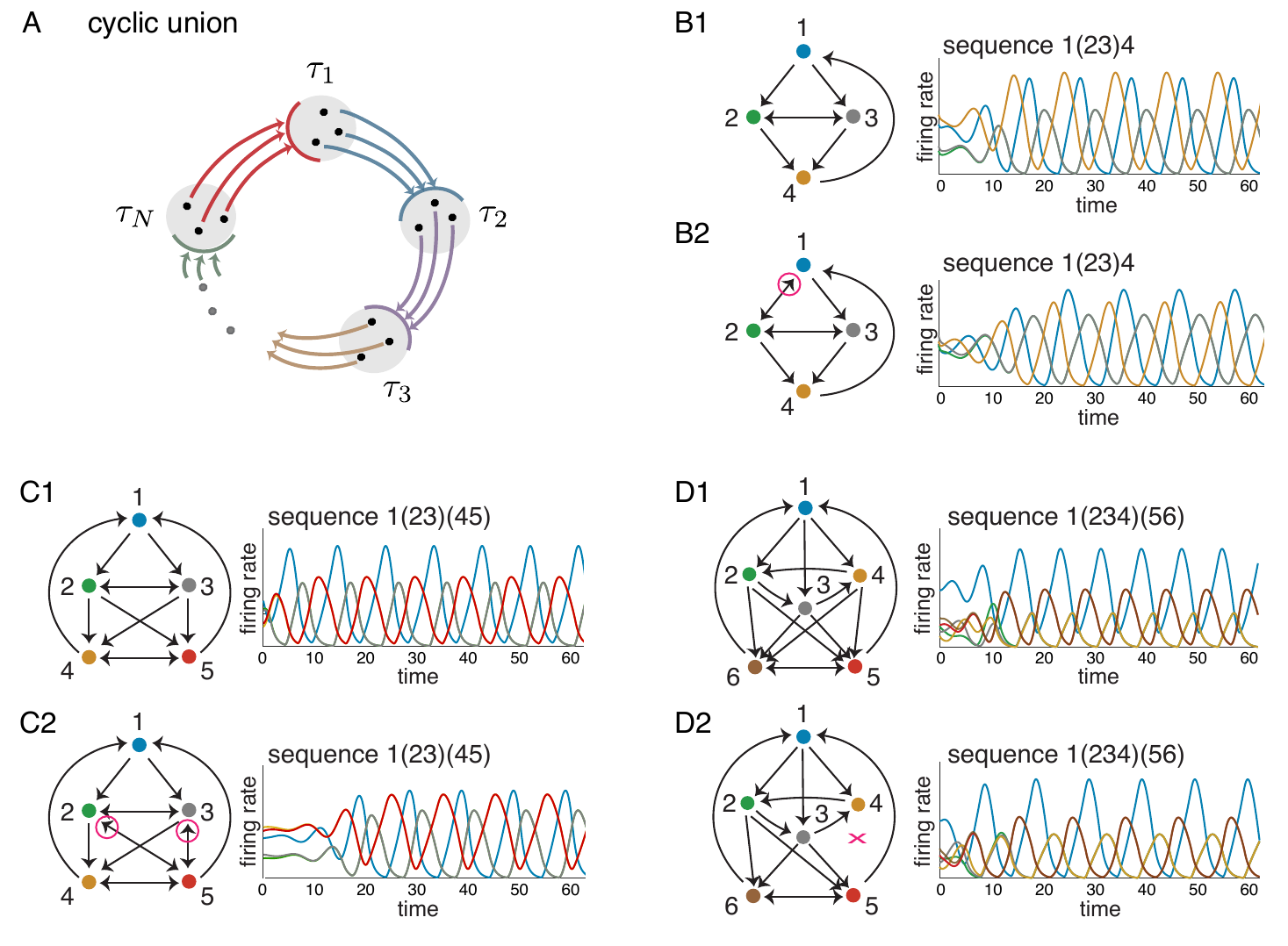}
\vspace{.05in}
\caption{{\bf Cyclic unions and related variations.}  (A) A cyclic union has component subgraphs with subsets of nodes $\tau_1, \ldots, \tau_N$, organized in a cyclic manner. While edges within each $G|_{\tau_i}$ can be arbitrary, edges between components are determined as follows: every node in $\tau_i$ sends an edge to every node $\tau_{i+1}$, with $\tau_N$ sending edges to $\tau_1$. (B1,C1,D1) Three cyclic unions with firing rate plots showing solutions to a corresponding CTLN.  Above each solution the associated sequence of firing rate peaks is given, with synchronously firing neurons denoted by parentheses.  (B2,C2,D2) These graphs are all variations on the cyclic unions above them, with some edges added or dropped (highlighted in magenta).  Solutions of the corresponding CLTNs qualitatively match the solutions of the corresponding cyclic unions. In each case, the sequence is identical.}
\label{fig:cycu-generalizations}
\end{center}
\vspace{-.1in}
\end{figure}

\paragraph{Cyclic unions\\}

The most straightforward generalization of a cycle is the {\it cyclic union}, an architecture first introduced in \cite{fp-paper}. Given a set of component subgraphs $G|_{\tau_1}, \ldots, G|_{\tau_N},$ on subsets of nodes $\tau_1, \ldots, \tau_N$, the \emph{cyclic union} is constructed by connecting these subgraphs in a cyclic fashion so that there are edges forward from every node in $\tau_i$ to every node in $\tau_{i+1}$ (cyclically identifying $\tau_N$ with $\tau_0$), and there are no other edges between components (see Figure~\ref{fig:cycu-generalizations}A).

The top graphs in Figure~\ref{fig:cycu-generalizations}B-D are examples of cyclic unions with three components.  All the nodes at a given height comprise a $\tau_i$ component, and we see that there are edges forward from every node in one component to each node in the next one.  Next to each graph is a solution to a corresponding CTLN, which is a global attractor of the network. Note that the activity traverses the components in cyclic order.   Cyclic unions are particularly well-behaved architectures where the fixed point supports can be fully characterized in terms of those of the components.  Specifically, in \cite{fp-paper} it was shown that the fixed points of a cyclic union $G$ are precisely the unions of supports of the component subgraphs, exactly one per component.

\begin{theorem}[cyclic unions (Theorem 13 in \cite{fp-paper})]\label{thm:cyclic-unions}
 Let $G$ be a cyclic union of component subgraphs $G|_{\tau_1}, \ldots, G|_{\tau_N}.$ For any $\sigma \subseteq [n]$, let $\sigma_i \od \sigma \cap \tau_i$.  Then
 \vspace{-.05in}
$$\sigma \in \FP(G) \quad \Leftrightarrow \quad \sigma_i \in \FP(G|_{\tau_i})~~\text{ for all } i \in [N].$$
\end{theorem}

The bottom graphs in Figure~\ref{fig:cycu-generalizations}B-D have very similar dynamics to the ones above them, but do not have a perfect cyclic union structure (each graph has some added back edges or dropped forward edges highlighted in magenta).  Despite deviations from the cyclic union architecture, these graphs produce sequential dynamics that similarly traverse the components in cyclic order. In fact, they are examples of a more general class of architectures: directional cycles.

\paragraph{Directional cycles\\}
In a cyclic union, if we restrict to the subnetwork consisting of a pair of consecutive components, $G|_{\tau_i \cup \tau_{i+1}}$, we find that activity initialized on $\tau_i$ flows forward and ends up concentrated on $\tau_{i+1}$.  Thus, there appears to be a directionality to the flow $\tau_i \to \tau_{i+1}$.  Moreover, the fixed points of $G|_{\tau_i \cup \tau_{i+1}}$ are all confined to live in $\tau_{i+1}$, and so the concentration of neural activity coincides with the subnetwork supporting the fixed points.  This is a phenomenon we have observed more generally that motivates us to define \emph{directional graphs}.  

We say that a graph is \emph{directional} whenever we have $\FP(G) \subseteq \FP(G|_\tau)$ for some $\tau \subsetneq [n]$.  In this case, we denote the complementary set as $\omega = [n] \setminus \tau$, and say that $G$ has direction $\omega \to \tau$.  We additionally require a more technical condition that allows us to prove that certain natural compositions, like chaining directional graphs together, produce a new directional graph (see Definition \ref{def:dir-graphs} for the full definition).  In simulations, we have seen that directional graphs have the desired directionality of neural activity, so that activity initialized on $\omega$ will flow through the network and become concentrated on the nodes of $\tau$. 

Note that while we predict that directional graphs have feedforward dynamics, they need not have a feedforward architecture.  In Figure~\ref{fig:cycu-generalizations}, each subgraph consisting of a pair of consecutive components is directional.  For example, the subgraph $G|_{\{1,2,3\}}$ in B2 is directional with direction $\{1\} \to \{2,3\}$, so that activity initialized on node $1$ tends to flow forward to nodes $2$ and $3$, despite the presence of the back edge $2 \to 1$.  Similarly, in C2, the subgraph $G|_{\{2,3,4,5\}}$ is directional with direction $\{2,3\} \to \{4,5\}$ despite the back edges $5 \to 2$ and $5 \to 3$.  The subgraph $G|_{\{2,3,4,5,6\}}$ in D2 is also directional with direction $\{2,3,4\} \to \{5,6\}$. Note that it is not necessary to have edges forward from every node in $\omega$ to every node in $\tau$.  

\begin{figure}[!ht]
\begin{center}
\includegraphics[width=.9\textwidth]{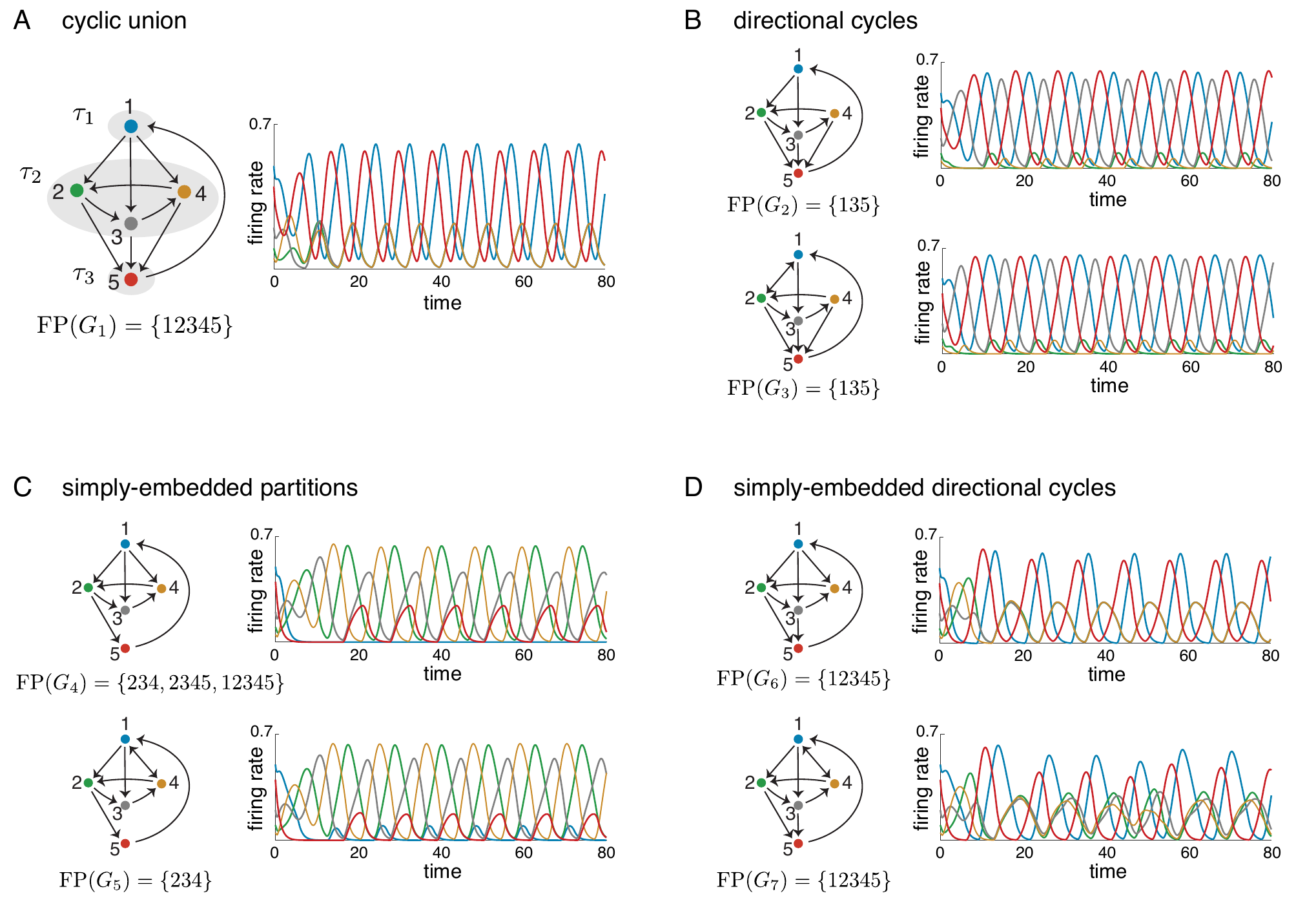}
\vspace{-.15in}
\end{center}
\caption{{\bf Example graphs generalizing cyclic union structure.} (A) A cyclic union of component subgraphs $G|_{\tau_1}, \ldots, G|_{\tau_3}$ together with its $\FP(G)$ and the global attractor of its corresponding CTLN.  
(B--D) Different generalizations of the cyclic union structure of the graph in A.  Each graph has the same component subgraphs $G|_{\tau_1}, \ldots, G|_{\tau_3}$, but different conditions on the edges between these components.  For each graph, $\FP(G)$ and the global attractor are shown. }
\label{fig:cyclic-union-dir-cycle-sa}
\end{figure}

With this broader notion of directional graph, we obtain our first generalization of cyclic unions, known as \emph{directional cycles}.  We define a \emph{directional cycle} as a graph with a partition of its nodes such that each $G|_{\tau_i \cup \tau_{i+1}}$ is directional with direction $\tau_i \to \tau_{i+1}$ (cyclically identifying $\tau_N$ with $\tau_0$).  We predict that these graphs will have a cyclic flow to their dynamics, hitting each $\tau_i$ component in cyclic order.  Figures~\ref{fig:cycu-generalizations}B-D (bottom) give examples of directional cycles and their dynamics, as do Figures~\ref{fig:cyclic-union-dir-cycle-sa}B,D.  While we have not been able to explicitly prove this property of the dynamics, we can prove that all the fixed point supports have such a cyclic structure.

\begin{theorem}[cyclic fixed points of directional cycles] \label{thm:dir-cycle}
Let $G$ be a directional cycle with components $\tau_1, \ldots, \tau_N$.  Then for any $\sigma \in \FP(G)$, the graph $G|_\sigma$ contains an undirected cycle\footnote{An \emph{undirected cycle} is a sequence of nodes connected by edges that form a cycle within the underlying undirected graph, in which the direction of edges is simply ignored.} that intersects every $\tau_i$ in cyclic order. 
\end{theorem}

Observe that unlike the case of cyclic unions, in directional cycles we do not have the property that fixed points $\sigma$ of the full network restrict to fixed points $\sigma_i\od \sigma \cap \tau_i$ of the component subnetworks $G|_{\tau_i}$.  For example, in 
Figure~\ref{fig:cyclic-union-dir-cycle-sa}B, $\sigma = \{1,3,5\} \in \FP(G_2)$, but $\sigma_2 = \{3\} \notin \FP(G_2|_{\tau_2})$, since the only fixed point of $G_2|_{\tau_2}$ is the full-support $\{2,3,4\}$.  But Theorem~\ref{thm:dir-cycle} does guarantee that $\sigma_i \neq \emptyset$ for all $i \in [N]$.    It turns out that there is a key structural property of cyclic unions that guarantees the fixed points are unions of component graph fixed points: such networks have what we call a \emph{simply-embedded partition}.

\begin{figure}[!ht]
\begin{center}
\includegraphics[width=5in]{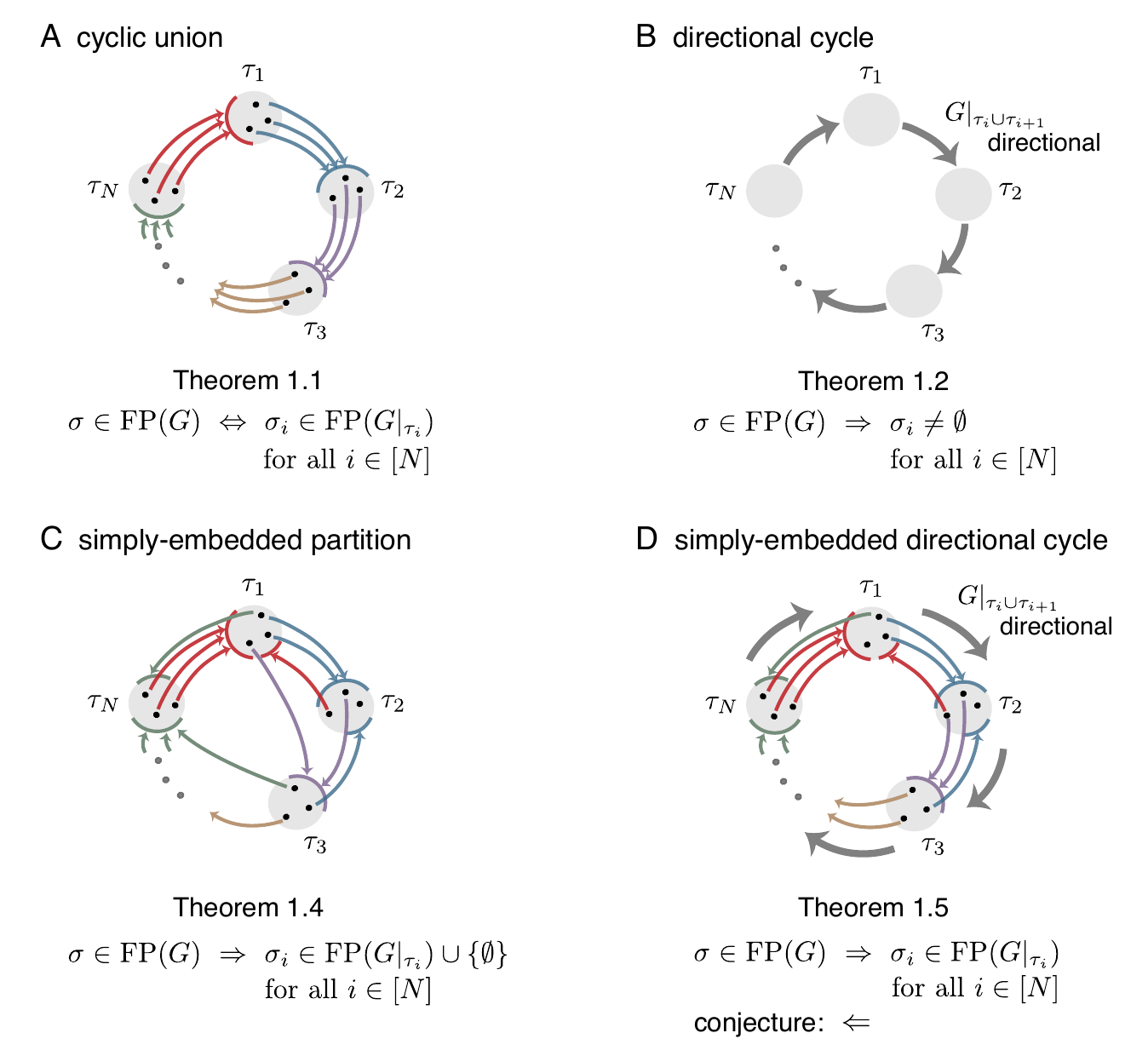}
\vspace{-.2in}
\end{center}
\caption{{\bf Summary of main results.} In each graph, colored edges from a node to a component indicate that the node projects edges out to all the nodes in the receiving component, as needed for a simply-embedded partition.  Thick gray edges indicate directionality of the subgraph $G|_{\tau_{i} \cup \tau_{i+1}}$.}
\label{fig:cyclic-generalizations-cartoon}
\vspace{-.2in}
\end{figure}

\paragraph{Simply-embedded partitions\\}

The notion of simply-added splits was introduced in \cite{fp-paper}, where it was shown that fixed points behave particularly nicely in networks that have this structure. Given a graph $G$ and a partition of its nodes into two components, $\{\omega | \tau\}$, we say that $\omega$ is \emph{simply-added} onto $\tau$ if for each $k \notin \omega$, either $k \to j$ for all $j \in \tau$ or $k \not\to j$ for all $j \in \tau$.  In this case, we say that $\tau$ is \emph{simply-embedded} in $G$. Here we introduce the more general notion of a simply-embedded partition. The key idea is that all nodes in a component $\tau_i$ are simply-embedded, and thus treated identically by each node outside that component.

\begin{definition}[simply-embedded partition]\label{def:sa-partition}
Given a graph $G$, a partition of its nodes $\{\tau_1|\cdots|\tau_N\}$ is called a \emph{simply-embedded partition} if every $\tau_i$ is simply-embedded in $G$.  In other words, for each $\tau_i$ and each $k \notin \tau_i$, either $k \to j$ for all $j \in \tau_i$ or $k \not\to j$ for all $j \in \tau_i$.  
\end{definition}

For the pair of graphs in Figure~\ref{fig:cyclic-union-dir-cycle-sa}C, $\{1\, |\, 2,3,4\, |\, 5\}$ is a simply-embedded partition: in each graph, the nodes $2, 3,$ and $4$ receive identical inputs from node $1$ as well as from node $5$.  For singleton components, the simply-embedded partition does not impose any constraints.  It turns out that this simply-embedded partition structure is sufficient to guarantee that all the fixed points of $G$ restrict to fixed points of the component subgraphs. This means that the fixed points of the components provide a kind of ``menu'' from which the fixed points of $G$ are made: each fixed point of the full network has support that is the union of component fixed point supports.

\begin{theorem}[$\FP(G)$ menu for simply-embedded partitions]\label{thm:menu}
Let $G$ have a simply-embedded partition $\{\tau_1|\cdots|\tau_N\}$.  For any $\sigma \subseteq [n]$, let $\sigma_i \od \sigma \cap \tau_i$.  Then 
$$\sigma \in \FP(G) \quad \Rightarrow \quad \sigma_i \in \FP(G|_{\tau_i})\cup \{\emptyset\}~~\text{ for all } i \in [N]. $$ 
In other words, every fixed point support of $G$ is a union of component fixed point supports $\sigma_i$, at most one per component.
\end{theorem}

\paragraph{Simply-embedded directional cycles\\}
While the simply-embedded partition generalizes one key property of $\FP(G)$ from cyclic unions, it does not guarantee that every fixed point intersects every component, nor does it guarantee the cyclic flow of the dynamics through the components, as we see in Figure~\ref{fig:cyclic-union-dir-cycle-sa}C.  But combining Theorems~\ref{thm:dir-cycle} and~\ref{thm:menu}, we immediately see that a \emph{simply-embedded directional cycle} has the desired fixed point properties while maintaining cyclic dynamics.

\begin{theorem}[simply-embedded directional cycles]\label{thm:sa-dir-cycle}
Let $G$ be a directional cycle whose components form a simply-embedded partition $\{\tau_1| \cdots|\tau_N\}$. For any $\sigma \subseteq [n]$, let $\sigma_i \od \sigma \cap \tau_i$.  Then 
\vspace{-.07in}
$$\sigma \in \FP(G) \quad \Rightarrow \quad \sigma_i \in \FP(G|_{\tau_i})~~\text{ for all } i \in [N].$$
In other words, every fixed point support of $G$ is a union of (nonempty) component fixed point supports, exactly one per component.
\end{theorem}

Figure~\ref{fig:cyclic-generalizations-cartoon} provides a simple visual summary of the different architectures generalizing the cyclic union, together with the main results on their fixed point supports.

We conjecture that the backwards direction of the statement in Theorem~\ref{thm:sa-dir-cycle} also holds, yielding an if and only if characterization of the fixed point supports. 

\begin{conjecture}\label{conj:sa-dir-cycle}
Let $G$ be a directional cycle whose components form a simply-embedded partition $\{\tau_1| \cdots|\tau_N\}$. Then 
\vspace{-.1in}
$$\sigma \in \FP(G) \quad \Leftrightarrow \quad \sigma_i \in \FP(G|_{\tau_i})~~\text{ for all } i \in [N].$$
In other words, $\FP(G)$ consists of all possible unions of (nonempty) component fixed point supports taking exactly one per component.
\end{conjecture}

\begin{figure}[!h]
\begin{center}
\includegraphics[width=.9\textwidth]{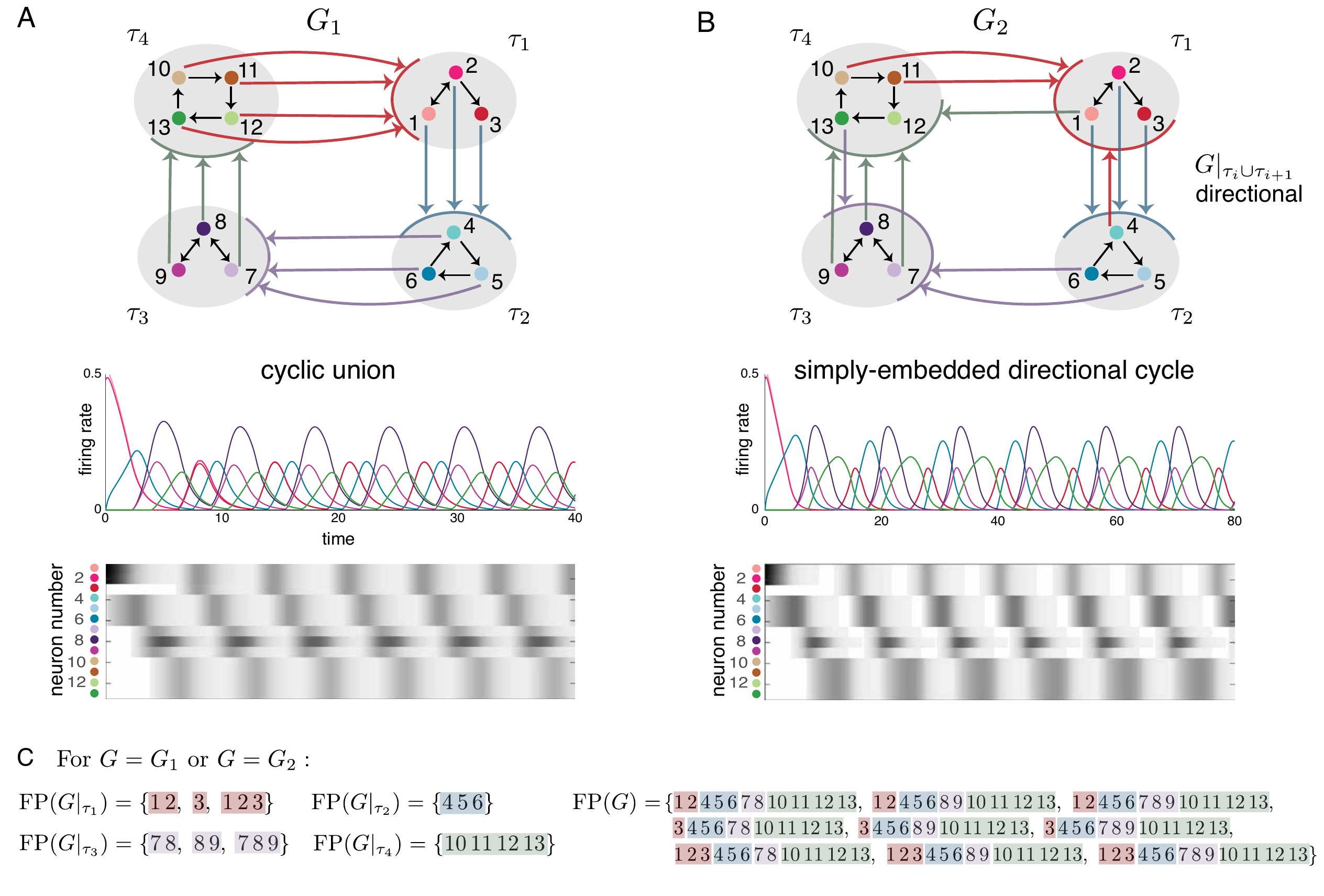}
\vspace{-.2in}
\end{center}
\caption{ \textbf{Cyclic union, simply-embedded directional cycle, their dynamics and $\boldsymbol{\FP(G)}$.} 
(A) (Top) A cyclic union $G_1$ with component subgraphs $G|_{\tau_1}, \ldots, G|_{\tau_4}$.  Thick colored edges from a node to a component indicate that the node projects edges out to all the nodes in the receiving component.  
(Bottom) A solution for the corresponding CTLN. 
(B) (Top) A simply-embedded directional cycle $G_2$ with the same component subgraphs as $G_1$.  (Bottom) A solution for the corresponding CTLN with the same initial condition as the network in panel A.  The solution of this network is qualitatively the same as that for the network in A except that the period is twice as long (note the different time axes).  (C) (Left) The fixed point supports $\FP(G|_{\tau_i})$ of each component subgraph.  (Right) $\FP(G)$ is identical for both $G_1$ and $G_2$: it consists of unions of component fixed point supports, exactly one per component. To highlight this structure of all the fixed point supports, the portion of each support that each $\tau_i$ is color-coded.
 }
\label{fig:dir-cycle-sa}
\vspace{-.15in}
\end{figure}

If the conjecture is true, then $\FP(G)$ for a simply-embedded directional cycle is identical to that of the cyclic union with the same component subgraphs.  While we cannot prove this conjecture in general, we have observed that it holds in computational analyses of over 10,000 simply-embedded directional cycles.  For example, Figure~\ref{fig:dir-cycle-sa} shows a larger example of a cyclic union and a simply-embedded directional cycle on the same component subgraphs, and $\FP(G)$ is identical for both networks.  Figure~\ref{fig:dir-cycle-sa} also shows an attractor for each network, and we see that the dynamics are qualitatively the same between the cyclic union and the simply-embedded directional cycle.  Interestingly, the activity of the simply-embedded directional cycle is significantly slower (see the differing time axes), with a period approximately twice as long as that of the perfect cyclic union.  

In the special case where $G$ is a simply-embedded directional cycle with a unique fixed point per component, Theorem~\ref{thm:sa-dir-cycle} shows that the only candidate fixed point support in $\FP(G)$ is the union of these component fixed point supports.  Since any network must have at least one fixed point \cite[Theorem 1]{fp-paper}, we immediately obtain the following result.

\begin{proposition}\label{prop:sa-dir-cycle}
Let $G$ be a directional cycle whose components form a simply-embedded partition $\{\tau_1| \cdots|\tau_N\}$, and suppose $\FP(G|_{\tau_i})$ has a unique fixed point for every $\tau_i$.  Then $\FP(G)$ has a unique fixed point with support $\sigma = \cup_{i=1}^N \sigma_i$, where $\sigma_i$ is the unique fixed point support in $\FP(G|_{\tau_i})$.
\end{proposition}

In particular, observe that Proposition~\ref{prop:sa-dir-cycle} implies that the conjecture holds in the special case where the components each have a unique fixed point.  As an illustration of this result, notice that the graphs in Figure~\ref{fig:cycu-generalizations}B-D and Figure~\ref{fig:cyclic-union-dir-cycle-sa}D are all simply-embedded directional cycles that have a unique fixed point with full support, which is the union of the unique full-support fixed points of each component.  For each of these graphs, their $\FP(G)$ coincides with that of the corresponding perfect cyclic union, and their dynamics are qualitatively identical.

\vspace{-.1in}
\paragraph{Roadmap\\}  
The remainder of the paper is organized as follows.  Section~\ref{sec:directional} focuses on constructions involving directional graphs. This includes \emph{directional chains}, which generalize synfire chains, as well as directional cycles.  The proof of Theorem~\ref{thm:dir-cycle} characterizing $\FP(G)$ for directional cycles is given in Section~\ref{sec:dir-cycle}.  

Section~\ref{sec:sa-graph-rules} is focused on simply-embedded partitions and the constraints they impose on $\FP(G)$.  The first section recaps Theorem~\ref{thm:menu} and illustrates it with some examples, then highlights some other interesting consequences about when a node is \emph{removable} from a network.  The remaining sections focus on graphs that have a simply-embedded partition together with some additional structure.  Section~\ref{sec:sa-dir-cycles} examines simply-embedded directional cycles and provides the proof of Theorem~\ref{thm:sa-dir-cycle}.  Section~\ref{sec:linear-chains} explores simple linear chains, which have a purely feedforward chain-like architecture, but without a guarantee of intrinsically feedforward activity (in contrast to directional chains).  Section~\ref{sec:bidir-sa-partitions} characterizes $\FP(G)$ for graphs with a \emph{strongly simply-embedded partition}.  The proofs of all the results in Section~\ref{sec:sa-graph-rules} require significant technical machinery, and are thus given in the Appendix: Sections~\ref{sec:menu-proof} --~\ref{sec:proof-bidir-sa} (except for the straightforward proof of Theorem~\ref{thm:sa-dir-cycle}).

Finally, in Section~\ref{sec:applications}, we analyze a number of networks of size $n=5$ to show how directional cycle graph architecture is predictive of the structure of corresponding sequential attractors, particularly when a graph has a simply-embedded directional cycle representation. From this analysis, we see that these architectures give insight into the sequences of neural activity that emerge, and not only the structure of fixed point supports.

\section{Directional graphs, chains, and cycles}\label{sec:directional}

In this section, we focus on generalizing cyclic unions in a way that preserves the cyclic dynamics of the associated attractor.  We refer to these networks as \emph{directional cycles}, which are built from component subgraphs where each consecutive pair form a \emph{directional graph}.  In order to make these notions more precise, we first provide a brief overview of key concepts about fixed points of CTLNs and some graph rules constraining the fixed point supports.  

\subsection{Preliminaries and prior graph rules} \label{sec:prior-graph-rules}
In this subsection we recall the results from \cite{fp-paper} that are relevant for this work.  A {\it fixed point} of a CTLN is simply a fixed point of the network equations~\eqref{eq:dynamics}. In other words, it is a vector $x^* \in \RR_{\geq 0}^n$ such that $\dfrac{dx_i}{dt}|_{x = x^*} = 0$ for all $i \in [n]$. The fixed points of CTLNs can be labeled by their {\it supports},
and for a given $G$ the set of all fixed point supports is denoted $\FP(G)=\FP(G, \varepsilon, \delta).$\footnote{As a slight abuse of notation, we typically omit the dependence of $\FP(G)$ on $\varepsilon$ and $\delta$ for simplicity.  Whenever a fixed point support can be determined using graph rules, its existence is independent of parameters, and thus this simplified notation is appropriate.}

In \cite{fp-paper}, multiple characterizations of $\FP(G)$ were developed for \emph{nondegenerate}\footnote{See Section~\ref{sec:fp-conditions} for the precise definition of nondegeneracy.} inhibitory threshold-linear networks in general as well as CTLNs specifically, including a variety of graph rules for CTLNs.  As an immediate consequence of one of these characterizations, it was shown that $\sigma$ is the support of a fixed point, i.e.\ $\sigma \in \FP(G)$, precisely when
\begin{enumerate}[label=(\arabic*)]
\item $\sigma \in \FP(G|_\sigma)$, and 
\item  $\sigma \in \FP(G|_{\sigma \cup \{k\}})$ for every $k \notin \sigma$.
\end{enumerate}
(See Appendix Section~\ref{sec:fp-conditions} and Corollary~\ref{cor:inheritance} for more details).  We say that $\sigma$ is a \emph{permitted motif} of $G$ when it is a fixed point of its restricted subnetwork, so that condition (1) holds.  And we say that a permitted motif $\sigma$ \emph{survives} to support a fixed point in the full network when condition (2) is satisfied.  Note that whether a subset $\sigma$ is permitted depends only on the subgraph $G|_\sigma$ (and potentially the choice of parameters $\varepsilon$ and $\delta$), while its survival will depend on the embedding of this subgraph in the full graph.  Importantly, condition (2) shows that survival can be checked one external node $k$ at a time.  Moreover, it turns out that the only aspect of the embedding that is relevant to survival is the edges from $\sigma$ \emph{out} to node $k$; the edges from $k$ back to $\sigma$ or to any other nodes in $G$ do not affect the survival of $\sigma$.

As our first example of permitted motifs, we consider {\it uniform in-degree} graphs. This family is particularly nice because the survival rules are parameter independent, and can be easily checked directly from the graph.

\begin{definition}
Let $G$ be a graph on $n$ nodes and $\sigma \subseteq [n]$.  We say that $G|_\sigma$ has {\it uniform in-degree} $d$ if every $i \in \sigma$ has in-degree $d_i^{\mathrm{in}} = d$ within $G|_\sigma$, i.e.\ every node $i\in \sigma $ has $d$ incoming edges from other nodes in $\sigma$.     
\end{definition}

\begin{figure}[!h]
\begin{center}
\vspace{-.1in}
\includegraphics[width=6in]{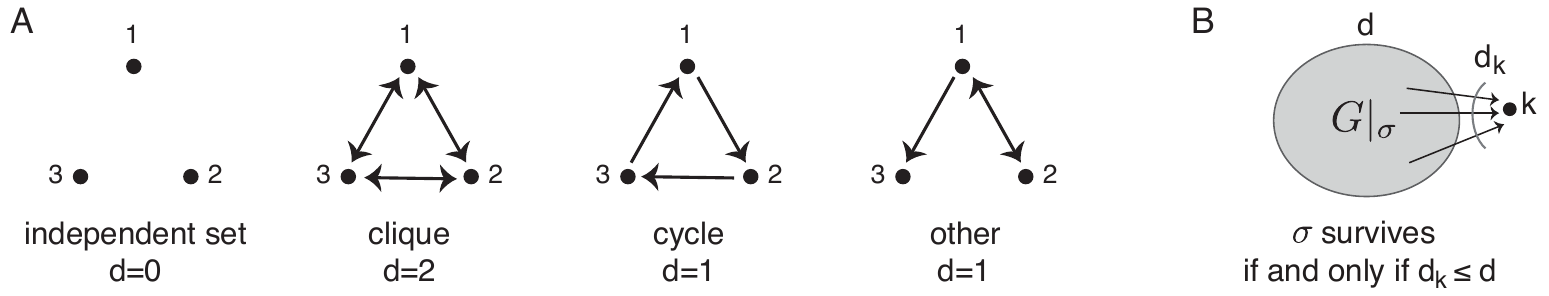}
\end{center}
\vspace{-.075in}
\caption{{\bf Uniform in-degree graphs.} (A) All $n=3$ graphs with uniform in-degree. (B) Cartoon showing survival rule for an arbitrary subgraph with uniform in-degree $d$.}
\label{fig:ufd-examples}
\vspace{-.1in}
\end{figure}

\begin{rules}[uniform in-degree \cite{fp-paper}]\label{rule:uniform-in-deg}
Let $G$ be a graph and suppose $G|_\sigma$ has uniform in-degree $d$.  For $k \notin \sigma$, let $d_k \od |\{i \in \sigma \mid i \to k\}|$ be the number of edges $k$ receives from $\sigma$. Then 
\vspace{-.05in}
$$\sigma \in \FP(G) \;\; \Leftrightarrow \;\; d_k \leq d \;\text{ for all }\; k \not\in \sigma.$$
\end{rules}

Figure~\ref{fig:ufd-examples}A shows all the uniform in-degree graphs of size $n=3$ together with some general graph theory terminology.  
Specifically, an {\it independent set} is a graph with uniform in-degree $d = 0$. A {\it $k$-clique} is an all-to-all bidirectionally connected graph with uniform in-degree $d = k-1$. An {\it $n$-cycle} is a graph with $n$ edges, $1 \to 2 \to \cdots \to n \to 1$, which has uniform in-degree $d = 1$.
Note that these families of uniform in-degree graphs are all cyclically symmetric; however, this is not necessary for uniform in-degree graphs in general, as the last graph in Figure~\ref{fig:ufd-examples}A shows.  Rule~\ref{rule:uniform-in-deg} guarantees that independent sets, cliques, and cycles all have a full-support fixed point. In fact, this fixed point is symmetric, with $x_i^* = x_j^*$ for all $i,j \in [n]$. This is true even for uniform in-degree graphs that are not symmetric.  Moreover, Rule~\ref{rule:uniform-in-deg} guarantees that these fixed points survive within a larger network whenever each external node receives only a limited number of inputs from the subnetwork.

More generally, fixed points can have very different values across neurons and their survival cannot be determined simply by the number of outgoing edges. However, there is some level of ``graphical balance'' that is required of $G|_\sigma$ for any fixed point support $\sigma$. For example, if $\sigma$ contains a pair of nodes $j,k$ that have the property that all nodes sending edges to $j$ also send edges to $k$, and $j \to k$ but $k \not\to j$, then $\sigma$ cannot be a fixed point support. This is because $k$ is receiving strictly more inputs than $j$, and this imbalance rules out their ability to coexist in the same fixed point support.  A similar analysis of relative inputs to different nodes can be used to determine fixed point survival in certain cases.  These ideas are made more precise below with the notion of \emph{graphical domination}, first introduced in \cite{fp-paper}.
	
\begin{definition}\label{def:graph-domination}
We say that $k$ \emph{graphically dominates} $j$ with respect to $\sigma$, and write $k >_\sigma j$, if $\sigma \cap \{j, k\} \neq \emptyset$ and the following three conditions all hold:
\begin{enumerate}[label=(\arabic*)]
\item for each $i \in \sigma \setminus \{j, k\}$, if $i \to j$ then $i \to k$,
\item if $j \in \sigma$, then $j \to k$, and 
\item if $k \in \sigma$, then $k \not\to j$.
\end{enumerate}
\end{definition}

\begin{figure}[!h]
\begin{center}
\includegraphics[width=5.1in]{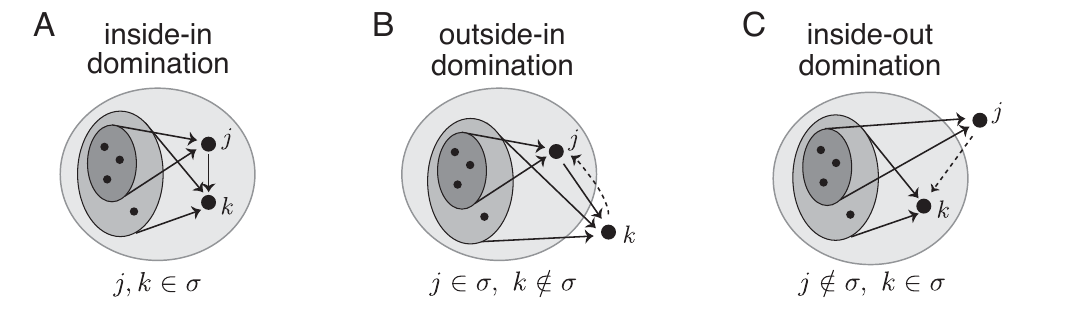}
\vspace{-.15in}
\end{center}
\caption{{\bf The three cases of graphical domination in Rule~\ref{rule:graph-domination}.} 
In each panel, $k$ graphically dominates $j$ with respect to $\sigma$ (the outermost shaded region). The inner shaded regions illustrate the subsets of nodes that send edges to $j$ and $k$. Note that the vertices sending edges to $j$ are a subset of those sending edges to $k$, but this
containment need not be strict. Dashed arrows indicate optional edges between $j$ and $k$.}
\label{fig:domination}
\vspace{-.1in}
\end{figure}

Figure~\ref{fig:domination} shows the three cases of domination in which we can conclude whether $\sigma$ supports a fixed point of the network.  Specifically, if there is inside-in domination (panel A), then $\sigma$ will not be a permitted motif, and thus $\sigma \notin \FP(G)$.  If there is outside-in domination by node $k$ (panel B), then $\sigma$ does not \emph{survive} the addition of node $k$, and so again $\sigma \notin \FP(G)$.  In contrast, if there is inside-out domination (panel C), then $\sigma$ is guaranteed to survive the addition of node $j$ whenever $\sigma$ is a permitted motif.  These cases were proven in \cite[Theorem 4]{fp-paper}, and are summarized below in Rule~\ref{rule:graph-domination}.

\begin{rules}[graphical domination \cite{fp-paper}]\label{rule:graph-domination}
Suppose $k$ graphically dominates $j$ with respect to $\sigma$.  Then the following statements all hold: 
\begin{enumerate}
 \item[a.] (inside-in) if $j,k \in \sigma$, then $\sigma \notin \FP(G|_\sigma)$, and thus $\sigma \notin \FP(G)$. 

 \item[b.] (outside-in) if $j \in \sigma$ and $k \notin \sigma$, then $\sigma \notin \FP(G|_{\sigma \cup \{k\}})$, and thus $\sigma \notin \FP(G)$. 

 \item[c.] (inside-out) if $k \in \sigma$ and $j \notin \sigma$, then $\sigma \in \FP(G|_{\sigma \cup \{j\}})$ if and only if $\sigma \in \FP(G|_\sigma)$.

 \end{enumerate}
\end{rules}

One case where graphical domination is guaranteed to exist is when $\sigma$ has a \emph{target}.  We say that $k$ is a \emph{target} of $\sigma$ if $i \to k$ for all $i \in \sigma \setminus \{k\}$.  Whenever $\sigma$ has a target node $k$, if $k \notin \sigma$, then we are guaranteed that $\sigma \notin \FP(G)$ by outside-in domination.  On the other hand, for $k \in \sigma$, if there is any node $j \in \sigma$ such that $k \not \to j$, then we have inside-in domination $k>_\sigma j$ and so again $\sigma \notin \FP(G)$.  

At the other extreme, if there is a $j \notin \sigma$ such that $j$ does not receive any edges from $\sigma$, then we are guaranteed that every $k \in \sigma$ inside-out dominates node $j$.  Thus, by Rule~\ref{rule:graph-domination}c, $\sigma$ survives the addition of node $j$ whenever $\sigma$ is a permitted motif.

Rules~\ref{rule:uniform-in-deg} and~\ref{rule:graph-domination} provide some graphical constraints on possible fixed point supports and on when a fixed point of a subnetwork survives to the full network.  Rule~\ref{rule:parity} provides one more constraint on $\FP(G)$.  Rule~\ref{rule:parity} is particularly useful for figuring out if there is a full-support fixed point whenever we know which proper subgraphs are permitted and which yield surviving fixed points.  Recall that these graph rules are for nondegenerate CTLNs.

\begin{rules}[parity \cite{fp-paper}]\label{rule:parity}
For any graph $G$, the total number of fixed points $|\FP(G)|$ is odd.
\end{rules}

\subsection{Directional graphs}
With this background in place, we can now precisely define directional graphs.  

\begin{definition}[directional graphs]\label{def:dir-graphs}
We say that a graph $G$ on $n$ nodes is \emph{directional, with direction $\omega \to \tau$}, if $\omega \cup \tau = [n]$ is a nontrivial partition of the nodes ($\omega, \tau \neq \emptyset$) such that the following property holds: for every $\sigma \not\subseteq \tau$, there exists some $j \in \sigma \cap \omega$ and $k \in [n]$ such that $k$ graphically dominates $j$ with respect to $\sigma$.  In particular, this property guarantees that $\FP(G) \subseteq \FP(G|_\tau)$ for all $\varepsilon, \delta$ in the legal range.\footnote{Note that we could guarantee $\FP(G) \subseteq \FP(G|_\tau)$ in a parameter-independent way without requiring that the dominated node $j \in \sigma$ live in $\omega$.  However, for Lemma~\ref{lemma:pairwise-chaining} (pairwise chaining) and Theorem~\ref{thm:dir-cycle} (directional cycles), we must further require $j \in \sigma \cap \omega$.}
\end{definition}

\begin{figure}[!ht]
\vspace{-.1in}
\begin{center}
\includegraphics[width=6.5in]{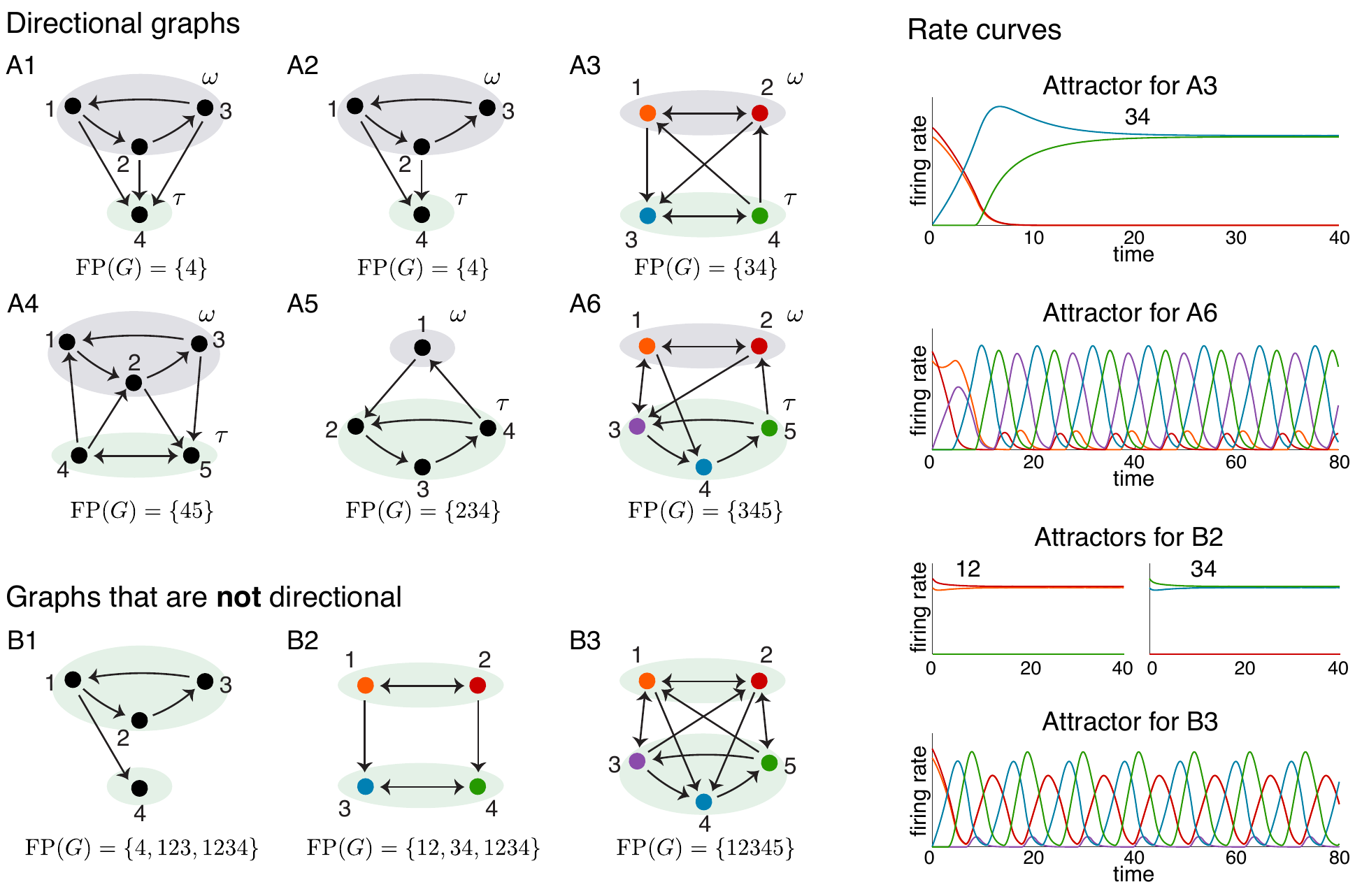}
\caption{{\bf Directional graphs: examples and non-examples.} (A) Example directional graphs and their $\FP(G)$.  On the right, solutions for A3 and A6 are shown where the activity was initialized on the nodes in $\omega$.  (B) Example non-directional graphs with their $\FP(G)$, as well as solutions for the networks in B2 and B3.
}
\label{fig:dir-graphs}
\end{center}
\vspace{-.25in}
\end{figure}

We call these graphs directional because in simulations we have observed that activity flows from $\omega$ to $\tau$, converging to an attractor concentrated on $\tau$.  The top panel of Figure~\ref{fig:dir-graphs} shows some example directional graphs.  Notice that each graph has a partition of the nodes $\omega \cup \tau$ such that all the fixed point supports are confined to $\tau$; moreover, each subset of nodes that intersects $\omega$ does not yield a fixed point as a result of graphical domination. 

\begin{example} 
Consider the graph $G$ in panel A3.  To see that $G$ is directional, observe that any $\sigma \subseteq \{1,2,3,4\}$ containing node 1 cannot support a fixed point because node 3 will dominate 1 with respect to $\sigma$ since (1) node 3 receives all the inputs that node 1 receives, (2) $1 \to 3$, and (3) $3 \not\to 1$.  Similarly, any $\sigma$ containing node 2 cannot support a fixed point since node 3 also dominates node 2.  Thus, every $\sigma \in \FP(G)$ must be a subset of $\tau$, and so $\FP(G) \subseteq \FP(G|_\tau)$ as a result of graphical domination.  

To the right of A3, we see the dynamics of the network when the activity has been initialized on the nodes in $\omega$.  The activity quickly flows from $\omega$ and converges to the stable fixed point supported on $\tau$, as predicted by the directionality of $G$.  This flow of activity occurs despite the multiple edges back from nodes in $\tau$ to nodes in $\omega$.  Similarly, graphs A4-A6 have equal numbers of forward and backward edges between $\omega$ and $\tau$, but in each case the dynamics flow towards $\tau$. In particular, the attractor for A6, obtained by initializing activity on $\omega$, is a sequential limit cycle supported on $\tau$.

In contrast, panels B1-B3 exhibit graphs that are \emph{not} directional: in particular, each one has a full-support fixed point. The graph in B2 is somewhat surprising, because it is similar to A3 but with a more obviously feedforward architecture. Dynamically, however, this graph is not directional and in fact supports two stable fixed point attractors that together involve all four nodes (see the attractors shown to the right).   Thus, feedforward architecture alone is not sufficient to guarantee feedforward dynamics.  Moreover, directional graphs can have feedforward dynamics even in the absence of feedforward architecture, as the graph in A3 demonstrates.
\end{example}

\begin{figure}[!ht]
\vspace{-.15in}
\begin{center}
\includegraphics[width=2.45in]{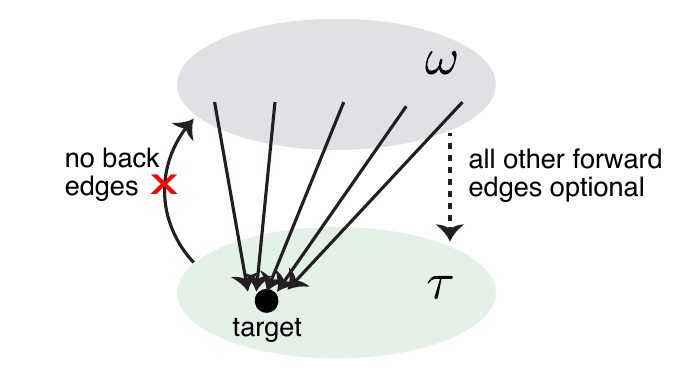}
\caption{{\bf Family of directional graphs.}}
\label{fig:dir-graph-family}
\end{center}
\vspace{-.25in}
\end{figure}

We can also see directional graphs inside cyclic unions.  For example, in Figure~\ref{fig:cyclic-union-dir-cycle-sa}A, the induced subgraphs $G|_{\tau_1 \cup \tau_2}$, $G|_{\tau_2 \cup \tau_3}$, and $G|_{\tau_3 \cup \tau_1}$ are all directional.  In fact, in any cyclic union, the induced subgraph on a pair of adjacent components is always directional.  Such subgraphs are a special case of the family of graphs in Figure~\ref{fig:dir-graph-family}, where the $\tau$ component contains a target of $\omega$ and there are no back edges from $\tau$ to $\omega$.

\begin{lemma}\label{lemma:dir-graph-family}
Suppose $G$ has a nontrivial partition of its nodes $\omega \cup \tau=[n]$ such that there are only forward edges from $\omega$ to $\tau$ and at least one node in $\tau$ is a target of $\omega$ (as in Figure~\ref{fig:dir-graph-family}).  Then $G$ is directional with direction $\omega \to \tau$.
\end{lemma}

\begin{proof}
Let $k \in \tau$ be a target of $\omega$.  Consider any $\sigma \not\subseteq \tau$ and let $j \in \sigma \cap \omega$.  We will show that the target node $k$ graphically dominates $j$ with respect to $\sigma$.  First we must show that for all $i \in \sigma \setminus \{j,k\}$, if $i \to j$ then $i \to k$.  Since there are no back edges from $\tau$ to $\omega$, the only $i \in \sigma$ with $i \to j$ are $i \in \sigma \cap \omega$.  But $k$ is a target of $\omega$, and so $i \in \omega$ implies that $i \to k$.  Thus condition 1 of graphical domination holds.   Moreover, $j \to k$ since $k$ is a target, and $k \not\to j$ since there are no back edges from $\tau$.  Thus, conditions 2 and 3 holds as well, and so $k$ dominates $j$ with respect to $\sigma$.  Therefore $G$ is directional with direction $\omega \to \tau$.
\end{proof}

\subsection{Directional chains}\label{sec:dir-chains}

\begin{figure}[!ht]
\vspace{-.1in}
\begin{center}
\includegraphics[height=2in]{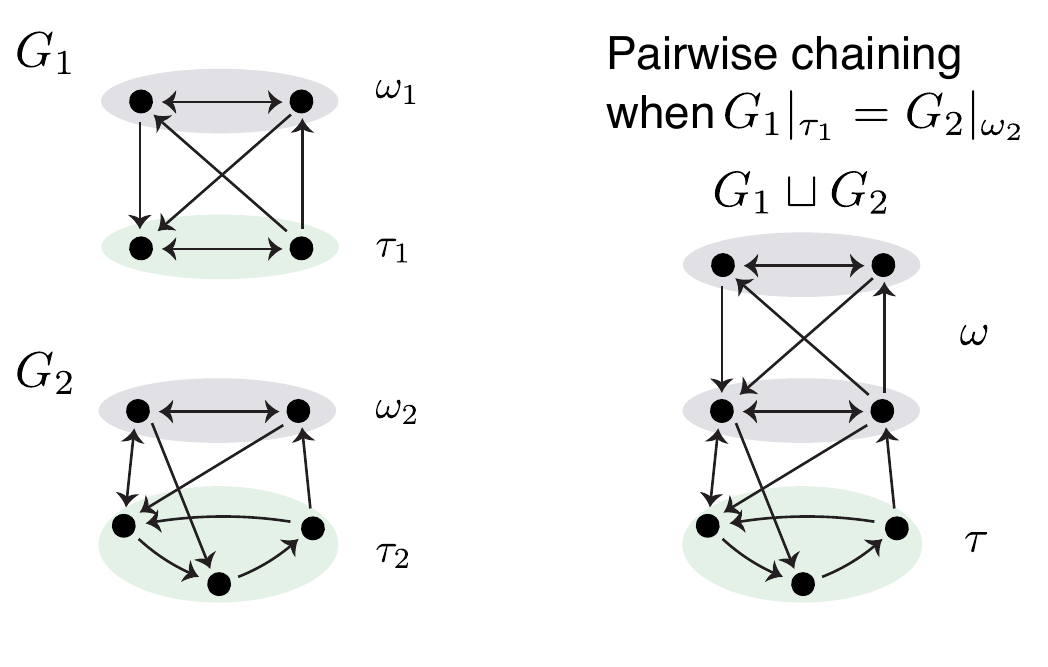}
\vspace{-.05in}
\caption{{\bf Pairwise chaining of two directional graphs.} (Left) Two directional graphs $G_1$ and $G_2$ with direction $\omega_i \to \tau_i$, where $G_1|_{\tau_1} = G_2|_{\omega_2}$.  (Right) The graph $G_1 \sqcup G_2$ formed from chaining together $G_1$ and $G_2$ by identifying ${\tau_1}$ with ${\omega_2}$.  By Lemma~\ref{lemma:pairwise-chaining}, $G_1 \sqcup G_2$ is directional for $\omega = \omega_1 \cup \omega_2$ (in gray) and $\tau=\tau_2$ (in green).}
\label{fig:pairwise-chaining}
\end{center}
\vspace{-.25in}
\end{figure}

One of the valuable features of directional graphs is that we can chain them together to produce new directional graphs. Namely,
if the graphs overlap so that the $\tau$ part of one graph coincides with the $\omega$ part of the next one, then the resulting ``chain" graph is also directional (see Figure~\ref{fig:pairwise-chaining}).

\begin{lemma}[pairwise chain] \label{lemma:pairwise-chaining}
Suppose $G_1$ and $G_2$ are directional graphs with directions $\omega_1 \to \tau_1$ and $\omega_2 \to \tau_2$, respectively, that satisfy $G_1|_{\tau_1} = G_2|_{\omega_2}$. Consider the pairwise chain $G_1 \sqcup G_2$ formed by identifying ${\tau_1}$ with ${\omega_2}$ (as in Figure~\ref{fig:pairwise-chaining}).
Then $G_1 \sqcup G_2$ is directional with $\omega=\omega_1 \cup \omega_2$ and $\tau = \tau_2$.  
\end{lemma}
\begin{proof}
Let $G \od G_1 \sqcup G_2$ be the pairwise chain with vertex set $[n] = \omega_1 \cup \omega_2 \cup \tau_2$ (where $\tau_1$ was identified with $\omega_2$), and let $\omega \od \omega_1 \cup \omega_2$, and $\tau \od \tau_2$.  Consider $\sigma \subseteq [n]$ with $\sigma \cap \omega \neq \emptyset$.  We will show that there exists a $j \in \sigma \cap \omega$ and $k \in [n]$ such that $k$ graphically dominates $j$ with respect to $\sigma$.  Observe that if $\sigma \subseteq \omega_1 \cup \omega_2 = \omega_1 \cup \tau_1$ and $\sigma \cap \omega_1 \neq \emptyset$, then such a $j$ and $k$ pair exist within $G_1$ since $G_1$ is directional; the same holds if $\sigma \subseteq \omega_2 \cup \tau_2$.  Thus we need only consider $\sigma$ that overlaps both $G_1$ and $G_2$, without being fully contained in either.  In other words, we have $\sigma \cap \omega_1 \neq \emptyset$ and $\sigma \cap \tau_2 \neq \emptyset$.  

Let $\sigma_1 \od \sigma \cap (\omega_1 \cup \tau_1)$ be $\sigma$ restricted to $G_1$.  By the directionality of $G_1$, there exists $j \in \sigma_1 \cap \omega_1$ and $k \in \omega_1 \cup \tau_1$ such that $k$ graphically dominates $j$ with respect to $\sigma_1$.  We will show that $k$ also dominates $j$ with respect to the full $\sigma$ in $G$.  First observe that conditions (2) and (3) of graphical domination are automatically satisfied for $\sigma$ by way of being satisfied for $\sigma_1$.  For condition (1), we must show that for all $i \in \sigma \setminus \{j,k\}$, if $i \to j$, then $i \to k$.  Since $j \in \omega_1$, the only possible nodes in $G$ that can send edges to $j$ are nodes in $G_1$, since $\omega_1$ is not in the overlap with $G_2$ so cannot receive edges from any nodes in $G_2$ outside of that overlap.  Thus, the only $i \in \sigma$ with $i \to j$ are nodes within $\sigma_1$, and for all $i \in \sigma_1 \setminus \{j,k\}$, whenever $i \to j$, we have $i \to k$ by the graphical domination relationship with respect to $\sigma_1$.  Therefore, condition (1) holds for all of $\sigma$, and so $k$ dominates $j$ with respect to $\sigma$ in $G$.  Thus, $G$ is directional with $\omega \to \tau$.  
\end{proof}

\begin{figure}[!ht]
\vspace{-.1in}
\begin{center}
\includegraphics[height=1.25in]{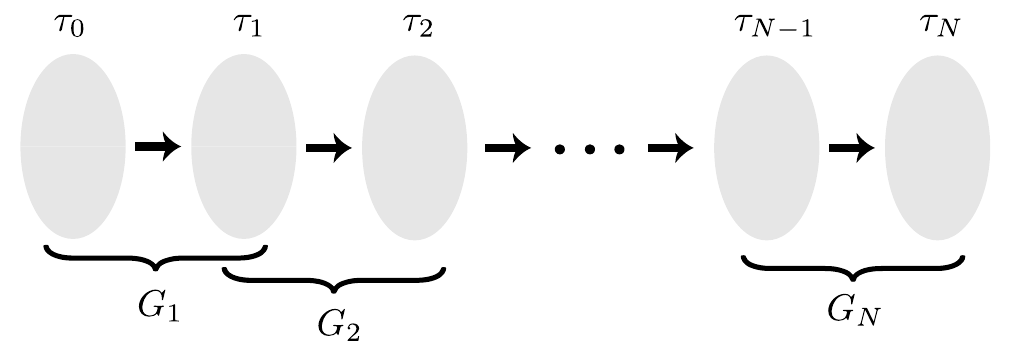}
\caption{{\bf Directional chain.} A cartoon of a directional chain with components $\tau_0, \tau_1, \ldots, \tau_N$.   For each $i=1, \ldots, N$, the induced subgraph $G_i \od G|_{\tau_{i-1} \cup \tau_i}$ is directional.  The arrows between components indicate the directionality $\tau_{i-1} \to \tau_i$.  Note there may be edges in both directions between adjacent components, as in the example directional graphs in Figure~\ref{fig:dir-graphs}, but there are no edges between non-adjacent components.
}
\label{fig:dir-chain-cartoon}
\end{center}
\vspace{-.25in}
\end{figure}

Lemma~\ref{lemma:pairwise-chaining} motivates the following definition of a \emph{directional chain}, obtained from iteratively chaining directional graphs together (see Figure~\ref{fig:dir-chain-cartoon}).  

\begin{definition}[directional chain]
Let $G$ be a graph with a partition of its nodes $\{\tau_0|\tau_1| \cdots|\tau_N\}$.  For each $i=1, \ldots, N$, let $G_i \od G|_{\tau_{i-1} \cup \tau_i}$ be the induced subgraph on adjacent components. We say that $G$ is a \emph{directional chain} if each $G_i$ is directional with direction $\tau_{i-1} \to \tau_i$, and every edge of $G$ is an edge in some $G_i$ (i.e., there are no edges between nonadjacent $\tau_i$ components).  
\end{definition}

Iteratively applying Lemma~\ref{lemma:pairwise-chaining}, we immediately obtain the following result showing that every directional chain is directional.

\begin{proposition}[directional chain]\label{prop:directional-chain}
Let $G$ be a directional chain with components $\tau_0, \tau_1, \ldots, \tau_N$ and directional graphs $G_i = G|_{\tau_{i-1} \cup \tau_i}$.
Then $G$ is directional with direction $\omega \to \tau$ for $\omega = \tau_0 \cup \cdots \cup \tau_{N-1}$ and $\tau = \tau_N$.  In particular, $\FP(G) \subseteq \FP(G|_{\tau_N}).$
\end{proposition}
\begin{proof}
For each $G_i \od G|_{\tau_{i-1} \cup \tau_i}$, denote the directional components of $G_i$ as $\omega_i$ and $\tau_i$, as in Lemma~\ref{lemma:pairwise-chaining}, so that $\omega_i = \tau_{i-1}$.  
Observe $G_{12} \od G_1 \sqcup G_2$ is a pairwise chain, so by Lemma~\ref{lemma:pairwise-chaining}, $G_{12}$ is directional with $\omega_{12} = \omega_1 \cup \omega_2$ and $\tau_{12} = \tau_2$.  Similarly, $G_{123} \od (G_1 \sqcup G_2) \sqcup G_3 = G_{12} \sqcup G_3$ is also a pairwise chain, and so by Lemma~\ref{lemma:pairwise-chaining}, $G_{123}$ is directional with $\omega_{123} = \omega_{12} \cup \omega_3 = \omega_1 \cup \omega_{2} \cup \omega_3 $ and $\tau_{123} = \tau_3$.  We can continue iterating in this fashion to see $G$ is a pairwise chain of directional graphs $G_{1 \cdots N-1} \sqcup G_N$, and thus by Lemma~\ref{lemma:pairwise-chaining}, $G$ is directional with direction $\omega \to \tau$ for $\omega = \omega_{1\cdots N-1} \cup \omega_N =  \tau_0 \cup \cdots \cup \tau_{N-1}$ and $\tau = \tau_N$.  
\end{proof}

\begin{figure}[!ht]
\vspace{-.1in}
\begin{center}
\includegraphics[width=5.8in]{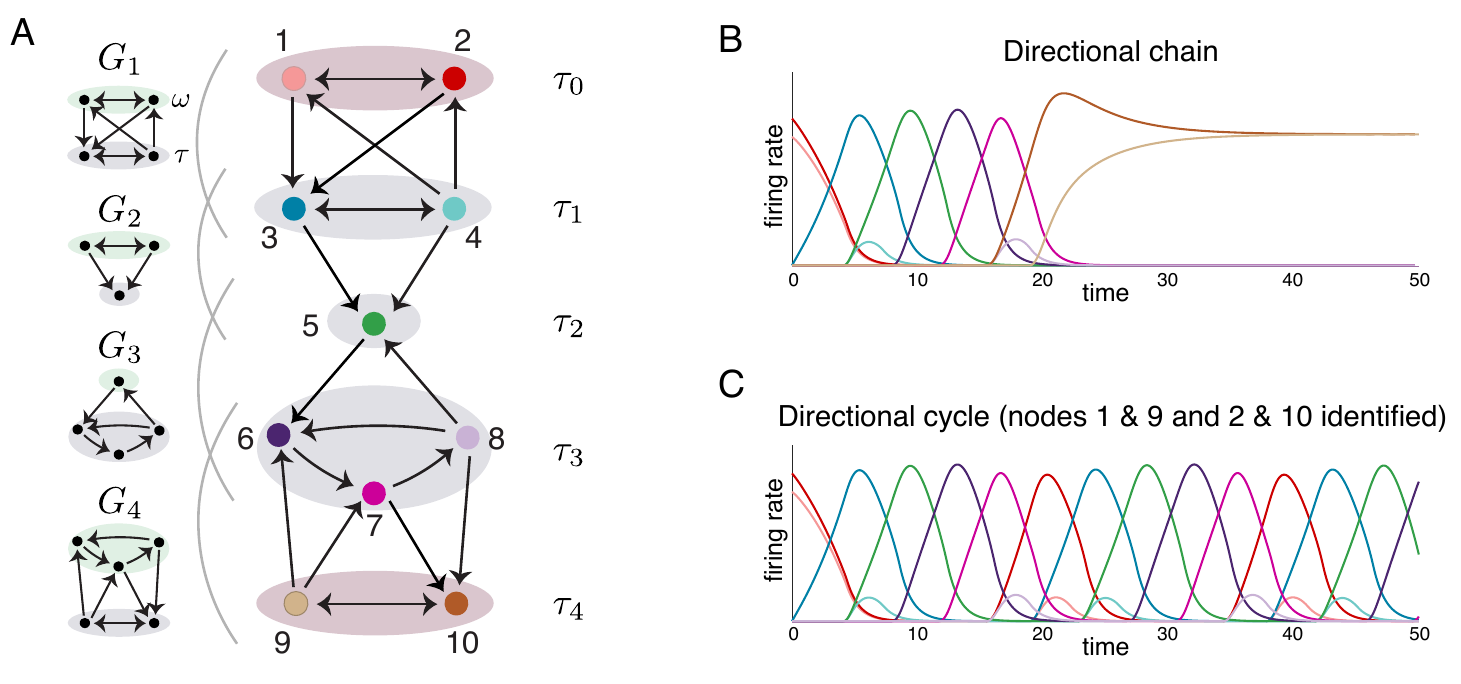}
\caption{{\bf Directional chain and directional cycle.} (A) A graph built from chaining together directional graphs $G_1, \ldots, G_4$, where $G_i = G|_{\tau_{i-1} \cup \tau_i}$. (B) A solution of the CTLN for the directional chain in A when the activity is initialized on nodes $1$ and $2$.  Activity flows through the chain, eventually stabilizing on the fixed point attractor of $\tau_4$. (C) A solution of the CTLN for the directional cycle obtained from the graph in A by identifying $\tau_4$ with $\tau_0$ to make the chain wrap cyclically wrap around.  The activity is initialized on nodes $1$ and $2$ and falls into a limit cycle hitting all the components $\tau_i$ in cyclic order. }
\label{fig:dir-chain-cycle}
\end{center}
\vspace{-.25in}
\end{figure}

By Proposition~\ref{prop:directional-chain}, we see that for any CTLN whose graph is a directional chain $G$, we must have $\FP(G) \subseteq \FP(G|_{\tau_N})$. In other words, all fixed points are confined to the last $\tau$ of the chain.  Figure~\ref{fig:dir-chain-cycle}A gives an example of such a chain 
built from directional graphs $G_1, \ldots, G_4$ where $G_i|_{\tau_i} = G_{i+1}|_{\tau_i}$ for each $i=1, \ldots, 3$. 
In Figure~\ref{fig:dir-chain-cycle}B, we see the resulting dynamics when the activity of the network is initialized on nodes $1$ and $2$, at the start of the chain. We see a clear sequence of activation, from $1$ and $2$ to $3, 5, 6,$ and $7$, and then stabilizing on the fixed point attractor for the clique $\{9,10\}$. In other words, the activity flows along the directional chain, generating a sequence that reflects the directionality of the construction. Note that the network behaves in the expected feedforward manner dynamically despite the existence of several feedback edges: $4 \to 1, 2$, $8 \to 5$, and $ 9 \to 7$.  

We see that directional chains produce sequences of neural activity in their transient dynamics, similar to that of synfire chains \cite{synfire-chain1, synfire-chain2, synfire-chain3}.  In contrast to synfire chains, though, directional chains can have recurrent connectivity throughout and do not rely on a purely feedforward architecture.

\subsection{Directional cycles}\label{sec:dir-cycle}

We can also chain directional graphs together in a cyclic manner, so the directional chain wraps around and $\tau_N$ is identified with $\tau_0$.   We call any graph $G$ that can be created in this way a \emph{directional cycle}. 

\begin{definition}[directional cycle]
Let $G$ be a graph with node partition $\{\tau_1| \cdots|\tau_N\}$.  For each $i=1, \ldots, N$, let $G_i \od G|_{\tau_{i-1} \cup \tau_i}$ be the induced subgraph on adjacent components (cyclically identifying $\tau_N = \tau_0$). We say that $G$ is a \emph{directional cycle} if each $G_i$ is directional with direction $\tau_{i-1} \to \tau_i$, and every edge of $G$ is an edge in some $G_i$ (i.e., there are no edges between nonadjacent $\tau_i$ components).
\end{definition}

For directional cycles, the chain has no beginning or end and so the fixed points cannot all lie in some final $\tau_N$. Instead, they become highly distributed across the network, intersecting each and every $\tau_i$.  In particular, in Theorem~\ref{thm:dir-cycle}, we show that every fixed point support of a directional cycle contains an \emph{undirected cycle}\footnote{To any directed graph $G$, we can associate a simple undirected graph $\widehat{G}$ by ignoring the direction on the edges. An \emph{undirected cycle} is a sequence of nodes connected by edges that form a cycle within the underlying undirected graph.  For example, $2458$ is an undirected cycle in the graph in Figure~\ref{fig:dir-chain-cycle}A when node 10 is identified with node 2.} that hits each $\tau_i$ in cyclic order.  

Figure~\ref{fig:dir-chain-cycle} provides an illustration of this. In the directional chain of panel A, suppose we identify $\tau_4$ with $\tau_0$, so nodes 1 and 9 are identified as are 2 and 10. Then the resulting network becomes a directional cycle. Figure~\ref{fig:dir-chain-cycle}C shows the activity obtained by initializing on nodes 1 and 2. We see a clear and repeating sequence of activity emerge, corresponding to the cycle 23567 in the graph, whose existence is predicted by Theorem~\ref{thm:dir-cycle}. Note that for this network 
 $\FP(G) = \{23567\}$ with the unique fixed point corresponding to the cycle motif giving rise to the sequence.
In other words, directional cycles produce periodic sequences of activity that cycle around the chain in the expected direction.

The remainder of this section is dedicated to the proof of Theorem~\ref{thm:dir-cycle} (reprinted below).\\

\begin{figure}[!ht]
\vspace{-.1in}
\begin{center}
\includegraphics[height=2.25in]{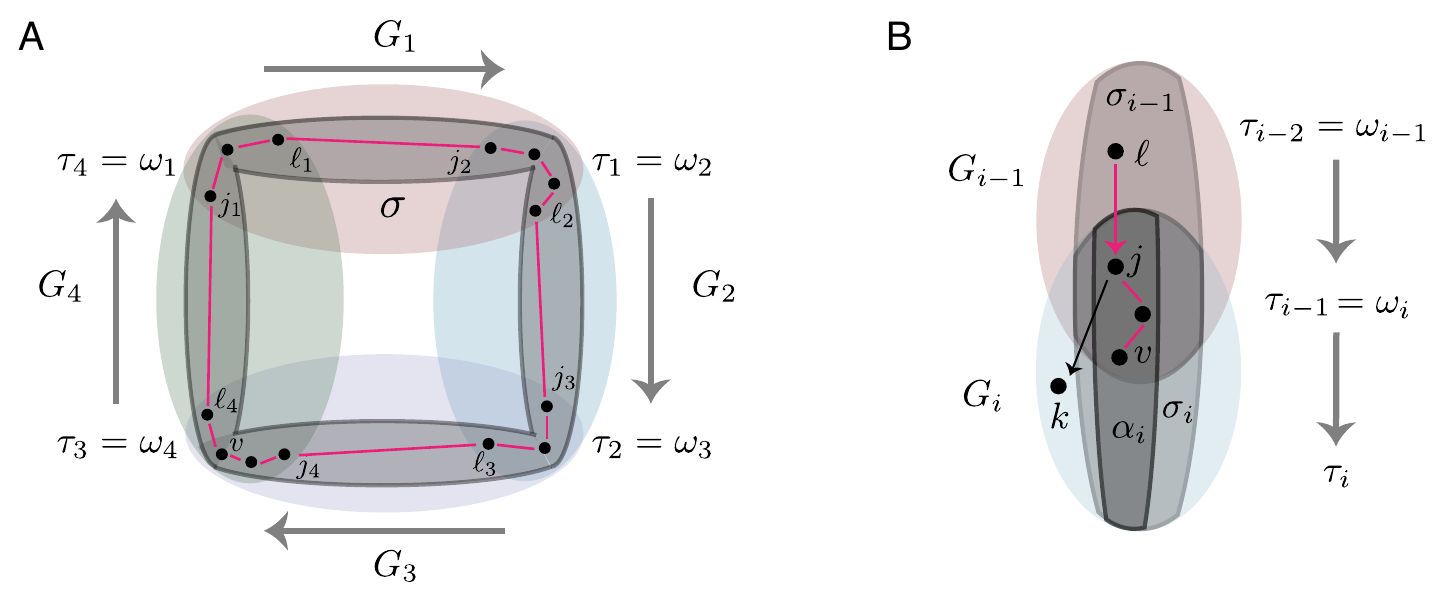}
\vspace{-.05in}
\caption{{\bf Illustrations for Theorem~\ref{thm:dir-cycle} and Lemma~\ref{lemma:pull-back}.} (A) A cartoon of a directional cycle; each pastel colored blob is a directional graph $G_i$ with direction $\omega_i=\tau_{i-1} \to \tau_i$ indicated by arrows along the outside.  Note that all vertices of $G$ lie within an overlap of adjacent $G_i$, but each $G_i$ has edges between the two overlaps $\tau_{i-1}$ and $\tau_i$.  Within the directional cycle, a fixed point support $\sigma \in \FP(G)$ is shown in dark gray.  Theorem~\ref{thm:dir-cycle} guarantees that $G|_\sigma$ contains an undirected cycle that hits all the $\tau_i$ in cyclic order (shown in magenta).  The vertices in the cycle are labeled following the notation for the proof of Theorem~\ref{thm:dir-cycle}. (B) Cartoon for set up of Lemma~\ref{lemma:pull-back}.  The pale pink and blue blobs depict overlapping directional graphs $G_{i-1}$ and $G_i$.  The restriction of fixed point support $\sigma \in \FP(G)$ to this subgraph is shown with its component subgraphs $\sigma_i \od \sigma \cap (\omega_i \cup \tau_i)$ denoted with light gray blobs.  The subgraph $G|_{\sigma_i}$ can be broken into its connected components, and $\alpha_i$ (dark gray) denotes one such component.  There exists a $j \in \alpha_i \cap \omega_i$ such that $j$ is dominated by some $k$ in $G_i$ with respect to $\alpha_i$.  Then  Lemma~\ref{lemma:pull-back} guarantees that there is some $\ell \in \sigma_{i-1} \cap \omega_{i-1}$ such that $\ell \to j$.}
\label{fig:directional-cycle-cartoon}
\end{center}
\vspace{-.25in}
\end{figure}

\noindent{\bf Theorem~\ref{thm:dir-cycle} }(cyclic fixed points of directional cycles).
Let $G$ be a directional cycle with components $\tau_1, \ldots, \tau_N$ and directional graphs $G_i = G|_{\tau_{i-1} \cup \tau_i}$ (cyclically identifying $\tau_N = \tau_0$).
Then for any $\sigma \in \FP(G)$, the graph $G|_\sigma$ contains an undirected cycle that intersects every $\tau_i$ in cyclic order (see illustration in Figure~\ref{fig:directional-cycle-cartoon}A).\\

To prove Theorem~\ref{thm:dir-cycle}, we first need the following lemma that shows that for any fixed point support $\sigma$ of a directional cycle, there is always an edge feeding into $\sigma_i$ ($\sigma$ restricted to the graph $G_i$) from the previous graph $G_{i-1}$.

\begin{lemma}\label{lemma:pull-back}
Let $G$ be a directional cycle with components $\tau_1, \ldots, \tau_N$ and directional graphs $G_i = G|_{\tau_{i-1} \cup \tau_i}$ (cyclically identifying $\tau_N = \tau_0$).   For each $G_i$, let $\omega_i = \tau_{i-1}$, so that $G_i$ has direction $\omega_i \to \tau_i$.
For $\sigma \in \FP(G)$, let $\sigma_i \od \sigma \cap (\omega_i \cup \tau_i)$ denote $\sigma$ restricted to graph $G_i$.  For any $v \in \sigma_i \cap \omega_i$, there exists $j \in \sigma_i \cap \omega_i$ ($j$ could equal $v$) and an $\ell \in \sigma_{i-1} \cap \omega_{i-1}$ such that $\ell \to j$ in $G$ and there is an undirected path from $v$ to $j$ in $\sigma_i$ (see illustration in Figure~\ref{fig:directional-cycle-cartoon}B).
\end{lemma}

\begin{proof}
Let $\sigma \in \FP(G)$, $\sigma_i = \sigma \cap (\omega_i \cup \tau_i)$, and let $\alpha_i$ be the connected component\footnote{In a slight abuse of language, we use connected component here to refer to the connected component of the  undirected graph associated to $G$.  Thus, a connected component consists of all nodes that are reachable by undirected paths, where the direction of edges in $G$ is ignored.} of $\sigma_i$ that contains $v$, so that $\alpha_i \cap \omega_i \neq \emptyset$.  Since $G_i$ is directional, there exists $j \in \alpha_i \cap \omega_i$ and $k \in \omega_i \cup \tau_i$ such that $k$ graphically dominates $j$ with respect to $\alpha_i$.  Since $\sigma \in \FP(G)$, we cannot have any $j$ graphically dominated by $k$ with respect to all of $\sigma$ in $G$, by Rule~\ref{rule:graph-domination}. Thus, there must exist some $\ell \in \sigma$ such that $\ell \to j$ but $\ell \not\to k$ (in order to violate condition (1) of the definition of graphical domination).  Moreover, we must have $\ell \in \tau_{i-2}$, $\tau_{i-1}$, or $\tau_i$, since $j \in \omega_i = \tau_{i-1}$ and $\ell \to j$, and there are no edges between nonadjacent components in a directional cycle.  
We cannot have $\ell$ in $G_i$, i.e., $\ell \notin \tau_{i-1} \cup \tau_i$, since there are no nodes in $\sigma_i \setminus \alpha_i$ that send edges into $\alpha_i$, by definition of connected component.  Thus, we must have $\ell \in \tau_{i-2} = \omega_{i-1}$. Therefore, we have $\ell \in \sigma_{i-1} \cap \omega_{i-1}$ and $j \in \alpha_i \cap \omega_i$ such that $\ell \to j$.
\end{proof}

We can now prove Theorem~\ref{thm:dir-cycle}, using Lemma~\ref{lemma:pull-back} to trace a path in $\sigma$ backwards through the directional cycle, demonstrating the existence of an undirected cycle in $\sigma$ that hits every $\tau_i$ in cyclic order.

\begin{proof}[\textbf{Proof of Theorem~\ref{thm:dir-cycle}}]
To set notation, for each $G_i = G|_{\tau_{i-1} \cup \tau_i}$, denote the directional components of $G_i$ as $\omega_i$ and $\tau_i$, so that $\omega_i = \tau_{i-1}$.   For $\sigma \subseteq [n]$, let $\sigma_i \od \sigma \cap (\omega_i \cup \tau_i)$ denote the restriction of $\sigma$ to the graph $G_i$.

Let $\sigma \in \FP(G)$ and let $v \in \sigma$.  Observe that $v \in \omega_i$ for some graph $G_i$ (since every node in $G$ is contained in some $\omega_i = \tau_{i-1}$).  Without loss of generality, let $v \in \omega_N$.  By Lemma~\ref{lemma:pull-back}, there exists a $j_{N}  \in \sigma_N \cap \omega_N$ and $\ell_{N-1} \in \sigma_{N-1} \cap \omega_{N-1}$ such that $\ell_{N-1} \to j_{N}$ and there is an undirected path from $v$ to $j_{N}$.  Next, consider $\ell_{N-1}$ playing the role of $v$ in $\sigma_{N-1} \cap \omega_{N-1}$.  We can again apply Lemma~\ref{lemma:pull-back} to obtain a $j_{N-1}  \in \sigma_{N-1} \cap \omega_{N-1} $ and $\ell_{N-2} \in  \sigma_{N-2} \cap \omega_{N-2}$ such that $\ell_{N-2} \to j_{N-1}$ and there is an undirected path from $\ell_{N-1}$ to $j_{N-1}$. Thus, we have an undirected path from $v$ to $ j_{N}$ to $ \ell_{N-1}$ to $ j_{N-1}$ and finally $\ell_{N-2}$ (see Figure~\ref{fig:directional-cycle-cartoon}A starting in the bottom left $\omega_4$).

Continuing in this manner, we see that  $G|_\sigma$ has an undirected path containing all these $j_{i} \in \omega_i =\tau_{i+1}$, and hitting each of the intersections $\tau_i$ in cyclic order.  To see that this path can eventually be closed to yield a cycle, notice that we can keep following this path backwards from $G_i$ to $G_{i-1}$ as it wraps around $G$, since every $\sigma_i$ on this path must have some edge into it from $\sigma_{i-1}$ that can be followed backwards.  Since each $\sigma_i$ has a finite number of connected components, by the pigeonhole principle, the path through $\sigma$ must at some point revisit a connected component $\alpha_i$ for some $i$.  Since $\alpha_i$ is connected, we can close our cycle by walking from the current node on the path through the component to the node previously visited in an earlier portion of the path.  Thus we have found an undirected path through $\sigma$ that starts and ends at the same point in some $\sigma_i$, yielding an undirected cycle that hits every $\tau_j$ in cyclic order.
\end{proof}

\section{Simply-embedded structure and graph rules}\label{sec:sa-graph-rules}
In this section, we focus on graphs with a \emph{simply-embedded partition}.  These graphs generalize cyclic unions in a way that preserves strong constraints on $\FP(G)$.  We begin by considering simply-embedded partitions in their full generality, and then move to some families of graphs that have additional structure.
Note that most of the results in this section require significant technical machinery to prove (initially developed in \cite{fp-paper}), and thus we save the proofs for the Appendix: Sections~\ref{sec:menu-proof} --~\ref{sec:proof-bidir-sa}.


\subsection{Simply-embedded partitions}\label{sec:simply-embedded}

Recall that a simply-embedded partition of a graph is a partition of the nodes such that all nodes within a single component are treated identically by any node outside of that component.  More precisely:\\

\noindent {\bf Definition~\ref{def:sa-partition}} (simply-embedded partition).
Given a graph $G$, a partition of its nodes $\{\tau_1|\cdots|\tau_N\}$ is called a \emph{simply-embedded partition} if every $\tau_i$ is simply-embedded in $G$.  In other words, for each $\tau_i$ and each $k \notin \tau_i$, either $k \to j$ for all $j \in \tau_i$ or $k \not\to j$ for all $j \in \tau_i$.   \\

 Notice that the definition is trivially satisfied in the cases where (a) there are no $k \notin \tau_i$ or (b) there is only a single $j\in \tau_i$ for every $i$.  Thus, every graph has two trivial simply-embedded partitions: one where all the nodes are in one component and one where every node is in its own component.  Neither of these partitions is useful for giving information about the structure of $G$.  But when a graph has a nontrivial simply-embedded partition, this structure is sufficient to dramatically constrain the possible fixed point supports of $G$ to unions of fixed points chosen from a \emph{menu} of component fixed point supports, $\FP(G|_{\tau_i})$.  
 \\

\noindent{\bf Theorem~\ref{thm:menu}} ($\FP(G)$ menu for simply-embedded partitions).
Let $G$ have a simply-embedded partition $\{\tau_1|\cdots|\tau_N\}$.  For any $\sigma \subseteq [n]$, let $\sigma_i \od \sigma \cap \tau_i$.  Then 
\vspace{-.075in}
$$\sigma \in \FP(G) \quad \Rightarrow \quad \sigma_i \in \FP(G|_{\tau_i})\cup \{\emptyset\}~~\text{ for all } i \in [N]. $$ 
In other words, every fixed point support of $G$ is a union of component fixed point supports $\sigma_i$, at most one per component.\\

Theorem~\ref{thm:menu} gives significant restrictions on the possible supports in $\FP(G)$ in terms of the component fixed point supports.  However, the converse is not true -- not every union of supports from the menu is guaranteed to yield a fixed point support in $\FP(G)$.  The following examples illustrate the range of $\FP(G)$ that can emerge from the same menu.

\begin{figure}[!h]
\begin{center}
\includegraphics[width=6.25in]{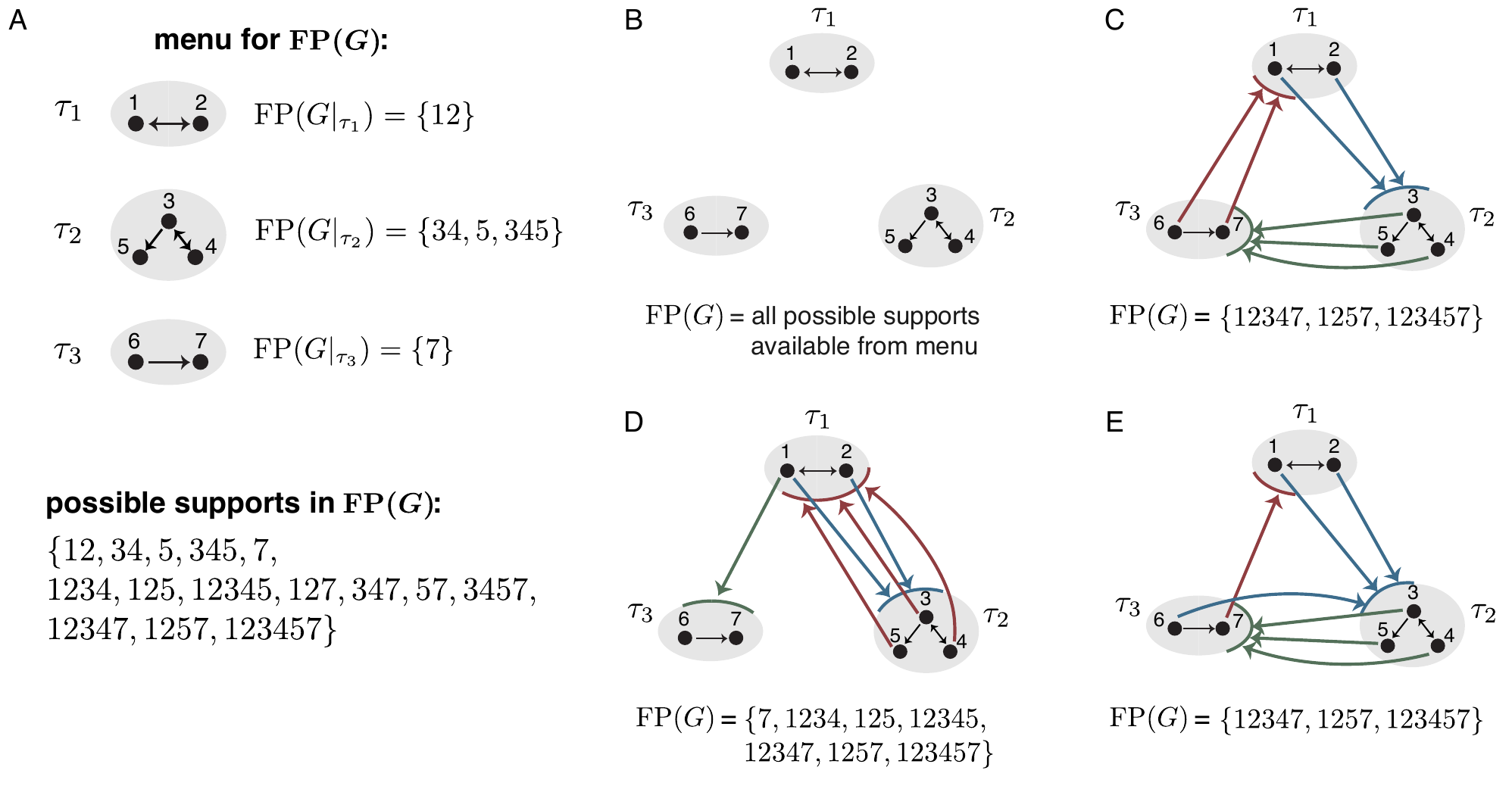}
\vspace{-.15in}
\end{center}
\caption{\textbf{Graphs with a simply-embedded partition from Example~\ref{ex:sa-partitions}.} (A) (Top) A collection of component subgraphs with their $\FP(G|_{\tau_i})$.  (Bottom) The set of possible fixed point supports for any graph that has these subgraphs in the simply-embedded partition. (B-E) Example graphs with a simply-embedded partition with the component subgraphs from A, together with their $\FP(G)$.  In (C-E), thick colored edges from a node to a component indicate that the node projects edges out to all the nodes in the receiving component.
}
\label{fig:sa-partition-examples}
\vspace{-.15in}
\end{figure}

\begin{example}\label{ex:sa-partitions}
Consider the component subgraphs shown in Figure~\ref{fig:sa-partition-examples}A together with their $\FP(G|_{\tau_i})$.  By Theorem~\ref{thm:menu}, any graph $G$ with a simply-embedded partition of these component subgraphs has a restricted menu for $\FP(G)$ consisting of the component fixed point supports (the set of all possible supports derived from this menu is shown on the bottom of panel A).  Note that an arbitrary graph on 7 nodes could have up to $2^7 -1 = 127$ possible fixed point supports, but the simply-embedded partition structure narrows the options to only 15 candidate fixed points.  Figure~\ref{fig:sa-partition-examples}B-E show four possible graphs with simply-embedded partitions of these component subgraphs, together with $\FP(G)$ for each of the graphs.  

Observe that the graph in Figure~\ref{fig:sa-partition-examples}B is a \emph{disjoint union} of its component subgraphs.  For this graph, $\FP(G)$ consists of all possible unions of at most one fixed point support per component subgraph (see \cite[Theorem 11]{fp-paper}).  Thus, every choice from the menu provided by Theorem~\ref{thm:menu} does in fact yield a fixed point for $G$.  

In contrast, the graph in Figure~\ref{fig:sa-partition-examples}C is a \emph{cyclic union} of the component subgraphs.  For this graph, $\FP(G)$ only has sets that contain a fixed point support from every component, i.e., $\sigma_i \neq \emptyset$ for all $i \in [N]$ (by Theorem~\ref{thm:cyclic-unions}).  Thus, any subset from the menu of Theorem~\ref{thm:menu} that does not intersect every $\tau_i$ does not produce a fixed point for $G$.

Meanwhile, the graph in Figure~\ref{fig:sa-partition-examples}D is a simply-embedded partition with heterogeneity in the outgoing edges from a component (notice different nodes in $\tau_1$ treat $\tau_3$ differently).  $\FP(G)$ has a mixture of types of supports: there are some $\sigma \in \FP(G)$ that do not intersect every component, and others that do. 

Finally, the graph in Figure~\ref{fig:sa-partition-examples}E is another simply-embedded partition with heterogeneity (notice different nodes in $\tau_3$ treat $\tau_1$ and $\tau_2$ differently). However, for this graph, there is a uniform rule for the fixed point supports: every fixed point consists of exactly one fixed point support per component subgraph (identical to $\FP(G)$ for the graph in panel C).

\end{example}

\vspace{-.05in}
Interestingly, some graphs have multiple nontrivial simply-embedded partitions, which can be analyzed in parallel to give further constraints on $\FP(G)$, and in some cases even fully nail down $\FP(G)$.

\begin{example}\label{ex:multiple-sa-partitions}
Consider the graph in Figure~\ref{fig:multiple-sa-partitions} with the two simply-embedded partitions $\{1\,|2,3,4|\,5\}$ and $\{2\,|1,3,5|\,4\}$. 
\begin{figure}[!h]
\begin{center}
\includegraphics[width=2.75in]{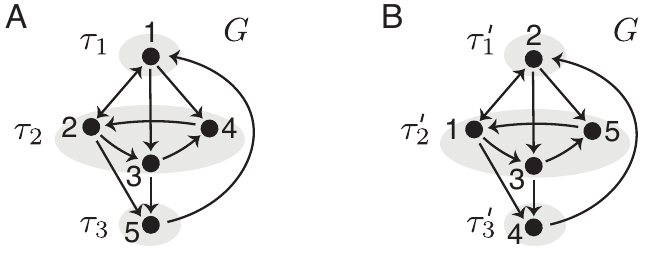}
\vspace{-.1in}
\end{center}
\caption{\textbf{Graph with two nontrivial simply-embedded partitions.} The panels show two drawings of the same graph $G$ to highlight two partitions: (A) the simply-embedded partition $\{1\,|2,3,4|\,5\}$, (B) the simply-embedded partition $\{2\,|1,3,5|\,4\}$.}
\label{fig:multiple-sa-partitions}
\vspace{-.15in}
\end{figure}
Since $234$ is the unique fixed point of $\tau_2$ in the partition in (A), Theorem~\ref{thm:menu} guarantees that any $\sigma \in \FP(G)$ that contains one of the nodes $2$, $3$, or $4$ must contain all three nodes because $\sigma \cap \tau_2 \in \FP(G|_{\tau_2})\cup \{\emptyset\}$.  
Similarly, since $135$ is the unique fixed point of $\tau_2'$ in the partition in (B), any $\sigma \in \FP(G)$ that contains one of the nodes $1$, $3$, or $5$ must contain all three.  Combining these two constraints, we see that we must have $\FP(G)=\{12345\}$.
\end{example}

In general though, Theorem~\ref{thm:menu} is not sufficient to fully determine $\FP(G)$, but it does significantly limit the menu of fixed point supports.  In particular, one direct consequence of Theorem~\ref{thm:menu} is that if there is some node $j \in \tau_i$ in $G$ that does not participate in any fixed points of $G|_{\tau_i}$, then $j$ cannot participate in any fixed point of the full graph $G$. Thus the supports of all the fixed points of $G$ are confined to $[n]\setminus \{j\}$.  For example, node $6$ in in Figure~\ref{fig:sa-partition-examples} does not appear in $\FP(G|_{\tau_3})$, and thus does not appear in any fixed point support for any of the graphs in panels B-E.  It turns out that if the removal of node $j$ does not change the fixed points of the component subgraph, i.e.\ if $\FP(G|_{\tau_i}) = \FP(G|_{\tau_i\setminus\{j\}})$, then we can actually remove $j$ from the full graph $G$ without changing $\FP(G)$.  Thus we have the following theorem.

\begin{theorem}[removable nodes]\label{thm:removables}
Let $G$ have a simply-embedded partition $\{\tau_1|\cdots|\tau_N\}$.  Suppose there exists a node $j \in \tau_i$ such that $\FP(G|_{\tau_i}) = \FP(G|_{\tau_i\setminus\{j\}})$.  Then $\FP(G) = \FP(G|_{[n]\setminus\{j\}}) $. 
\end{theorem}

Theorem~\ref{thm:removables} shows that if a node $j$ is locally removable without altering fixed points of its component, then node $j$ is also globally removable without altering the fixed points of the full graph $G$.  This result gives a new tool for determining that two graphs have the same collection of fixed points. 

\begin{corollary}\label{cor:removables}
Let $G$ have a simply-embedded partition $\{\tau_1|\cdots|\tau_N\}$ and suppose there exists $j\in\tau_i$ such that $\FP(G|_{\tau_i}) = \FP(G|_{\tau_i\setminus\{j\}})$. Let $G'$ be any graph that can be obtained from $G$ by deleting or adding outgoing edges from $j$ to any other component without altering the simply-embedded structure of $G$.  Then $\FP(G') = \FP(G)$.
\end{corollary}

As an illustration of Corollary~\ref{cor:removables}, let $G$ be the graph from Figure~\ref{fig:sa-partition-examples}C and $G'$ be the graph from Figure~\ref{fig:sa-partition-examples}E.  It is easy to check that $\FP(G|_{\tau_3}) = \FP(G|_{\tau_3\setminus \{6\}})$, and so node $6$ is removable.   Since $G$ and $G'$ differ only in edges out from node $6$ to other components, and the simply-embedded partition is maintained, Corollary~\ref{cor:removables} guarantees that $\FP(G)= \FP(G')$.

While Theorems~\ref{thm:menu} and~\ref{thm:removables} give significant constraints on $\FP(G)$, the simply-embedded partition structure alone is not sufficient to nail down $\FP(G)$.  In the following subsections, we consider a variety of families of graphs that have additional structure that enables us to draw stronger conclusions about $\FP(G)$.

\subsection{Directional cycles with a simply-embedded partition}\label{sec:sa-dir-cycles}
We begin by considering simply-embedded directional cycles, i.e.\ directional cycles where the directional partition of the nodes into components $\{\tau_1|\cdots|\tau_N\}$ is also a simply-embedded partition.  We expect these graphs to be similar to the corresponding cyclic union of the same component subgraphs, both in terms of their dynamics (as a result of the directionality property) and in terms of their $\FP(G)$ (as a result of the simply-embedded partition).  
Theorem~\ref{thm:sa-dir-cycle} guarantees that every fixed point of a simply-embedded directional cycle has the same component structure as those of the corresponding cyclic union.\\

\noindent{\bf Theorem~\ref{thm:sa-dir-cycle}} (simply-embedded directional cycles).
Let $G$ be a directional cycle whose components form a simply-embedded partition $\{\tau_1| \cdots|\tau_N\}$. For any $\sigma \subseteq [n]$, let $\sigma_i \od \sigma \cap \tau_i$.  Then 
\vspace{-.05in}
$$\sigma \in \FP(G) \quad \Rightarrow \quad \sigma_i \in \FP(G|_{\tau_i})~~\text{ for all } i \in [N].$$
In other words, every fixed point support of $G$ is a union of (nonempty) component fixed point supports, exactly one per component.

\begin{proof}
By Theorem~\ref{thm:menu}, for any $\sigma \in \FP(G)$, we have $ \sigma_i \in \FP(G|_{\tau_i}) \cup \{\emptyset\}$.  By Theorem~\ref{thm:dir-cycle}, every $\sigma \in \FP(G)$ contains a cycle that intersects every $\tau_i$, and so $\sigma_i\neq\emptyset$ for all $i\in[N]$.  Thus, $ \sigma_i \in \FP(G|_{\tau_i})$ for all $i \in [N]$.  
\end{proof}

We conjecture that the backwards direction of the statement in Theorem~\ref{thm:sa-dir-cycle} also holds, yielding an if and only if characterization of the fixed point supports.  If this were true, then the fixed point supports of a simply-embedded directional cycle would be identical to those of the corresponding cyclic union. To prove this characterization, it is natural to try to mimic the proof of Theorem~\ref{thm:cyclic-unions} (Theorem 13 in \cite{fp-paper}), which is the analogous result for cyclic unions.  The key to that proof is to induct on the size of the cyclic union by analyzing relevant subgraphs $G|_\sigma$.  Since each $G|_\sigma$ is itself a cyclic union, the inductive hypothesis can then be applied.  Unfortunately, when we consider the analogous subgraphs of a simply-embedded directional cycle, they need not be directional cycles themselves.  Thus, induction cannot be used to prove the conjecture. However, computational analyses of over 10,000 simply-embedded directional cycles give us reasonable confidence that the conjecture holds.  Specifically, we randomly sampled 1000s of simply-embedded directional cycles with 3, 4 or 5 components where the component subgraphs had up to 4 nodes each.  For every one of these simply-embedded directional cycles, we found that $\FP(G) = \{ \bigcup_{i=1}^N \sigma_i~|~\sigma_i \in \FP(G|_{\tau_i}) \text{ for all } i \in [N]\}$, as predicted.
Moreover, as we saw in Figures~\ref{fig:cycu-generalizations} and~\ref{fig:dir-cycle-sa}, the dynamics of simply-embedded directional cycles tend to mimic those of the corresponding cyclic union.

\subsection{Simple linear chains}\label{sec:linear-chains}
In the previous subsection, we saw that when we have a simply-embedded partition on top of a directional cycle structure, this adds significant constraints on $\FP(G)$.  It is natural to ask what happens when we cut such a cyclic structure between components and are left with just a directional chain.  Does the added structure of a simply-embedded partition similarly give a stronger handle on $\FP(G)$ for a directional chain?  

Recall from Proposition~\ref{prop:directional-chain} that a directional chain is provably directional onto the last component, and so $\FP(G) \subseteq \FP(G|_{\tau_N})$.  Simply-embedded partitions only add the constraint that for each $\sigma \in \FP(G)$, we have $\sigma_i \in \FP(G|_{\tau_i}) \cup \{\emptyset\}$.  But this gives no new information since for directional chains, we are already guaranteed that $\sigma_i = \emptyset$ for all $i \neq N$ and $\sigma_N \in \FP(G|_{\tau_N})$.  
\begin{figure}[!h]
\begin{center}
\includegraphics[width=6.25in]{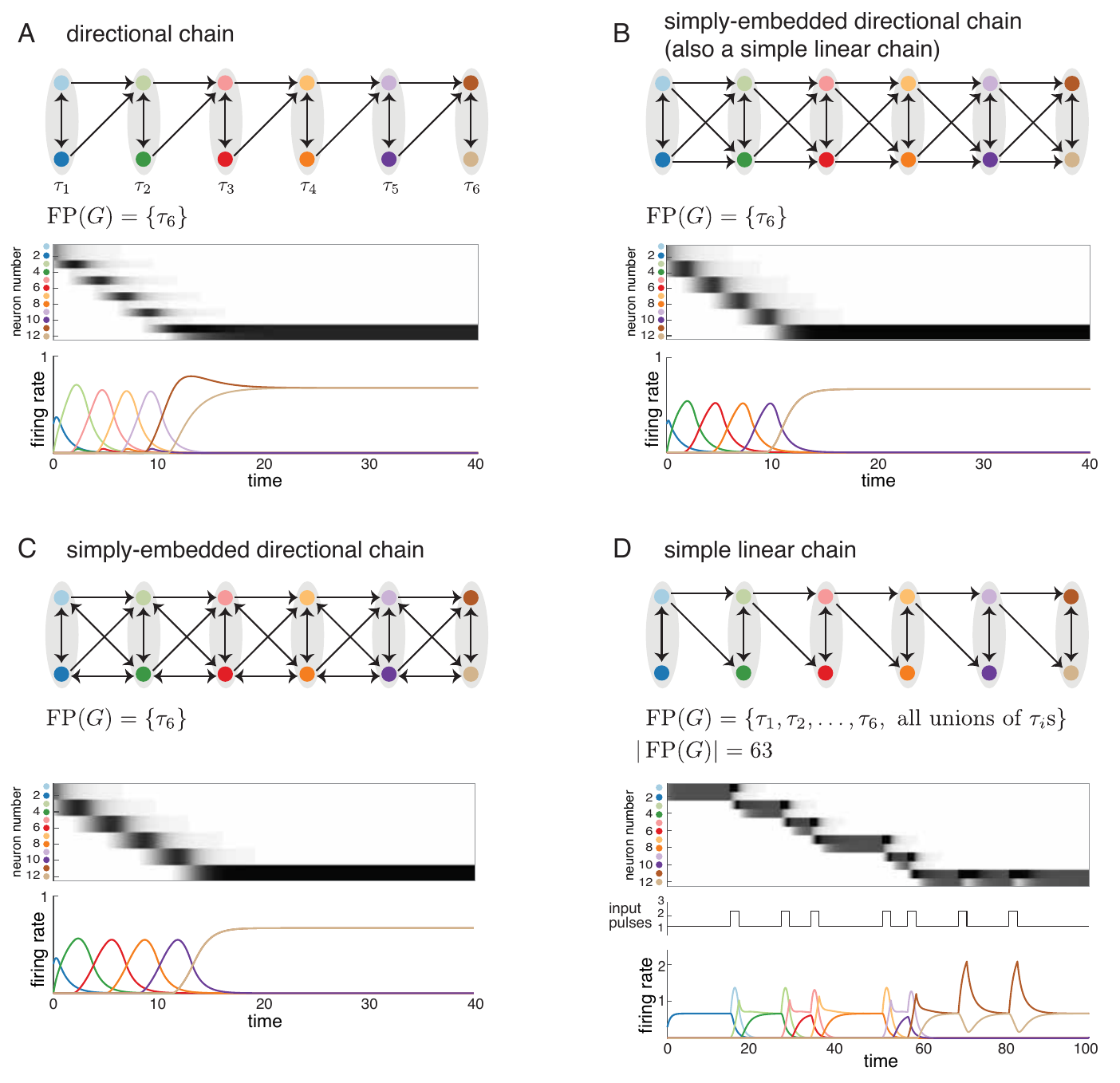}
\vspace{-.25in}
\end{center}
\caption{ \textbf{Directional chains vs. simple linear chains.} (A-C) Graphs that are directional chains.  Activity initialized on $\tau_1$ flows through the chain, hitting each component in sequence, and converging on the nodes of $\tau_6$.  (D) A simple linear chain that is not directional.  Each component clique supports a stable fixed point of the network.  Unions of these component fixed point supports also yield fixed points.  Activity initialized on $\tau_1$ would stay indefinitely at the corresponding stable fixed point.  Small kicks to the $\theta$ input (labeled as input pulses on the plot) can cause the activity to fall out of the current stable fixed point and move forward to converge onto $\tau_{i+1}$.  At time 60, the activity has converged to $\tau_6$.  After this point, all additional input pulses lead to increases in the activity of the nodes in $\tau_6$, but the activity can never escape the final stable fixed point of the chain.}
\label{fig:dir-vs-linear-chain}
\vspace{-.1in}
\end{figure}
Figure~\ref{fig:dir-vs-linear-chain}A-C shows examples of directional chains both with and without simply-embedded structure.  
Notice that in all three of these graphs, $\FP(G)$ is identical, and it is fully predicted by Proposition~\ref{prop:directional-chain} since $\FP(G) \subseteq \FP(G|_{\tau_6}) = \{\tau_6\}$.   Moreover, the dynamics progress forward down the directional chain and converge to the stable fixed point on $\tau_6$, irrespective of any simply-embedded structure.

Combining simply-embedded partitions with directional chain structure does not yield any new information about $\FP(G)$.  But what about simply-embedded partitions in graphs that have a weaker chain-like structure, where all the edges feed forward between components, but there are not enough forward edges to guarantee directionality?  For example, consider the graph in Figure~\ref{fig:dir-vs-linear-chain}D.  All the edges between components feed forward following a chain-like architecture; moreover, each $j \in \tau_i$ treats the nodes in $\tau_{i+1}$ identically, and so $\{\tau_1|\cdots|\tau_6\}$ is a simply-embedded partition, but it is not directional.  We refer to graphs with this chain-like architecture on a simply-embedded partition as \emph{simple linear chains}.

\begin{definition}[simple linear chain] \label{def:linear-chain}
Let $G$ be a graph with node partition $\{\tau_1|\cdots|\tau_N\}$.  We say that $G$ is a \emph{simple linear chain} if the following two conditions hold:
\begin{enumerate}
\item the only edges between components go from nodes in $\tau_i$ to $\tau_{i+1}$, and 
\item for every $j \in \tau_i$, either $j \to k$ for every $k \in \tau_{i+1}$ or $j \not\to k$ for every $k \in \tau_{i+1}$.
\end{enumerate}
\end{definition}

With simple linear chains, we are no longer guaranteed that the fixed points all collapse onto the last component.  For example, in Figure~\ref{fig:dir-vs-linear-chain}D, each $\tau_i \in \FP(G)$ since each clique survives in $G|_{\tau_i \cup \tau_{i+1}}$.  Additionally, every union of $\tau_i$s is also a fixed point support.  Since each surviving clique yields a stable fixed point, we see that the network dynamics in panel D do not naturally progress through the chain, but rather stabilize on an individual component.  Interestingly, though, if we transiently kick all the neurons in the network by temporarily increasing $\theta$, then the dynamics can escape from the current $\tau_i$, and the activity flows forward and stabilizes on $\tau_{i+1}$. The following theorem shows that the structure of $\FP(G)$ illustrated in Figure~\ref{fig:dir-vs-linear-chain}D holds for simple linear chains more generally.

\begin{theorem}[simple linear chains]\label{thm:linear-chain}
Let $G$ be a simple linear chain with components $\tau_1, \ldots, \tau_N$.  
\begin{enumerate}
\item[(i)] If $\sigma \in \FP(G)$, then $\sigma_i \in \FP(G|_{\tau_i}) \cup \{\emptyset\}$ for all $i \in [N]$, where $\sigma_i = \sigma \cap \tau_i$. 
\item[(ii)] Consider a collection $\{\sigma_i\}_{i \in [N]}$ of $\sigma_i \in \FP(G|_{\tau_i}) \cup \{\emptyset\}$.  If additionally $\sigma_i \in \FP(G|_{\tau_i\cup\tau_{i+1}}) \cup \{\emptyset\}$ for all $i \in [N]$, then \\
\vspace{-.25in}
$$\bigcup_{i \in [N]} \sigma_i \in \FP(G).$$
\end{enumerate}
\vspace{-.075in}
In other words, $\FP(G)$ is closed under unions of component fixed point supports that survive in $G|_{\tau_i\cup\tau_{i+1}}$.
\end{theorem}

\begin{figure}[!h]
\begin{center}
\includegraphics[width=5.5in]{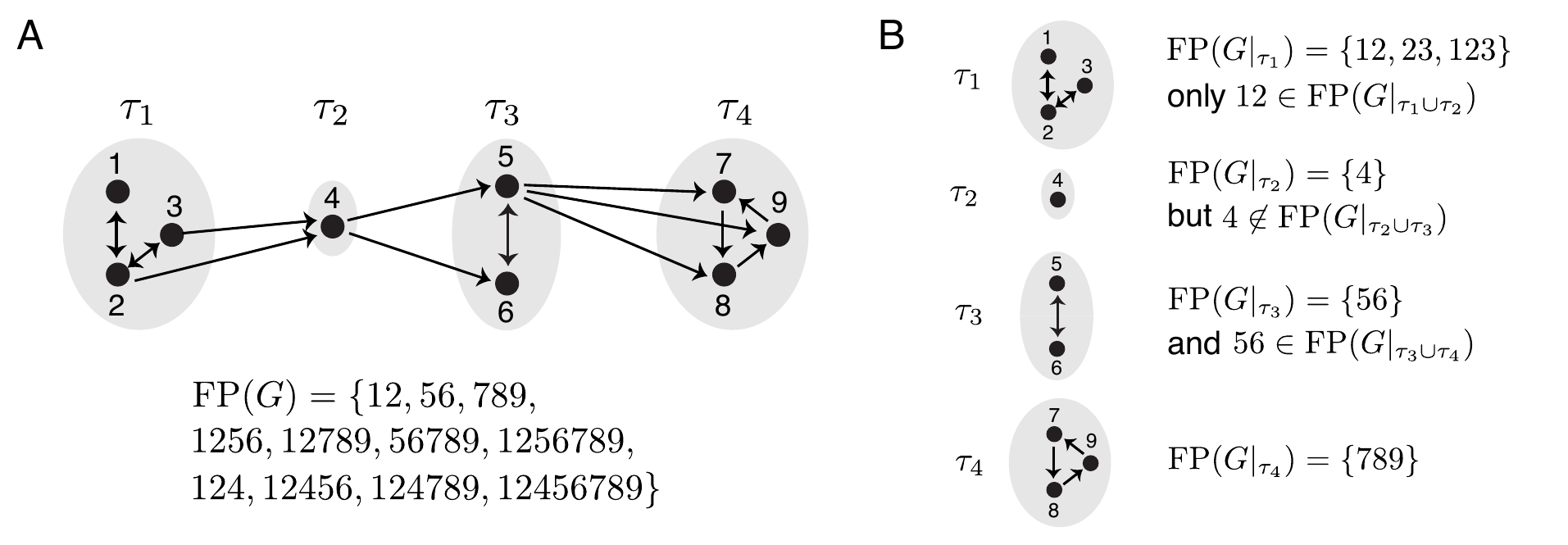}
\vspace{-.2in}
\end{center}
\caption{ \textbf{Simple linear chain.} (A) An example simple linear chain together with its $\FP(G)$.  The first row of $\FP(G)$ gives the surviving fixed points from each component subgraph; the second row shows that all unions of these component fixed points are also in $\FP(G)$ (Theorem~\ref{thm:linear-chain}(ii)); the third row shows the additional fixed point supports in $\FP(G)$ that arise from the broader menu (Theorem~\ref{thm:linear-chain}(i)).  (B) $\FP(G|_{\tau_i})$ for each component subgraph from A, and the list of which of these supports survive the addition of the next component in the chain.}
\label{fig:linear-chain-thm-example}
\vspace{-.05in}
\end{figure}

Figure~\ref{fig:linear-chain-thm-example} illustrates Theorem~\ref{thm:linear-chain} with an example simple linear chain.  By Theorem~\ref{thm:linear-chain}(i), every fixed point support in $\FP(G)$ restricts to a fixed point in $\FP(G|_{\tau_i})$.  Next consider a collection of $\sigma_i$ such that $\sigma_i \in \FP(G|_{\tau_i\cup\tau_{i+1}}) \cup \emptyset$ for all $i \in [N]$.  First observe that each $\sigma_i \in \FP(G|_{\tau_i \cup \tau_{i+1}})$ actually survives to the full network, and so $\sigma_i \in \FP(G)$.  This is guaranteed because $\sigma_i$ has no outgoing edges to nodes outside of $\tau_i \cup \tau_{i+1}$ (Rule~\ref{rule:graph-domination}C).  Moreover, by Theorem~\ref{thm:linear-chain}(ii), we see that every union of surviving component fixed points yields a fixed point of the full network, but additional fixed point supports are also possible.

\begin{figure}[!h]
\begin{center}
\vspace{-.2in}
\includegraphics[height=1.6in]{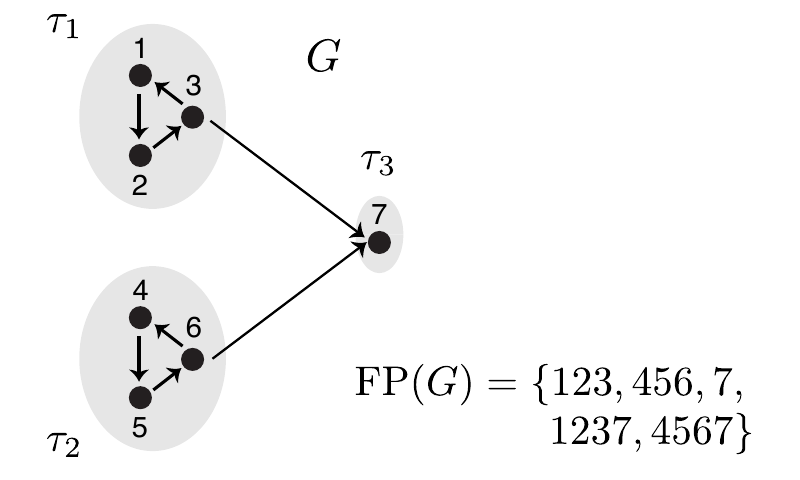}
\vspace{-.1in}
\end{center}
\caption{ \textbf{Simple feedforward network.} A feedforward network generalizing the conditions of the simple linear chain. Notice that $\FP(G)$ is \emph{not} closed under unions of surviving component fixed points, since $123,\, 456 \in \FP(G)$ but $123456\notin\FP(G)$. }
\label{fig:feedforward-example}
\vspace{-.2in}
\end{figure}

A natural generalization of simple linear chains is \emph{simple feedforward networks} where $G$ consists of ordered component subgraphs such that the only edges allowed between components are from a smaller numbered component to a larger one, and again we require that for any pair $\tau_i$ and $\tau_k$ with $k>i$, each $j \in \tau_i$ either sends edges to every node in $\tau_k$ or to no nodes in $\tau_k$.  Given that these simple feedforward networks have such similar structure to that of the simple linear chains, we might hope that an analogous result to Theorem~\ref{thm:linear-chain} holds for these networks.  These simple feedforward networks do have a simply-embedded partition structure, and so Theorem~\ref{thm:linear-chain}(i) holds for these networks as well (as an immediate corollary of Theorem~\ref{thm:menu}).  But an analogue of Theorem~\ref{thm:linear-chain}(ii) does \emph{not} hold.  Specifically, survival of component fixed points does not guarantee that the union of these component supports will yield a fixed point.  Figure~\ref{fig:feedforward-example} provides an explicit counterexample: we see that the $3$-cycles $123$ and $456$ both survive to $\FP(G)$ (by Rule~\ref{rule:uniform-in-deg}), but their union $123456 \notin\FP(G)$ since it is uniform in-degree 1 with two outgoing edges to node $7$.
%

\subsection{Strongly simply-embedded partitions}\label{sec:bidir-sa-partitions}
Recall that $\{\tau_1|\cdots|\tau_N\}$ is a simply-embedded partition of a graph $G$ if each component $\tau_i$ is simply-embedded in $G$ so that every node in $\tau_i$ is treated identically by the rest of the graph; specifically, if any node outside of $\tau_i$ sends an edge to one node in $\tau_i$, then it sends edges to \emph{every} node in $\tau_i$.  In this context, there is still freedom allowing nodes to treat different components differently, e.g.\ node $k$ may send edges to all nodes in $\tau_i$, but send no edges to nodes in $\tau_j$.  In this subsection, we consider graphs with a more rigid partition structure, which we call a \emph{strongly simply-embedded partition}.  In these graphs, each node must treat all the components identically.  More precisely, we have:

\begin{definition}[strongly simply-embedded partition]
Let $G$ be a graph with a partition of its nodes $\{\tau_1|\cdots|\tau_N\}$.  The partition is called \emph{strongly simply-embedded} if for every 
node $j$ in $G$, either $j \to k$ for all $k \notin \tau_i$ or $j \not\to k$ for all $k \notin \tau_i$, where $\tau_i$ is the component containing $j$.
\end{definition}

\begin{figure}[!hb]
\begin{center}
\includegraphics[width=5.4in]{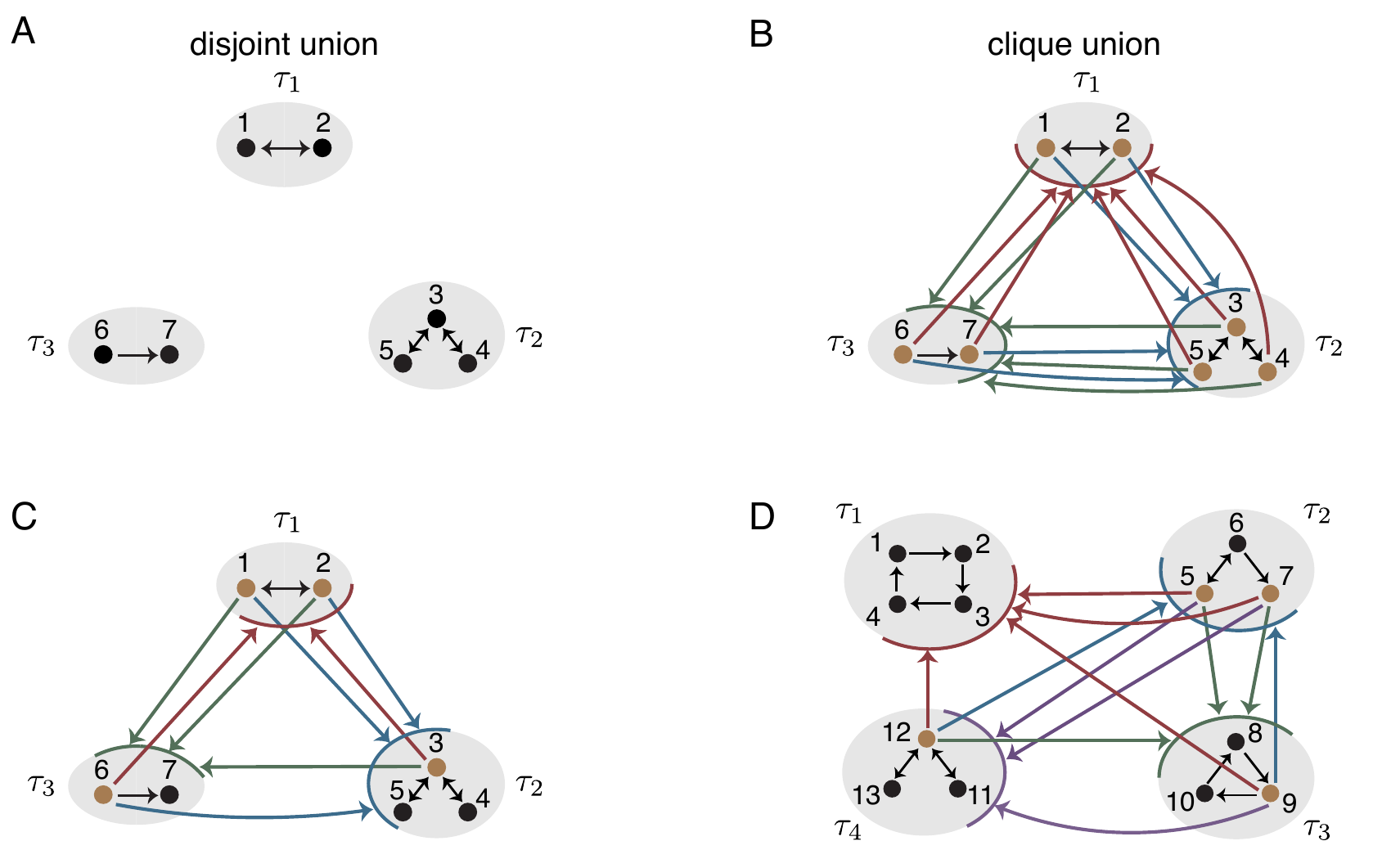}
\vspace{-.2in}
\end{center}
\caption{ \textbf{Strongly simply-embedded partitions.} Four example graphs with a strongly simply-embedded partition, characterized by the fact that each node treats all the other components identically.  Thus, any node that sends an edge out to one component, must in fact send edges out to every component (i.e., it must be a projector onto the rest of the graph).  Projector nodes are colored brown.  (A) A disjoint union.  (B) A clique union.  (C-D) Example graphs with a mix of projector and nonprojector nodes within each component. }
\label{fig:bidir-sa-part-examples}
\end{figure}

Notice that in a strongly simply-embedded partition, each node $j$ either projects edges onto every other node outside its component $\tau_i$ (in which case, we say that $j$ is a \emph{projector} onto $[n] \setminus \tau_i$) or it does not project any edges to nodes outside its component (in which case, we say that $j$ is a \emph{nonprojector} onto $[n] \setminus \tau_i$).  The simplest examples of graphs with a strongly simply-embedded partition are \emph{disjoint unions} and \emph{clique unions}, which are building block constructions first studied in \cite{fp-paper}.  In a \emph{disjoint union} of component subgraphs $G|_{\tau_1}, \ldots, G|_{\tau_N}$, there are no edges between components (see Figure~\ref{fig:bidir-sa-part-examples}A).  In this case, every node in $G$ is a nonprojector onto the rest of the graph.  At the other extreme, a \emph{clique union} has bidirectional edges between every pair of nodes in different components.  In a clique union, every node is a projector onto the rest of the graph (see Figure~\ref{fig:bidir-sa-part-examples}B).  More generally, strongly simply-embedded partitions can have a mix of projector and nonprojector nodes even within the same component, as shown in Figure~\ref{fig:bidir-sa-part-examples}C and D (projector nodes are colored brown and have outgoing edges to every component).

Similar to simple linear chains, it turns out that strongly simply-embedded partitions also have the property that $\FP(G)$ is closed under unions of surviving fixed point supports of the component subgraphs.  With the added structure of the strongly simply-embedded partition, though, we can actually say something stronger -- $\FP(G)$ can be fully determined from knowledge of the component fixed point supports together with knowledge of which of those component fixed points survive in the full network.  This complete characterization of $\FP(G)$ is given in Theorem~\ref{thm:bidir-sa-partition} below.

\begin{theorem}\label{thm:bidir-sa-partition}
Suppose $G$ has a strongly simply-embedded partition $\{\tau_1|\dots|\tau_N\}$, and let $\sigma_i \od \sigma \cap \tau_i$ for any $\sigma \subseteq [n]$.  
Then $\sigma \in \FP(G)$ if and only if $\sigma_i \in \FP(G|_{\tau_i}) \cup \{\emptyset\}$ for each $i \in [N]$, and either
\begin{enumerate}
\item[(a)] every $\sigma_i$ is in $\FP(G) \cup \{ \emptyset\}$, or 
\item[(b)] none of the $\sigma_i$ are in $\FP(G) \cup \{ \emptyset\}$.
\end{enumerate}
In other words, $\sigma \in \FP(G)$ if and only if $\sigma$ is either a union of surviving fixed points $\sigma_i$, at most one per component, or it is a union of dying fixed points, exactly one from every component.
\end{theorem}

A key to the proof of Theorem~\ref{thm:bidir-sa-partition} is the significant additional constraints on the simply-embedded structure imposed by the strongly simply-embedded partition.  Specifically, with a strongly simply-embedded partition, not only is the original partition $\{\tau_1|\cdots|\tau_N\}$ simply-embedded, but also \emph{every coarsening} of the partition (where the components are unions of the $\tau_i$) is simply-embedded.  Notice this property does not hold in general for simply-embedded partitions.  For example, given a cyclic union on $\{\tau_1|\cdots|\tau_N\}$, the coarser partition $\{\tau_{1} \cup \tau_2~|~ \tau_{3} \cup \cdots \cup \tau_N\}$ is not a simply-embedded partition since not all nodes in $\tau_{1} \cup \tau_2$ are treated identically by the rest of the graph: the nodes in $\tau_1$ receive edges from $\tau_N$, while the nodes in $\tau_2$ do not.  
The guarantee of the simply-embedded property for every coarser partition enables an inductive proof to fully nail down $\FP(G)$ for strongly simply-embedded partitions.

As an application of Theorem~\ref{thm:bidir-sa-partition}, we can immediately recover characterizations of the fixed points of disjoint unions and clique unions previously given in \cite[Theorems 11 and 12]{fp-paper}.  In a disjoint union, every component fixed point support survives to the full network since it has no outgoing edges (by Rule~\ref{rule:graph-domination}: inside-out domination).  Thus, for a disjoint union, $\FP(G)$ consists of all the fixed points of type (a) from Theorem~\ref{thm:bidir-sa-partition}: unions of (surviving) component fixed points $\sigma_i$, at most one per component.  In contrast, in a clique union, every component fixed point support dies in the full network since it has a target that outside-in dominates it (in fact, every node outside of $\tau_i$ is a target of any subset of $\tau_i$).  Thus, for a clique union, $\FP(G)$ consists of all the fixed points of type (b): unions of (dying) component fixed points $\sigma_i$, exactly one from every component.  Both the disjoint union and clique union characterizations of $\FP(G)$ \cite[Theorems 11 and 12]{fp-paper} are now immediate corollaries of Theorem~\ref{thm:bidir-sa-partition}, and the earlier proofs of these results in \cite{fp-paper} have a similar flavor to the proof of Theorem~\ref{thm:bidir-sa-partition}, which we provide in Appendix Section~\ref{sec:proof-bidir-sa}.

\begin{corollary}\label{cor:disjoint-clique-unions}
Let $G$ be a graph with partition $\{\tau_1|\cdots|\tau_N\}$.  
\begin{itemize}
\item[(a)] If $G$ is a disjoint union of $G|_{\tau_1}, \ldots, G|_{\tau_N}$, then $\sigma \in \FP(G)$ if and only if $\sigma_i \in \FP(G|_{\tau_i}) \cup \{\emptyset\}$ for all $i \in [N]$.
\item[(b)] If $G$ is a clique union of $G|_{\tau_1}, \ldots, G|_{\tau_N}$, then $\sigma \in \FP(G)$ if and only if $\sigma_i \in \FP(G|_{\tau_i})$ for all $i \in [N]$.
\end{itemize}
\end{corollary}

\begin{figure}[!hb]
\begin{center}
\vspace{-.1in}
\includegraphics[width=.8\textwidth]{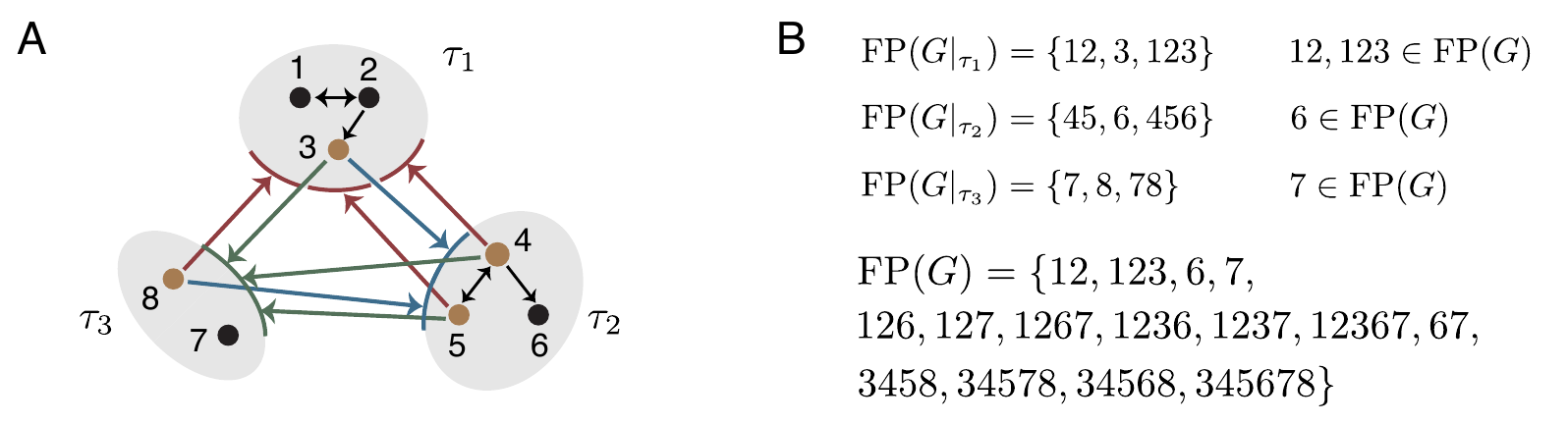}
\vspace{-.1in}
\end{center}
\caption{ \textbf{Strongly simply-embedded partition with $\boldsymbol{\FP(G)}$.} (A) A graph with a strongly simply-embedded partition $\{\tau_1|\tau_2|\tau_3\}$.  Projector nodes are colored brown. (B) (Top) $\FP(G|_{\tau_i})$ for each component subgraph together with the supports from each component that survive within the full graph.  (Bottom) $\FP(G)$ for the strongly simply-embedded partition graph.  The first two lines of $\FP(G)$ consist of unions of surviving fixed points, at most one per component.  The third line gives the fixed points that are unions of dying fixed point supports, exactly one from every component. }
\label{fig:bidir-sa-part-FP}
\vspace{-.2in}
\end{figure}

More generally, though, a strongly simply-embedded partition can have a mix of surviving and dying component fixed points, so that $\FP(G)$ has a mix of both type (a) and type (b) fixed point supports.  Figure~\ref{fig:bidir-sa-part-FP}A gives an example strongly simply-embedded partition, and panel B shows both the set of component fixed point supports, $\FP(G|_{\tau_i})$, and the subset of those that survive to yield fixed points of the full network.  Since there are dying fixed points in every component, we see that $\FP(G)$ has a mix of both type (a) and type (b) fixed point supports.  
%


\section{Applications to sequential attractor prediction}\label{sec:applications}

In this section, we consider a number of networks of size $n=5$ to show how directional cycle graph architecture is predictive of the structure of corresponding sequential attractors, particularly when a graph has a simply-embedded directional cycle representation.  From this analysis, we see that these graph structures are useful not only for predicting the set of fixed points of a network, but also for explicitly connecting architecture to the pattern of sequential neural activity that emerges in attractors.  

We focus our analysis on a subset of the graphs of size $n=5$ with the special property that $\FP(G)=\{12345\}$, i.e., the graphs have a unique fixed point, which has full support.  In prior work \cite{rule-of-thumb}, these types of networks, known as \emph{core motifs}, were observed to produce attractors in which all the neurons are highly active, as opposed to the activity being concentrated on only a subnetwork of neurons.  As a result, the global connectivity is relevant to shaping the activity of all attractors in these networks, and thus they are particularly well suited to analysis with the directional cycle framework.  The networks analyzed here were taken from \cite{n5-github}, which provided a comprehensive analysis of all the attractors of CTLNs of size $n=5$ with $\varepsilon=0.51,\, \delta=1.76,\, \theta=1$; we follow the numbering of graphs used there and focus on the sequential attractors that emerge for that choice of parameters.

\begin{figure}[!hb]
\begin{center}
\includegraphics[width=6.25in]{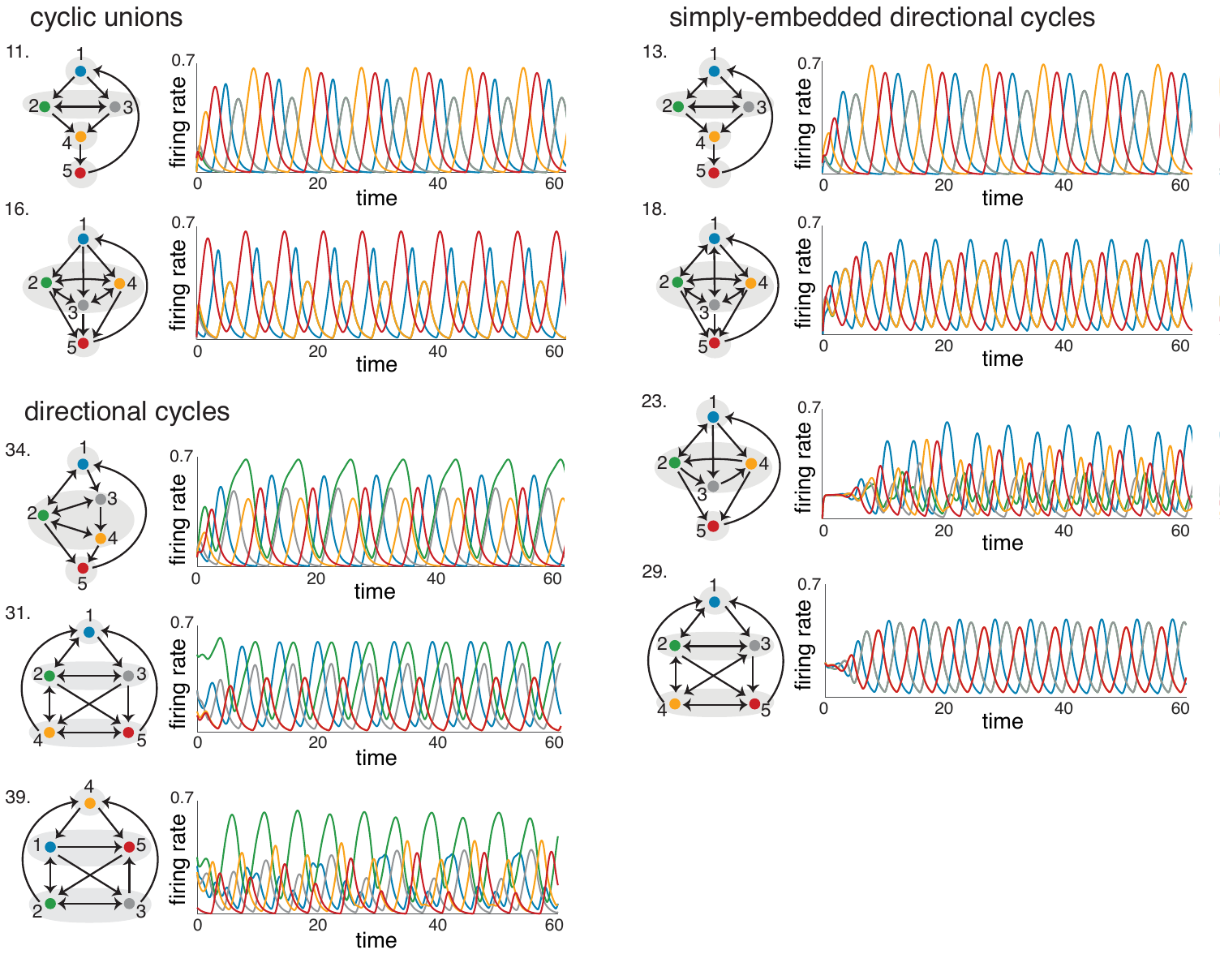}
\vspace{-.2in}
\end{center}
\caption{{\bf Example graphs of size $\mathbf{5}$.}  A collection of example $n=5$ graphs of different types and their corresponding dynamic attractors for $\varepsilon=0.51, \delta=1.76, \theta=1$.  The graphs are numbered following the ordering given in \cite{n5-github}, which extensively catalogued $\FP(G)$ and the dynamic attractors for all graphs of size $5$ for this parameter choice.}
\label{fig:n5-core-examples}
\vspace{-.1in}
\end{figure}

Figure~\ref{fig:n5-core-examples} shows a sampling of graphs drawn in a way that highlights their directional cycle structure.  Observe that for the cyclic unions and the more general simply-embedded directional cycles, the corresponding attractor has activity that flows through the components in precisely the cyclic order dictated by the directional cycle.  For components that are cliques, the activity of the nodes in the clique is perfectly synchronized and thus only the activity of the highest-numbered node is visible in the plots.  
Also notice that the activity of nodes in a synchronous component is typically lower than that of the singleton components because there is competition among the nodes within a component, and the total population activity is bounded.  When a component is not a clique, as in graph 23, we see that the activity may cycle through the nodes in the component before the activity flows on to the next component in the directional cycle.  

When a graph only has directional cycle representations that do not coincide with a simply-embedded partition, this architecture can still give insight into the sequential attractor, but this is not guaranteed.  For example, for graph 34 in Figure~\ref{fig:n5-core-examples}, we see that the neural activity cycles through the components in the cyclic order of the directional cycle, despite the absence of a simply-embedded partition.  In this attractor, there is interesting activity among nodes $2, 3,$ and $4$ (the second component): the firing rate curve of node $2$ forms an ``envelope" over those of $3$ and $4$ as a result of the bidirectional edges $2 \leftrightarrow 3$ and $2 \leftrightarrow 4$.  For graph 31, we see that the activity also appears to cycle through the components in cyclic order, but the activity in the second component is not synchronized, despite this component being a clique; specifically, node $2$ fires at a higher rate than $3$, although they both peak at roughly the same time.  This is likely because node $2$ receives an extra input from node $4$ that node $3$ does not receive, which is precisely why this directional cycle representation does not correspond to a simply-embedded partition.  Finally, graph 39 has a single directional cycle representation, but this structure does not appear to give any insight into the pattern of neural activity in the corresponding attractor.

\begin{figure}[!hb]
\begin{center}
\includegraphics[width=6.35in]{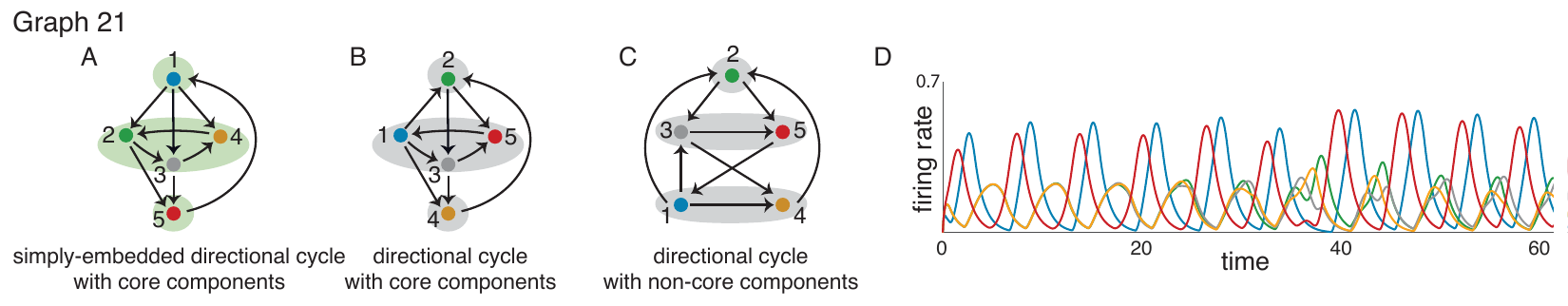}
\vspace{-.1in}
\end{center}
\caption{{\bf The three directional cycle representations of graph 21.} (A) The simply-embedded directional cycle representation of graph 21.  (B-C) Directional cycle representations that are not from a simply-embedded partition.  (D) The global attractor of the CTLN for graph 21 with $\varepsilon=0.51, \delta=1.76, \theta=1$.  The sequence of neural firing matches that of both of the directional cycle representations from A and B.  But the structure of the attractor (with neurons $1$ and $5$ high firing) is best represented by the simply-embedded directional cycle from A, since singleton components yield the high-firing neurons in directional cycles.  
}
\label{fig:dynamics-graph21}
\end{figure}

It is worth noting that a given graph may have multiple directional cycle representations.   Figure~\ref{fig:dynamics-graph21} shows the 3 directional cycle representations of graph 21 together with its sequential attractor.  We see that the simply-embedded directional cycle is the best predictor of the attractor since it not only reflects the sequence in which the nodes will fire, but it also predicts that nodes $1$ and $5$ will be the highest firing, as they are the singleton components. In Figure~\ref{fig:n5-core-examples}, we have the same phenomenon for graph 13: it has seven directional cycle representations, but only the one shown is simply-embedded, and this is the only 
representation that reflects the activity in the attractor.

From these examples, it appears that directional cycle architecture and simply-embedded partitions are useful for understanding the structure of the sequential attractors of a network.  The role of simply-embedded partitions is somewhat surprising given that this structure was primarily proposed for its utility in narrowing the set of possible fixed point supports, rather than for any explicit expected impact on neural activity.  One possible explanation for the importance of simply-embedded partitions in predicting dynamics is that in such a partition, every node within a component receives identical inputs from the rest of the graph.  This makes it more likely that nodes within a simply-embedded component will fire together or at least in close sequence in the corresponding attractor.  Thus, whenever a graph has a nontrivial simply-embedded partition, only directional cycle structures that respect that partition are likely to predict the sequential structure of the attractor.


\section{Conclusion}
In this work, we investigated different architectures that give rise to sequential attractors.  We considered cyclic unions and two generalizations of cyclic union architecture: directional cycles and simply-embedded partitions.  Computationally, we have seen that directional cycles yield sequential attractors that traverse the components in cyclic order.  While we cannot yet prove that directional cycles produce these cyclic dynamics, Theorem~\ref{thm:dir-cycle} showed that every fixed point support contains a cycle that hits every component in cyclic order.  Moreover, we have proven even stronger results about the fixed points of cyclic unions, simply-embedded partitions, and simply-embedded directional cycles (Theorems~\ref{thm:cyclic-unions},~\ref{thm:menu}, and~\ref{thm:sa-dir-cycle}).  

Additionally, in Section~\ref{sec:applications}, we analyzed the sequential attractors of various networks of size $n=5$ through the lens of directional cycle and simply-embedded architectures.  We found that whenever a graph has simply-embedded directional cycle structure, this architecture nicely predicts the sequential activity of the dynamic attractor.  Moreover, directional cycles can provide insight into the expected sequential activity even when the partition is not simply-embedded.   Thus, we conclude that directional cycle and simply-embedded architectures provide a useful framework for predicting (as well as engineering) persistent sequential activity in threshold-linear networks.

\paragraph{Acknowledgements.}
This work was supported by NIH R01 EB022862 (CC \& KM), NIH R01 NS120581 (CC), NSF DMS-1951165 (CC), and NSF DMS-1951599 (KM). 

\bibliographystyle{unsrt}
\bibliography{CTLN-refs}

\newpage
\section{Appendix: Supplemental Materials}
\subsection{Background on fixed points and simply-added splits} \label{sec:fp-conditions}
\paragraph{Characterizations of fixed point supports.\\}

To exploit previous characterizations of fixed points in terms of their supports \cite{fp-paper}, we will restrict consideration to CTLNs that are {\bf nondegenerate}, as defined below.

\begin{definition} \label{def:nondegenerate}
We say that a CTLN $W=W(G, \varepsilon, \delta)$ is {\it nondegenerate} if 
\begin{itemize}
\item $\det(I-W_\sigma) \neq 0$ for each $\sigma \subseteq [n]$, and 
\item for each $\sigma \subseteq [n]$ and all $i \in \sigma$,  the corresponding Cramer's determinant is nonzero: $\det((I-W_\sigma)_i;\theta) \neq 0$. 
\end{itemize}
\end{definition}
\noindent Note that almost all CTLNs are nondegenerate, since having a zero determinant is a highly fine-tuned condition.  The notation $\det(A_i; b)$ denotes the determinant obtained by replacing the $i^{\text{th}}$ column of $A$ with the vector $b$, as in Cramer's rule.   In the case of a restricted matrix, $((A_\sigma)_i;b_\sigma)$ denotes the matrix obtained from $A_\sigma$ by replacing the column corresponding to the index $i \in \sigma$ with $b_\sigma$ (note that this is not typically the $i^{\text{th}}$ column of $A_\sigma$).

When a CTLN is nondegenerate, there can be at most one fixed point per support.  Specifically, if $x^*$ is a fixed point with support $\sigma$, then for all $i \in \sigma$, we have $x^*_i=x^\sigma_i$ where
\begin{equation}\label{eq:xsigma}
x^\sigma \od \theta (I-W_\sigma)^{-1} 1_\sigma, \quad
\end{equation}
and for all $k \notin \sigma$, we have $x^*_k=0$.  (Note that $1_\sigma$ denotes the vector of all ones with length $|\sigma|$.) To check if a given subset $\sigma \subseteq [n]$ is the support of a fixed point of a CTLN $W=W(G, \varepsilon, \delta)$, one method is to compute the putative value of the fixed point via Equation~\eqref{eq:xsigma} and see if it actually satisfies the TLN equations.  Specifically, we see that $\sigma$ is the support of a fixed point of $W$ if and only if 
\begin{itemize}
\item[(i)] $x_i^\sigma > 0$ for all $i \in \sigma$ (``on"-neuron conditions), and
\item[(ii)]  $\displaystyle\sum_{i\in\sigma} W_{ki}x_i^\sigma+\theta \leq 0$ for all $k \notin \sigma$ (``off"-neuron conditions). 
\end{itemize}
(This is straightforward, but see \cite{pattern-completion} for more details.) Intuitively, $\sigma$ is the support of a fixed point of the CTLN if the fixed point $x^\sigma$ of the linear system restricted to $\sigma$ has only positive entries, so that all the neurons in $\sigma$ are ``on" at the fixed point, and if the inputs to all the external nodes are sufficiently inhibitory (negative) to ensure that those external neurons remain ``off".   Since condition (i) above  only depends on $W_\sigma$, a necessary condition 
 for $\sigma \in \FP(G)$ is that $\sigma \in \FP(G|_\sigma),$ where $G|_\sigma$ refers to the subgraph of $G$ obtained by restricting to the vertices of $\sigma$ and the edges between them.  A fixed point $\sigma \in \FP(G|_\sigma)$ \emph{survives} the addition of other nodes $k \notin \sigma$ precisely when condition (ii) is satisfied.

Unfortunately, the ``on" and ``off"-neuron characterization of fixed point supports relies on actually solving for a fixed point using $(I-W_\sigma)^{-1}$, and thus is difficult to directly connect to the graph structure encoded in $W = W(G, \varepsilon, \delta)$. In \cite{fp-paper}, an alternative characterization was developed in terms of Cramer's determinants (which are directly related to the values of $x_i^\sigma$ by Cramer's rule).  Specifically, for any $\sigma \subseteq [n],$ we define $s_i^\sigma$ to be the relevant Cramer's determinant:
\begin{equation}\label{eq:s_i}
s_i^\sigma \od \det((I-W_{\sigma\cup\{i\}})_i;b_{\sigma\cup\{i\}}), \;\; \text{for each} \;\; i \in [n].
\end{equation} 

\noindent In \cite[Lemma 2]{fp-paper}, a formula for $s_k^\sigma$ was proven that directly connects it to the relevant quantity in the ``off"-neuron condition:

\begin{equation}\label{eq:s_k^sigma}
s_k^\sigma = \sum_{i \in \sigma} W_{ki}s_i^\sigma + \theta \det(I-W_\sigma) \text{ for any } k \in [n].
\end{equation}

Combining this with Cramer's rule, it was shown that $\FP(G)$ can be fully characterized in terms of the \emph{signs} of the $s_i^\sigma$.  It turns out these signs are also connected to the \emph{index} of a fixed point. For each fixed point of a CTLN $W= W(G, \varepsilon, \delta)$, labeled by its support $\sigma \in \FP(G)$, we define the {\em index} as
$$\idx(\sigma) \od \sgn \det(I-W_\sigma).$$
Since we assume our CTLNs are nondegenerate, $\det(I-W_\sigma) \neq 0$ and thus $\idx(\sigma) \in \{\pm 1\}$.  

\begin{theorem}[sign conditions (Theorem 2 in \cite{fp-paper})] \label{thm:sgn-conditions}
Let $G$ be a graph on $n$ neurons and $W=W(G, \varepsilon, \delta)$ be a CTLN with graph $G$.  For any nonempty $\sigma~\subseteq~[n]$,
$$\sigma \text{ is a permitted motif }  \;\; \Leftrightarrow \;\; \sgn s_i^\sigma = \sgn s_j^\sigma
\text{ for all } i,j \in \sigma.$$
When $\sigma$ is permitted, 
$\sgn s_i^\sigma = \sgn\det(I-W_\sigma) = \idx(\sigma)$ for all $i \in \sigma$.\\
Furthermore,
\vspace{-.15in}
$$\sigma \in \FP(G)  \;\; \Leftrightarrow \;\; \sgn s_i^\sigma = \sgn s_j^\sigma = -\sgn s_k^\sigma
\text{ for all } i,j \in \sigma,\; k \not\in \sigma.$$
 \end{theorem} 

From this result, we immediately obtain the following corollary. 

\begin{corollary}[Corollary 2 in \cite{fp-paper}]\label{cor:inheritance}
Let $\sigma\subseteq [n]$. The following are equivalent:
\begin{enumerate}
	\item $\sigma \in \FP(G)$
	\item $\sigma \in \FP(G|_{\tau})$ for all $\sigma \subseteq \tau \subseteq [n]$
	\item $\sigma \in \FP(G|_{\sigma})$ and $\sigma \in \FP(G|_{\sigma \cup k})$ for all $k \notin \sigma$
	\item $\sigma \in \FP(G|_{\sigma \cup k})$ for all $k \notin \sigma$
\end{enumerate}
\end{corollary}

This shows that for $\sigma$ to support a fixed point of the full network, it must support a fixed point in its own subnetwork, as well as every other subnetwork in between.  Moreover, by (3), it is possible to check survival just one external node $k$ at a time.  Note that survival of an added node $k$ is fully determined by $\sgn s_k^\sigma$ by Theorem~\ref{thm:sgn-conditions}.  Moreover, since  $s_k^\sigma = \sum_{i \in \sigma} W_{ki}s_i^\sigma + \theta \det(I-W_\sigma)$, we see that $\sgn s_k^\sigma$ only depends on the \emph{outgoing edges} from $\sigma$ to $k$ (captured in $W_{ki}$ values) as well as the edges within $\sigma$ (reflected in $s_i^\sigma$ and $\det(I-W_\sigma)$).  Thus, only the outgoing edges from $\sigma$ are relevant to its survival in a larger network.

\paragraph{Background on simply-added splits.}

It turns out that the $s_i^\sigma$ are easy to compute when a graph has simply-added structure.  Recall that in a simply-embedded partition, every node within a component receives identical incoming edges from the rest of the graph.  This is a special case of the more general notion of a \emph{simply-added split}.

\begin{definition}[simply-added split]
Let $G$ be a graph on $n$ nodes.  For any nonempty $\omega,~ \tau\subseteq [n]$ such that $\omega \cap \tau= \emptyset$, we say $\omega$ is \emph{simply-added} onto $\tau$ if for each $j \in \omega$, either $j$ is a \emph{projector} onto $\tau$, i.e., $j \to k$ for all $k \in \tau$, or $j$ is a \emph{nonprojector} onto $\tau$, so $j \not\to k$ for all $k \in \tau$. In this case, we say that $\tau$ is \emph{simply-embedded} in $G$, and we say that $(\omega, \tau)$ is a \emph{simply-added split} of the subgraph $G|_\sigma$, for $\sigma = \omega\cup\tau $.
\end{definition}

Note that when a graph has a simply-embedded partition $\{\tau_1|\cdots|\tau_N\}$, we have a simply-added split for every $\tau_i$; specifically, $[n] \setminus \tau_i$ is simply-added onto $\tau_i$, since by definition, $\tau_i$ is simply-embedded in $G$.   In \cite{fp-paper}, it was shown that whenever a simply-added split exists, we can understand many of the $s_i^\sigma$ values as scalings of $s_i^\tau$ from the smaller component subgraph $G|_\tau$.

\begin{theorem}[Theorem 3 in \cite{fp-paper}]\label{thm:simply-added}
Let $G$ be a graph on $n$ nodes, and let $\omega,~ \tau\subseteq [n]$ be such that $\omega$ is simply-added to $\tau$. For $\sigma \subseteq \omega \cup \tau$, define $\sigma_\omega \od \sigma \cap \omega$ and $\sigma_\tau \od \sigma \cap \tau$.  Then 
$$s_i^\sigma = \frac{1}{\theta}s_i^{\sigma_\omega} s^{\sigma_\tau}_i = \alpha s^{\sigma_\tau}_i \quad\text{for each $i \in \tau$,}$$ 
where $\alpha = \frac{1}{\theta}s_i^{\sigma_\omega}$ has the same value for every $i \in \tau$.
\end{theorem}


\subsection{Proofs of Theorems~\ref{thm:menu} and other results on simply-embedded partitions}\label{sec:menu-proof}

Theorem~\ref{thm:simply-added} can immediately be leveraged for simply-embedded partitions to connect the $s_j^\sigma$ values to the $s_j^{\sigma_i}$ values from the component subgraphs.  This will be key to the proof of Theorem~\ref{thm:menu}.

\begin{lemma}\label{lemma:simply-added}
Let $G$ have a simply-embedded partition $\{\tau_1|\cdots|\tau_N\}$, and consider $\sigma \subseteq [n]$. Let $\sigma_i \od \sigma \cap \tau_i$.  Then for any $\sigma_i \neq \emptyset$,
\vspace{-.05in}
$$\sgn s_j^\sigma = \sgn s^\sigma_k \quad \Leftrightarrow \quad \sgn s_j^{\sigma_i} = \sgn s_k^{\sigma_i}, \quad\text{for all $j,k \in \tau_i$}.$$
\end{lemma}
\begin{proof}
By definition of simply-embedded partition, $G$ has a simply-added split where $[n]\setminus \tau_i$ is simply-added onto $\tau_i$ (and thus also onto $\sigma_i$). Thus by Theorem~\ref{thm:simply-added}, $s_j^\sigma = \alpha s_j^{\sigma_i}$, where $\alpha = \frac{1}{\theta} s_j^{\sigma \setminus \sigma_i}$  is identical for all $j\in \tau_i$.  Hence, for all $j,k \in \tau_i$, we have that $\sgn s_j^\sigma = \sgn s^\sigma_k$ if and only if  $\sgn \alpha s_j^{\sigma_i} = \sgn \alpha s_k^{\sigma_i}$  if and only if $\sgn s_j^{\sigma_i} = \sgn s_k^{\sigma_i}$.

\end{proof}

Theorem~\ref{thm:menu} (reprinted below) now follows directly from Lemma~\ref{lemma:simply-added} together with the sign conditions characterization of fixed point supports (Theorem~\ref{thm:sgn-conditions}).\\

\noindent{\bf Theorem~\ref{thm:menu}} ($\FP(G)$ menu for simply-embedded partitions).
Let $G$ have a simply-embedded partition $\{\tau_1|\cdots|\tau_N\}$.  For any $\sigma \subseteq [n]$, let $\sigma_i \od \sigma \cap \tau_i$.  Then 
$$\sigma \in \FP(G) \quad \Rightarrow \quad \sigma_i \in \FP(G|_{\tau_i})\cup \{\emptyset\}~~\text{ for all } i \in [N]. $$ 
In other words, every fixed point support of $G$ is a union of component fixed point supports $\sigma_i$, at most one per component.

\begin{proof}
For $\sigma \in \FP(G)$, we have 
$$\sgn s_j^\sigma = \sgn s_k^\sigma = -\sgn s_l^\sigma$$
 for any $j,k \in \sigma_i$ and $l \in \tau_i\setminus\sigma_i$, by Theorem~\ref{thm:sgn-conditions} (sign conditions). Then by Lemma \ref{lemma:simply-added}, we see that whenever $\sigma_i \neq \emptyset$, 
 $$\sgn s_j^{\sigma_i} = \sgn s_k^{\sigma_i} = -\sgn_l^{\sigma_i},$$  and so $\sigma_i$ satisfies the sign conditions in $G|_{\tau_i}$. Thus $\sigma_i \in \FP(G|_{\tau_i})$ for every nonempty $\sigma_i$.
\end{proof}

Next we prove that whenever a graph $G$ has a simply-embedded partition and there is a locally removable node (i.e. a node whose removal does not affect its component $\FP(G|_{\tau_i})$), then that node is also globally removable with no impact on $\FP(G)$ (Theorem~\ref{thm:removables} reprinted below for convenience).\\  

\noindent{\bf Theorem~\ref{thm:removables}} (removable nodes). 
Let $G$ have a simply-embedded partition $\{\tau_1|\cdots|\tau_N\}$.  Suppose there exists a node $j \in \tau_i$ such that $\FP(G|_{\tau_i}) = \FP(G|_{\tau_i\setminus\{j\}})$.  Then $\FP(G) = \FP(G|_{[n]\setminus\{j\}}) $. 

\begin{proof}
To see that $\FP(G)\subseteq \FP(G|_{[n]\setminus\{j\}})$, notice that for all $\sigma \in \FP(G)$, we have $\sigma \subseteq [n] \setminus \{j\}$ by Theorem~\ref{thm:menu}.  Then by Corollary \ref{cor:inheritance}(2), we must have $\sigma \in \FP(G|_{[n]\setminus\{j\}})$, and so $\FP(G)\subseteq \FP(G|_{[n]\setminus\{j\}})$.  

For the reverse containment, we will show that every fixed point in $\FP(G|_{[n]\setminus\{j\}})$ survives the addition of node $j$ by appealing to Theorem~\ref{thm:sgn-conditions} (sign conditions).   There are two cases to consider: $\sigma_i = \emptyset$ and $\sigma_i \neq \emptyset$, where $j \in \tau_i$ and $\sigma_i \od \sigma \cap \tau_i$.

\noindent\underline{Case 1:} $\sigma_i = \emptyset$.  Since $j$ is not contained in the support of any fixed point of $G|_{\tau_i}$, there must be at least one other node $k$ in $\tau_i$, since $\FP(G|_{\tau_i})$ cannot be empty.  Since $G$ is a simply-embedded partition, we have that $[n] \setminus \tau_i$ is simply-embedded onto $\tau_i$ meaning that every node in $\tau_i$ receives identical inputs from the rest of the graph. Recall from Equation~\eqref{eq:s_k^sigma}, that $s_j^\sigma = \sum_{\ell \in \sigma} W_{j\ell} s_\ell + \theta\det(I-W_\sigma)$.  Then since $\sigma \subseteq [n] \setminus \tau_i$, we have that $j$ and $k$ receive identical inputs from $\sigma$, so $W_{j\ell} = W_{k \ell}$ for all $\ell \in \sigma$, and thus $s_j^\sigma = s_k^\sigma$.  Since $\sigma \in \FP(G|_{[n]\setminus\{j\}})$, we have $\sgn s^\sigma_k = -\sgn s^\sigma_\ell$ for all $\ell \in \sigma$ by Theorem~\ref{thm:sgn-conditions} (sign conditions).  Thus, we also have $\sgn s^\sigma_j = -\sgn s^\sigma_\ell$ and $\sigma$ survives the addition of node $j$, so $\sigma \in \FP(G)$.    

\noindent\underline{Case 2:} $\sigma_i \neq \emptyset$.  First observe that $G|_{[n]\setminus\{j\}}$ has the same simply-embedded partition structure as $G$, but with $\tau_i \setminus\{j\}$ rather than $\tau_i$.  Thus $\sigma \in \FP(G|_{[n]\setminus\{j\}})$ implies that $\sigma_i \in \FP(G|_{\tau_i \setminus \{j\}})$ by Theorem~\ref{thm:menu} (menu).  By hypothesis, $\FP(G|_{\tau_i \setminus \{j\}}) = \FP(G|_{\tau_i})$, and so $\sigma_i \in \FP(G|_{\tau_i})$.  Then by Theorem~\ref{thm:sgn-conditions} (sign conditions), since $j \not\in \sigma_i$, we have $\sgn s_j^{\sigma_i}= - \sgn s_\ell^{\sigma_i}$ for all $\ell \in \sigma_i$.  And by Lemma \ref{lemma:simply-added}, this ensures $\sgn s_j^{\sigma}= - \sgn s_\ell^{\sigma}$ for all $\ell \in \sigma_i$.  Since $\sigma \in \FP(G|_{[n]\setminus\{j\}})$, we have that $\sgn s_\ell^\sigma$ is identical for all $\ell \in \sigma$, not just $\ell \in \sigma_i$, and so $\sgn s_j^{\sigma}= - \sgn s_\ell^{\sigma}$ for all $\ell \in \sigma$.  Thus by Theorem~\ref{thm:sgn-conditions} (sign conditions), $\sigma$ survives the addition of node $j$, so $\sigma \in \FP(G)$.    
\end{proof}

\noindent{\bf Corollary~\ref{cor:removables}.} Let $G$ have a simply-embedded partition $\{\tau_1|\cdots|\tau_N\}$ and suppose there exists $j\in\tau_i$ such that $\FP(G|_{\tau_i}) = \FP(G|_{\tau_i\setminus\{j\}})$. Let $G'$ be any graph that can be obtained from $G$ by deleting or adding all the outgoing edges from $j$ to any component $\tau_k$ with $k \neq i$.  Then $\FP(G') = \FP(G)$.

\begin{proof}
Observe that by deleting all the outgoing edges from $j$ to a component $\tau_k$, node $j$ has simply changed from a projector onto $\tau_k$ to a nonprojector.  Alternatively, by adding all the outgoing edges to $\tau_k$, node $j$ switches from being a nonprojector onto $\tau_k$ to being a projector.  In either case, $j$ is still simply-added onto $\tau_k$, and so $G'$ has the same simply-embedded partition $\{\tau_1|\cdots|\tau_N\}$ as $G$ had.  Additionally, since no edges within $\tau_i$ have been altered, we have that 
$\FP(G'|_{\tau_i}) = \FP(G|_{\tau_i})= \FP(G|_{\tau_i\setminus\{j\}}) = \FP(G'|_{\tau_i\setminus\{j\}}).$
Thus both $G$ and $G'$ satisfy the hypotheses of Theorem~\ref{thm:removables}.  Moreover, $G|_{[n]\setminus\{j\}} = G'|_{[n]\setminus\{j\}}$ since the only differences between $G$ and $G'$ were in edges involving node $j$, which has been removed.  Thus, by Theorem~\ref{thm:removables},  $\FP(G) = \FP(G|_{[n]\setminus\{j\}}) = \FP(G').$
\end{proof}


\subsection{Background on bidirectional simply-added splits}

In order to prove the properties of $\FP(G)$ for simple linear chains and strongly simply-embedded partitions, we first need to review some background from \cite{fp-paper} on \emph{bidirectional simply-added splits}.  These are partitions into two components in which each component is simply-added onto the other component (so the simply-added property is bidirectional).

	\begin{definition}[bidirectional simply-added split]
Let $G$ be a graph on $n$ nodes.  For any nonempty $\omega,~ \tau\subseteq [n]$ such that $[n] = \omega \cup \tau$ and $\omega \cap \tau= \emptyset$, we say that $G$ has a \emph{bidirectional simply-added split} $(\omega, \tau)$ if $\omega$ is simply-added onto $\tau$ and $\tau$ is simply-added onto $\omega$.  In other words, for all $j \in \omega$, either $j \to k$ for all $k \in \tau$ or $j \not\to k$ for all $k \in \tau$, \emph{and} for all $k \in \tau$, either $k \to j$ for all $j \in \omega$ or $k \not\to j$ for all $j \in \omega$.  
\end{definition}

\begin{figure}[!h]
\begin{center}
\vspace{-.1in}
\includegraphics[height=1.15in]{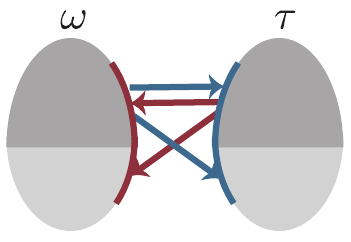}
\vspace{-.15in}
\end{center}
\caption{\textbf{Bidirectional simply-added split.} In this graph $\omega$ is simply-added to $\tau$ and vice versa.  Thus $\omega$ is composed of two classes of nodes: projectors onto $\tau$ (top dark gray region) and nonprojectors onto $\tau$ (bottom light gray region).  Similarly, $\tau$ can be decomposed into projectors and nonprojectors onto $\omega$.  The thick colored arrows indicate that every node of a given region sends an edge to every node in the other region. The edges within $\omega$ and $\tau$ can be arbitrary. }
\label{fig:bidirectional-sa}
\vspace{-.1in}
\end{figure}

Note that a simply-embedded partition consisting of just two components $\{\tau_1~|~\tau_2\}$ is a bidirectional simply-added split.  But with larger simply-embedded partitions, $\{\tau_1|\cdots |\tau_N\}$, it is not generally true that $(\tau_i, [n] \setminus \tau_i)$ is a bidirectional simply-added split.  However, \emph{strongly simply-embedded partitions} will always satisfy that $(\tau_i, [n] \setminus \tau_i)$ is a bidirectional simply-added split.  This is because in a strongly simply-embedded partition, any $j \in \tau_i$ treats all the other components identically, so it is either a projector or a non-projector onto all of $[n] \setminus \tau_i$.  

In \cite{fp-paper}, it was shown that $\FP(G)$ is fully determined by the fixed points of the component subgraphs $G|_\omega$ and $G|_\tau$ when $(\omega, \tau)$ is a bidirectional simply-added split.  To make this characterization precise, we first need some notation.  For any $\omega \subseteq [n]$, let $S_\omega$ denote the fixed point supports of $G|_\omega$ that survive to be fixed points of $G$, and let $D_\omega$ denote the non-surviving (dying) fixed points:
$$S_\omega \od \FP(G|_\omega) \cap \FP(G), \quad \text{and} \quad D_\omega \od \FP(G|_\omega)\setminus S_\omega.$$

\begin{theorem}[Theorem 14 in \cite{fp-paper}]\label{thm:bidir-sa}
Let $G$ be a graph with bidirectional simply-added split $[n] = \omega \cup \tau$. For any nonempty $\sigma \subseteq [n]$, let $\sigma = \sigma_\omega \cup \sigma_\tau$ where
$\sigma_\omega \od \sigma \cap \omega$ and $\sigma_\tau \od \sigma \cap \tau$. Then
$\sigma \in \FP(G)$ if and only if one of the following holds:
\begin{itemize}
\item[(i)]$ \sigma_\tau \in S_\tau \cup \{\emptyset\}\;\; \text{and}\;\; \sigma_\omega \in S_\omega \cup \{\emptyset\}, \;\; \text{or}$
\item[(ii)] $\sigma_\tau \in D_\tau \;\; \text{and}\;\; \sigma_\omega \in D_\omega.$
\end{itemize}
In other words, $\sigma \in \FP(G)$ if and only if $\sigma$ is either a union of surviving fixed points $\sigma_i$, at most one from $\omega$ and at most one from $\tau$, or it is a union of dying fixed points, exactly one from $\omega$ and one from $\tau$.
\end{theorem}

We will see that both simple linear chains and strongly simply-embedded partitions have bidirectional simply-added splits within them, and so Theorem~\ref{thm:bidir-sa} will be key to the proofs characterizing their $\FP(G)$.  First, though, we take a brief detour to explore the special case of bidirectional simply-added splits with singletons in a component, in order to see some special internal structure of $\FP(G)$ in these cases.

\subsection{Internal structure of $\FP(G)$ with singletons}

A special case of a bidirectional simply-added split occurs whenever a graph contains a node that is projector/nonprojector onto the rest of the graph.  Specifically, since any subset is always simply-added onto a single node $j$ trivially, we see that we have a bidirectional simply-added split $(\{j\}, [n]\setminus \{j\})$ whenever $j$ is either a projector or a nonprojector onto the rest of the graph.  Recall that if $j$ is a nonprojector onto $[n]\setminus \{j\}$, then $j$ has no outgoing edges in $G$, and so it is a \emph{sink}.  Moreover, we have seen that sinks are the only single nodes that can support fixed points since a singleton $\{j\}$ is trivially uniform in-degree 0, and thus only survives when it has no outgoing edges, by Rule~\ref{rule:uniform-in-deg}.  Combining this observation with the bidirectional simply-added split for a sink, we see there is certain internal structure that must be present in $\FP(G)$ whenever it contains any singleton sets.  

\begin{proposition}\label{prop:FP-with-singleton}
Let $G$ be a graph such that there is some singleton $\{j\} \in \FP(G)$.  Then for any $\sigma \in \FP(G)$ (with $\sigma \neq \{j\}$), 
\begin{enumerate}[label=(\arabic*)]
\item If $j \notin \sigma$, then $\sigma \cup \{j\} \in \FP(G)$; i.e., $\FP(G)$ is closed under unions with singletons.
\item If $j \in \sigma$, then $\sigma \setminus \{j\} \in \FP(G)$; i.e., $\FP(G)$ is closed under set differences with singletons.
\end{enumerate}
\end{proposition}

\begin{proof}
First notice that since $\{j\} \in \FP(G)$, $j$ is a sink in $G$ by Rule \ref{rule:uniform-in-deg} (since a singleton is trivially uniform in-degree 0, and thus survives exactly when it has no outgoing edges), and therefore $(\{j\}, [n]\setminus \{j\})$ is a bidirectional simply-added split. 

To prove (1), suppose $j\notin\sigma$. Since $(\{j\}, [n]\setminus \{j\})$ is a bidirectional simply-added split, Theorem~\ref{thm:bidir-sa} guarantees that $\sigma\cup \{j\} \in \FP(G)$ if and only if $\{j\}, \sigma$ both survive or both die.  By assumption, both sets are in $\FP(G)$, so both survive.  Thus,  $\sigma \cup \{j\} \in \FP(G)$.

To prove (2), suppose $j \in \sigma$. By Theorem \ref{thm:bidir-sa}, $\sigma \in \FP(G)$ if and only if $\{j\}, \sigma \setminus \{j\}$ both survive or both die. By assumption, $\{j\} \in \FP(G)$, and so $\sigma \setminus \{j\} \in \FP(G)$ as well.
\end{proof}

\begin{corollary}
Let $G$ be a graph such that $\FP(G)$ contains singleton sets  $\{j_1\}, \{j_2\}, \ldots, \{j_\ell\}$, and let $\S=\{j_1, \ldots, j_\ell\}$ be the set of singletons. Then for any $\sigma \in \FP(G)$ and any $\omega \subseteq \S$
\vspace{-.1in}
$$\sigma \cup \omega \in \FP(G).$$
Moreover, let $\tau = [n] \setminus  \mathcal{S}$.  Then $\FP(G)$ has the direct product structure:
$$\FP(G) \cup \{\emptyset\}  \cong  \left(\{\sigma \in \FP(G|_{\tau})~|~ \sigma\in \FP(G) \} \cup \{\emptyset\} \right) \times \mathcal{P}( \mathcal{S}),$$
where $\mathcal{P}( \mathcal{S})$ denotes the power set of $\S$.  In other words, every fixed point support in $\FP(G)$ has the form $\sigma \cup \omega$ where $\sigma \in \FP(G|_{\tau}) \cup \{\emptyset\}$ and $\omega \subseteq  \S$.
\end{corollary}

\begin{proof}
The first statement follows by iterating Proposition~\ref{prop:FP-with-singleton}(1) $|\omega|$ times for each of the added singletons in $\omega$. To prove the second statement, we will show that every $\nu \in \FP(G)$ is the union of a surviving fixed point $\sigma \subseteq \tau$ (or the empty set) with a subset of $\S$ (including empty set); moreover, every such union yields a fixed point (other than $\emptyset \cup \emptyset$).  The direct product structure of $\FP(G)$ immediately follows from this decomposition of the fixed point supports.  By the first result, we see that every such union is contained in $\FP(G)$.  Thus, all that remains to show is that every element of $\FP(G)$ is such a union.  Let $\nu \in \FP(G)$ and let $\sigma = \nu \cap \tau$ and $\omega = \nu \cap \S$, so that $\nu = \sigma \cup \omega$.  If $\sigma$ or $\omega$ are empty, then we're done, so suppose both are nonempty.  Then we can iteratively apply Proposition~\ref{prop:FP-with-singleton}(2) $|\omega|$ times to see that $\sigma \in \FP(G)$.  Thus, every fixed point support arises as a union of some $\sigma \subseteq \tau$ with an arbitrary subset of $\S$, where $\sigma\in \FP(G)\cup \{\emptyset\}$ (and for every $\sigma \in \FP(G)$, we have $\sigma \in \FP(G|_\tau)$ as well by Corollary~\ref{cor:inheritance}(2)). 
\end{proof}


\subsection{Simple linear chain proofs}\label{sec:linear-chains-proofs}
In this section, we prove Theorem~\ref{thm:linear-chain} showing that $\FP(G)$ for a simple linear chain is closed under unions of component fixed points $\sigma_i$ that survive in $G|_{\tau_i \cup \tau_{i+1}}$.  The proof relies on the existence of a bidirectional simply-added split within a simple linear chain between the first $N-1$ components of the chain and $\tau_N$.  

Another key to the proof is the fact that if $\sigma_i \in \FP(G|_{\tau_i \cup \tau_{i+1}})$, then it turns out that $\sigma_i \in \FP(G)$; in other words, survival of the addition of the next component is sufficient to guarantee survival in the full network.  This occurs because $\sigma_i$ has no outgoing edges to any nodes outside of $\tau_i \cup \tau_{i+1}$.  Lemma~\ref{lemma:no-out-edges} shows that whenever a permitted motif has no outgoing edges to a node $k$, then it is guaranteed to survive the addition of node $k$.

\begin{lemma} \label{lemma:no-out-edges}
Let $G$ be a graph on $n$ nodes,  let $\sigma \subseteq [n]$ be nonempty, and $k \in [n]\setminus \sigma$.  If $i \not\to k$ for all $i \in \sigma$, then 
\vspace{-.1in}
$$\sigma \in \FP(G|_{\sigma \cup \{k\}}) \quad \Leftrightarrow \quad \sigma \in \FP(G|_{\sigma}).$$
In other words, if $\sigma$ has no outgoing edges to node $k$ then $\sigma$ is guaranteed to survive the addition of node $k$ whenever $\sigma$ is a permitted motif.  
\end{lemma}

\begin{proof}
For any $j \in \sigma$, we have that $j$ inside-out dominates $k$. Thus by Rule~\ref{rule:graph-domination}c,  $\sigma \in \FP(G|_{\sigma \cup \{k\}})$ if and only if $\sigma \in \FP(G|_{\sigma})$.
\end{proof}

The proof of Lemma~\ref{lemma:no-out-edges} illustrated how inside-out graphical domination can be used to guarantee survival of a permitted motif.  The presence of such a graphical domination relationship is a sufficient condition to guarantee survival, but unfortunately it is not a necessary condition, so the absence of such a relationship does not guarantee that a permitted motif does \emph{not} survive.  It turns out though, that graphical domination is a special case of \emph{general domination}, and the presence/absence of a general domination relationship does precisely characterize survival of a fixed point support.  To complete the proof of Theorem~\ref{thm:linear-chain}, we must appeal to general domination, and so we briefly review that concept here and the complete characterization of fixed point supports that it provides.  (For a more detailed discussion of general domination, see Section 6 of \cite{fp-paper}).  

Recall that Theorem~\ref{thm:sgn-conditions} (sign conditions) gives a complete characterization of when a subset $\sigma$ supports a fixed point in terms of the \emph{signs} of the Cramer's determinants $s_i^\sigma$.  For general domination, these Cramer's determinants again play a key role, but in this case it will be the \emph{magnitudes} of $s_i^\sigma$ that are relevant, irrespective of their signs.  Specifically, for any $j \in [n]$, we define the relevant domination quantity:
$$w_j^{\sigma} = \sum_{i \in \sigma}\widetilde{W}_{ji}|s_i^\sigma|,$$
where $\widetilde{W} = -I+W$, so that $\widetilde{W}_{ji} = W_{ji}$ if $j \neq i$ and $\widetilde{W}_{ji} = -1$ if $j = i$.  

We say that \emph{$k$ dominates $j$ with respect to $\sigma$}, if $w_k^\sigma > w_j^\sigma$.  It turns out that $\sigma \in \FP(G)$ precisely when these domination quantities are perfectly balanced within $\sigma$, so that $\sigma$ is \emph{domination-free}, and when every external node $k \notin \sigma$ is inside-out dominated by nodes inside $\sigma$:

\begin{theorem}[general domination ([Theorem 15 in \cite{fp-paper})] \label{thm:gen-domination}
Let $G$ be a graph on $n$ neurons and $W=W(G, \varepsilon, \delta)$ be a CTLN with graph $G$, and consider $\sigma \subseteq [n]$. Let $\widetilde{W} = -I+W$ and $w_j^{\sigma}$ be as above. Then 
$$\sigma \in \FP(G|_{\sigma}) \quad \Leftrightarrow \quad w_i^\sigma = w_j^\sigma \text{ for all } {i,j\in\sigma}.$$ 
If $\sigma \in \FP(G|_{\sigma})$, then $\sigma \in \FP(G)$ if and only if for each $k \notin \sigma$, there exists $j \in \sigma$ such that $w_j^\sigma > w_k^\sigma$, i.e.\ such that $j$ inside-out dominates $k$.
\end{theorem}

It turns out that the simply-embedded partition structure of the simple linear chain with the added restriction that $\tau_i$ does not send edges to any $\tau_k$ other than $\tau_{i+1}$ gives significant structure to the values of $s_i^\sigma$ and thus to the domination quantities $w_j^\sigma$.  This structure is the key to the proof of Theorem~\ref{thm:linear-chain}.\\

\noindent{\bf Theorem~\ref{thm:linear-chain}} (simple linear chains).
Let $G$ be a simple linear chain with components $\tau_1, \ldots, \tau_N$.  
\begin{enumerate}
\item[(i)] If $\sigma \in \FP(G)$, then $\sigma_i \in \FP(G|_{\tau_i}) \cup \{\emptyset\}$ for all $i \in [N]$, where $\sigma_i = \sigma \cap \tau_i$. 
\item[(ii)] Consider a collection $\{\sigma_i\}_{i \in [N]}$ of $\sigma_i \in \FP(G|_{\tau_i}) \cup \{\emptyset\}$.  If additionally $\sigma_i \in \FP(G|_{\tau_i\cup\tau_{i+1}}) \cup \{\emptyset\}$ for all $i \in [N]$, then 
$$\bigcup_{i \in [N]} \sigma_i \in \FP(G).$$
\end{enumerate}
\vspace{-.05in}
\noindent In other words, $\FP(G)$ is closed under unions of component fixed point supports that survive in $G|_{\tau_i\cup\tau_{i+1}}$.

\begin{proof}
(i) follows directly from Theorem~\ref{thm:menu} by noting that the simple linear chain structure endows $G$ with a simply-embedded partition: 
for every $\tau_i$, the nodes in $\tau_{i-1}$ are each either a projector or nonprojector onto $\tau_i$, while all nodes outside of $\tau_{i-1}$ are all nonprojectors onto $\tau_i$.

To prove (ii), consider $\{\sigma_i\}_{i \in [N]}$ where $\sigma_i\in\FP(G|_{\tau_i \cup \tau_{i+1}}) \cup \{\emptyset\}$ for all $i \in [N]$.   Notice that by Lemma \ref{lemma:no-out-edges}, the fact that $\sigma_i\in\FP(G|_{\tau_i \cup \tau_{i+1}})$ implies that $\sigma_i \in \FP(G)$ since $\sigma_i$ has no outgoing edges to any external node $k$ outside of $\tau_i \cup \tau_{i+1}$.   Thus, we may assume $\sigma_i \in \FP(G) \cup\{\emptyset\}$ for all $i \in [N]$.  We will prove that this guarantees that $\displaystyle \cup_{i \in [N]} \sigma_i \in \FP(G)$ by induction on the number $N$ of components of the simple linear chain. 

For $N = 1$, the result is trivially true. For $N = 2$, observe that the simple linear chain on $\{\tau_1~|~\tau_2\}$ actually has the structure of a bidirectional simply-embedded split $(\tau_1, \tau_2)$, and thus Theorem~\ref{thm:bidir-sa} gives the complete structure of $\FP(G)$ in terms of the surviving fixed points of the component subgraphs $S_{\tau_i}$ and the dying fixed points $D_{\tau_i}$.  The sets of interest here, $\sigma_i \subseteq \tau_i$ with $\sigma_i \in \FP(G)$, are precisely the elements of $S_{\tau_i}$.  Theorem~\ref{thm:bidir-sa}(1) then guarantees that $\sigma_1 \cup \sigma_2 \in \FP(G)$ whenever $\sigma_i \in \FP(G)$, and so the result holds when $N=2$.  

Now, suppose the result holds for any simple linear chain with $N-1$ components. For ease of notation, denote $\sigma_{1\cdots N-1}\od\sigma_1\cup\cdots \cup\sigma_{N-1}$ and let $\sigma \od \cup_{i \in [N]}\sigma_i$.  We will show the result holds for any simple linear chain $G$ with $N$ components.  

Observe that if $\sigma_N = \emptyset$, we have $\sigma = \sigma_{1 \cdots N-1} \in \FP(G|_{\tau_{1 \cdots N-1}})$ by the inductive hypothesis, and we need only show that this implies that $\sigma_{1\cdots N-1} \in \FP(G)$.  On the other hand, if $\sigma_N \neq \emptyset$, then $\sigma = \sigma_{1\cdots N-1} \cup \sigma_N$, where $\sigma_N \in \FP(G)$ by Lemma~\ref{lemma:no-out-edges}, since $\sigma_N \in \FP(G|_{\tau_N})$ and $\sigma_N$ has no outgoing edges to any external nodes outside of $\tau_N$.  Notice that the simple linear chain structure of $G$ ensures that $(\tau_{1 \cdots N-1}, \tau_N)$ is a bidirectional simply-embedded split.  Thus by Theorem~\ref{thm:bidir-sa}, since $\sigma_N$ is a surviving fixed point support, $\sigma_{1\cdots N-1} \cup \sigma_N \in \FP(G)$ if and only if $\sigma_{1\cdots N-1} \in \FP(G)$.  Therefore for any $\{\sigma_i\}_{i \in [N]}$, it suffices to show that $\sigma_{1\cdots N-1} \in \FP(G)$, and the result will follow.  

Notice that by the inductive hypothesis, $\sigma_{1 \cdots N-1} \in \FP(G|_{\tau_{1 \cdots N-1}})$, and thus to show $\sigma_{1\cdots N-1} \in \FP(G)$, we need only show that $\sigma_{1\cdots N-1}$ survives the addition of the nodes in $\tau_N$.  There are two cases to consider here based on whether $\sigma_{1\cdots N-1}$ intersects $\tau_{N-1}$ or not.  Observe that if $\sigma_{1\cdots N-1} \cap \tau_{N-1} = \emptyset$, then $\sigma_{1\cdots N-1}$ has no outgoing edges to $\tau_N$ since only nodes in $\tau_{N-1}$ can send edges forward to $\tau_N$ by the linear chain structure.  In this case, we have $i \not\to k$ for all $i \in \sigma_{1\cdots N-1}$ and all $k \in \tau_N$, and so Lemma~\ref{lemma:no-out-edges} guarantees that $\sigma_{1\cdots N-1} \in \FP(G)$ since we already had $\sigma_{1 \cdots N-1} \in \FP(G|_{\tau_{1 \cdots N-1}})$.  

For the other case where $\sigma_{1\cdots N-1} \cap \tau_{N-1} \neq \emptyset$, we will prove $\sigma_{1\cdots N-1} \in \FP(G)$ by appealing to Theorem~\ref{thm:gen-domination} (general domination) and demonstrating that each $k \in \tau_N$ is \emph{inside-out dominated} by some node $j \in \sigma_{1\cdots N-1}$.  First notice that $\sigma_{1\cdots N-1} = \sigma_{1\cdots N-2} \cup \sigma_{N-1}$ and by the simple linear chain structure of $G$, we have that $\tau_{1\cdots N-2}$ is simply-embedded onto $\tau_{N-1}$.  Thus by Theorem~\ref{thm:simply-added},
\begin{equation}\label{eqn:bidir-sa-split}
s_i^{\sigma_{1\cdots N-1}} = \frac{1}{\theta} s_i^{\sigma_{1\cdots N-2}} s_i^{\sigma_{N-1}} = \alpha s_i^{\sigma_{N-1}} \textrm{ for all } i \in \sigma_{N-1},
\end{equation}
where $\alpha = \frac{1}{\theta}s_i^{\sigma_{1\cdots N-2}}$ has the same value for every $i \in \sigma_{N-1}$.  Using this, we can now compute the domination quantities $w_j^{\sigma_{1\cdots N-1}}$ and $w_k^{\sigma_{1\cdots N-1}}$ for $j \in \sigma_{N-1}$ and $k \in \tau_N$.  For $j \in \sigma_{N-1}$, we have:
\begin{align*} 
w_j^{\sigma_{1\cdots N-1}} &\od \sum_{i \in \sigma_{1\cdots N-1}} \widetilde{W}_{ji} |s_i^{\sigma_{1\cdots N-1}}|\\ 
&=  \sum_{i \in \sigma_{1\cdots N-2}} \widetilde{W}_{ji} |s_i^{\sigma_{1\cdots N-1}}| + \sum_{i \in \sigma_{N-1}} \widetilde{W}_{ji} |s_i^{\sigma_{1\cdots N-1}}| \\
&=  \sum_{i \in \sigma_{1\cdots N-2}} \widetilde{W}_{ji} |s_i^{\sigma_{1\cdots N-1}}| + \sum_{i \in \sigma_{N-1}} \widetilde{W}_{ji} |\alpha s_i^{\sigma_{N-1}}| \quad\text{by \eqref{eqn:bidir-sa-split}}\\
&=  \sum_{i \in \sigma_{1\cdots N-2}} \widetilde{W}_{ji} |s_i^{\sigma_{1\cdots N-1}}| + |\alpha|\sum_{i \in \sigma_{N-1}} \widetilde{W}_{ji} |s_i^{\sigma_{N-1}}| \\
&=  \sum_{i \in \sigma_{1\cdots N-2}} \widetilde{W}_{ji} |s_i^{\sigma_{1\cdots N-1}}| + |\alpha|w_j^{\sigma_{N-1}}
\end{align*}
On the other hand, for $k \in \tau_N$ we have the following formula for $w_k^{\sigma_{1\cdots N-1}}$, where we use the fact that $\widetilde{W}_{ki} = -1-\delta$ for all $i \in \sigma_{1\cdots N-2}$ since there are no edges from nodes in $\tau_{1 \cdots N-2}$ to $\tau_N$:
\begin{align*} 
w_k^{\sigma_{1\cdots N-1}} &\od  \sum_{i \in \sigma_{1\cdots N-1}} \widetilde{W}_{ki} |s_i^{\sigma_{1\cdots N-1}}|\\ 
&=  \sum_{i \in \sigma_{1\cdots N-2}} \widetilde{W}_{ki} |s_i^{\sigma_{1\cdots N-1}}| + \sum_{i \in \sigma_{N-1}} \widetilde{W}_{ki} |s_i^{\sigma_{1\cdots N-1}}| \\
&=  \sum_{i \in \sigma_{1\cdots N-2}} (-1-\delta)|s_i^{\sigma_{1\cdots N-1}}| + \sum_{i \in \sigma_{N-1}} \widetilde{W}_{ki} |\alpha s_i^{\sigma_{N-1}}| \\
&=  \sum_{i \in \sigma_{1\cdots N-2}} (-1-\delta) |s_i^{\sigma_{1\cdots N-1}}| + |\alpha|\sum_{i \in \sigma_{N-1}} \widetilde{W}_{ki} |s_i^{\sigma_{N-1}}|\\
&=  \sum_{i \in \sigma_{1\cdots N-2}} (-1-\delta) |s_i^{\sigma_{1\cdots N-1}}| + |\alpha|w_k^{\sigma_{N-1}}.
\end{align*}  
Moreover, since $\sigma_{N-1} \in \FP(G)$, we have that $j \in \sigma_{N-1}$ must inside-out dominate the external node $k$, so $ w_j^{\sigma_{N-1}} >  w_k^{\sigma_{N-1}}$.  Combining this with the fact that $\widetilde{W}_{ji} \geq -1-\delta$, we see that
\begin{align*} 
w_k^{\sigma_{1\cdots N-1}} &\leq  \sum_{i \in \sigma_{1\cdots N-2}} \widetilde{W}_{ji} |s_i^{\sigma_{1\cdots N-1}}| + |\alpha|w_k^{\sigma_{N-1}} \\
&< \sum_{i \in \sigma_{1\cdots N-2}} \widetilde{W}_{ji} |s_i^{\sigma_{1\cdots N-1}}| + |\alpha|w_j^{\sigma_{N-1}} = w_j^{\sigma_{1\cdots N-1}}
\end{align*}  
Thus $w_j^{\sigma_{1\cdots N-1}} > w_k^{\sigma_{1\cdots N-1}}$ and so $j$ inside-out dominates $k$ for all $k \in \tau_N$.  Thus by Theorem~\ref{thm:gen-domination}, $\sigma_{1\cdots N-1} \in \FP(G)$, and so $\cup_{i \in [N]} \sigma_i  = \sigma_{1\cdots N-1} \cup \sigma_N \in \FP(G)$ as desired.  
\end{proof}

\subsection{Proofs for strongly simply-embedded partitions}
In this section we prove Theorem~\ref{thm:bidir-sa-partition}, characterizing $\FP(G)$ for strongly simply-embedded partitions.  First, we prove Lemma~\ref{lemma:full-factorization} which shows that the strongly simply-embedded structure guarantees a complete factorization of the $s_j^\sigma$ values in terms of the $s_j^{\sigma_i}$ of the component fixed point supports.  Moreover, the $s_j^{\sigma_i}$ values are fully determined by whether $\sigma_i$ is a surviving or a dying fixed point of $G|_{\tau_i}$.  Recall that we denote the sets of surviving and dying fixed points as:
$$S_{\tau_i} \od \FP(G|_{\tau_i}) \cap \FP(G) \quad \text{and} \quad D_{\tau_i} \od \FP(G|_{\tau_i}) \setminus S_{\tau_i}.$$

\begin{lemma}\label{lemma:full-factorization}
Let $G$ be a graph on $n$ nodes with a strongly simply-embedded partition $\{\tau_1|\dots|\tau_N\}$. For any $\sigma \subseteq [n]$, denote $\sigma_i \od \sigma \cap \tau_i$, and $\sigma_{i_1\dotsi_k} \od \sigma_{i_1} \cup \dots \cup \sigma_{i_k}$ and let $I = \{i \in [N]~|~ \sigma_i \neq \emptyset\}$. Then for every $j\in [n]$, $$s_j^\sigma = \frac{1}{\theta^{|I|-1}}\prod_{i\in I} s_j^{\sigma_i},$$ 
where $s_j^{\sigma_i}$ has the same value for every $j \in [n]\setminus \tau_i$.\\
Moreoever, for any $\sigma_i \in \FP(G|_{\tau_i})$ and $j \in \tau_i$:
 $$\sgn s_j^{\sigma_i} = \begin{cases} 
      \phantom{-}\idx(\sigma_i) & \text{if } j \in \sigma_i \\
      -\idx(\sigma_i) & \text{if } j \in \tau_i\setminus \sigma_i
   \end{cases}$$ 
while for any $k\notin \tau_i$, 
$$\sgn s_k^{\sigma_i} = \begin{cases} 
      -\idx(\sigma_i) & \text{if } \sigma_i \in S_{\tau_i}\\
      \phantom{-}\idx(\sigma_i) & \text{if } \sigma_i \in D_{\tau_i}
   \end{cases}$$

\end{lemma}
\smallskip
\begin{proof}
Since $\{\tau_1|\dots|\tau_N\}$ is a strongly simply-embedded partition of $G$, we have $[n]\setminus\tau_1$ simply-added onto $\tau_1$, and so $$s_j^\sigma = \frac{1}{\theta} s_j^{\sigma_{2\dots N}}s_j^{\sigma_1} \text{ for all } j \in \tau_1$$
by Theorem~\ref{thm:simply-added}.  On the other hand, since $\tau_1$ is also simply-added onto $[n]\setminus\tau_1$, we also have
$$s_j^\sigma = \frac{1}{\theta} s_j^{\sigma_1} s_j^{\sigma_{2\dots N}} \text{ for all } j \in [n]\setminus\tau_1.$$   
Therefore, the above factorization holds for all $j \in [n]$. Similarly, since $[n]\setminus\tau_2$ is simply-added to $\tau_2$ and vice versa, 
$$s_j^{\sigma_{2\dots N}}= \frac{1}{\theta} s_j^{\sigma_2} s_j^{\sigma_{3\dots N} \text{ for all } j \in [n]}$$ 
by Theorem~\ref{thm:simply-added}, and so $s_j^\sigma = \frac{1}{\theta^2} s_j^{\sigma_1} s_j^{\sigma_2} s_j^{\sigma_{3\dots N}}$.  Continuing in this fashion, we see that for any $j \in [n]$, 
$$s_j^{\sigma} =\frac{1}{\theta^{N-1}} s_j^{\sigma_1}\dots s_j^{\sigma_N}.$$ 
Note that if $\sigma_i = \emptyset$, then $s_j^{\sigma_i} = s_j^{\emptyset} = s_j^{\{j\}} = \theta$, and thus for all $j \in [n]$, 
$$s_j^{\sigma} = \frac{\theta^{N-|I|}}{\theta^{N-1}}\prod_{i \in I}s_j^{\sigma_i} = \frac{1}{\theta^{|I|-1}}\prod_{i\in I} s_j^{\sigma_i}.$$
The fact that $s_j^{\sigma_i}$ has the same value for every $j \in [n]\setminus \tau_i$ is a direct consequence of Theorem \ref{thm:simply-added} since $\tau_i$ is simply-added onto $[n]\setminus \tau_i$.

Finally, to prove the last statements about the signs of $s_j^{\sigma_i}$, observe that for $j \in \tau_i$, the values of $\sgn s_j^{\sigma_i}$ are fully determined by Theorem \ref{thm:sgn-conditions} (sign conditions) since $\sigma_i \in \FP(G|_{\tau_i})$ by hypothesis.   In particular, if $\sigma_i \in S_{\tau_i}$, then $\sigma_i$ survives the addition of every $k \notin \tau_i$, and so $\sgn s_k^{\sigma_i} = -\idx(\sigma_i)$ by Theorem \ref{thm:sgn-conditions} (sign conditions).  On the other hand, if $\sigma_i \in D_{\tau_i}$ then $\sigma_i$ dies in $G$ and so there is some $k \notin \tau_i$ for which $\sgn s_k^{\sigma_i} = \idx(\sigma_i)$.  But by the first part of the theorem, all the $s_k^{\sigma_i}$ values are identical for $k \in [n] \setminus \tau_i$, and thus $\sgn s_k^{\sigma_i} = \idx(\sigma_i)$ for all such $k$.  
\end{proof}

With Lemma~\ref{lemma:full-factorization}, it is now straightforward to prove Theorem~\ref{thm:bidir-sa-partition} (reprinted below).  This theorem generalizes Theorem~\ref{thm:bidir-sa}, characterizing every element of $\FP(G)$ in terms of the sets of surviving and dying component fixed points supports, $S_{\tau_i}$ and $D_{\tau_i}$.  Notice that in the statement of Theorem~\ref{thm:bidir-sa-partition}, all the fixed point supports of type (a) have the form $\bigcup_{i \in I} \sigma_i$ for $\sigma_i \in S_{\tau_i}$ and $I \subseteq [N]$, while those of type (b) have the form $\bigcup_{i=1}^N \sigma_i$ for $\sigma_i \in D_{\tau_i}$.\\

\noindent{\bf Theorem~\ref{thm:bidir-sa-partition}.} 
Suppose $G$ has a strongly simply-embedded partition $\{\tau_1|\dots|\tau_N\}$, and let $\sigma_i \od \sigma \cap \tau_i$ for any $\sigma \subseteq [n]$.  Then $\sigma \in \FP(G)$ if and only if $\sigma_i \in \FP(G|_{\tau_i}) \cup \{\emptyset\}$ for each $i \in [N]$, and either
\begin{enumerate}
\item[(a)] every $\sigma_i$ is in $\FP(G) \cup \{ \emptyset\}$, or 
\item[(b)] none of the $\sigma_i$ are in $\FP(G) \cup \{ \emptyset\}$.
\end{enumerate}
In other words, $\sigma \in \FP(G)$ if and only if $\sigma$ is either a union of surviving fixed points $\sigma_i$, at most one per component, or it is a union of dying fixed points, exactly one from every component.

\begin{proof} \label{sec:proof-bidir-sa}
First notice that since $G$ has a strongly simply-embedded partition $\{\tau_1|\dots|\tau_N\}$, by Lemma~\ref{lemma:full-factorization}, for all $j \in [n]$, we have
$$s_j^\sigma = \prod_{i \in I} s_j^{\sigma_i}$$
where $I \od \{i ~|~ \sigma_i \neq \emptyset \}$, and we have set $\theta=1$, without loss of generality.  Moreover, $s_j^{\sigma_i}$ is constant across $j \in [n]\setminus\tau_i$ for each $i \in [N]$. 

\noindent ($\Rightarrow$) Suppose $\sigma \in \FP(G).$ Since $G$ has a simply-embedded partition, Theorem~\ref{thm:menu} (menu) guarantees $\sigma_i \in \FP(G|_{\tau_i})$ for every $i \in I$. Thus we can use the values of $\sgn s_j^{\sigma_i}$ given in Lemma~\ref{lemma:full-factorization} to examine the sign conditions for $\sigma$.
For any $j \in \sigma$, there exists $i \in I$ such that $j \in \sigma_i$, and then
\begin{equation}\label{eq:sgn-s_j}
\sgn s_j^\sigma =\ \  \idx(\sigma_i) \hspace{-.25in}\prod_{\{a \in I\setminus \{i\}~|~\sigma_a\in S_a\}} \hspace{-.35in} -\idx(\sigma_a)  \hspace{-.25in}\prod_{\{b \in I\setminus \{i\}~|~\sigma_b\in D_b\}} \hspace{-.3in} \idx(\sigma_b) = (-1)^{|\S \setminus \{i\}|}\prod_{\ell\in I} \idx(\sigma_\ell),
\end{equation}
 where $\S \od \{a\in I~|~\sigma_a \in S_a\}$. \\
 Now, observe that if $\sigma$ contained a mix of $\sigma_a \in S_a$ and $\sigma_b \in D_b$, then there would be $i, j \in \sigma$ such that $i \in \sigma_a$ for some $a\in \S$, while $j \in \sigma_b$ for some $b \notin \S$. In this case, 
 $$\sgn s_i^\sigma = (-1)^{|\S|-1}\prod_{\ell \in I}\idx(\sigma_\ell) = -(-1)^{|\S|} \prod_{\ell \in I} \idx(\sigma_\ell) = -\sgn s_j^{\sigma}.$$ 
 But by Theorem~\ref{thm:sgn-conditions} (sign conditions), $\sigma \in \FP(G)$ implies that $\sgn s_i^\sigma = \sgn s_j^\sigma$ for all $i, j \in \sigma$, yielding a contradiction. Thus, we must have either $\sigma_i \in S_{\tau_i}$ for all $i \in I$, as in (a), or $\sigma_i \in D_{\tau_i}$ for all $i \in I$ as in (b). 
 
Next we show that in case (b) when $\sigma_i \in D_{\tau_i}$ for all $i \in I$, we must have $I = [N]$, so that $\sigma$ takes a dying fixed point from every component.  Assume to the contrary that $I \subsetneq [N]$ so that there is some $m \in [N]$ such that $\tau_m \cap \sigma = \emptyset$.  Then, for $k\in\tau_m$ (so $k \notin \sigma$), we have $\sgn s_k^{\sigma_\ell} = \idx(\sigma_\ell)$ for all $\ell \in I$, by Lemma~\ref{lemma:full-factorization}, since $\sigma_\ell \in D_{\tau_\ell}$.  Thus
$$\sgn s_k^\sigma = \prod_{\ell \in I} \sgn s_k^{\sigma_\ell} = \prod_{\ell \in I} \idx(\sigma_\ell).$$ 
Meanwhile, for all $j \in \sigma$ we have $j \in \tau_i$ for some $i \in I$, and Equation~\eqref{eq:sgn-s_j} gives
 $$\sgn s_j^{\sigma} =(-1)^{|\S \setminus \{i\}|}\prod_{\ell\in I} \idx(\sigma_\ell) = \prod_{\ell\in I} \idx(\sigma_\ell)$$ 
 since $\S = \emptyset$ because $\sigma_\ell \in D_{\tau_\ell}$ for all $\ell \in I$. Thus, 
 $$\sgn s_k^{\sigma} = \prod_{\ell\in I} \idx(\sigma_\ell) = \sgn s_j^{\sigma}$$ 
 for some $j \in \sigma$ and $k \notin \sigma$, contradicting the sign conditions for $\sigma \in \FP(G)$.  Therefore, we must have $I = [N]$.\\
 
\noindent ($\Leftarrow$) First consider case (a) where $\sigma_i \in S_{\tau_i}$ for all $i \in I$. We will show that $\sigma \od \bigcup_{i \in I} \sigma_i \in \FP(G)$ by checking the sign conditions.   For any $j \in \sigma$, there exists $i \in I$ such that $j \in \tau_i$.  Then by Equation~\eqref{eq:sgn-s_j}, we have 
$$\sgn s_j^{\sigma} = (-1)^{|\S \setminus \{i\}|}\prod_{\ell\in I} \idx(\sigma_\ell) = (-1)^{|I|-1}\prod_{\ell\in I}\idx\sigma_\ell,$$ 
since $\S = I$ in this case.  
On the other hand, for $k \notin \sigma$, we have $\sgn s_k^{\sigma_\ell} = -\idx \sigma_\ell$ for all $\ell \in I$, by Lemma~\ref{lemma:full-factorization}, since $\sigma_\ell \in S_{\tau_\ell}$.  Thus
$$\sgn s_k^{\sigma} = \prod_{\ell\in I}(-\idx\sigma_\ell) = (-1)^{|I|}\prod_{\ell\in I}\idx\sigma_\ell = -\sgn s_j^\sigma.$$ 
Therefore $\sigma \in \FP(G)$ by Theorem \ref{thm:sgn-conditions} (sign conditions). \\
   
Next, consider case (b) where $\sigma_\ell \in D_{\tau_\ell}$ for all $\ell \in [N]$ (so $I = [N]$). 
Then for any $j \in \sigma$, there is $i \in [N]$ such that $j \in \sigma_i$ and by Equation~\eqref{eq:sgn-s_j}, we have
   $$\sgn s_j^{\sigma}=(-1)^{|\S \setminus \{i\}|}\prod_{\ell\in [N]} \idx(\sigma_\ell) = \prod_{\ell\in [N]} \idx(\sigma_\ell),$$ 
since $\S=\emptyset$.  
Meanwhile, for any $k \notin \sigma$ there is some $m$ such that $k \in \tau_m$ with $\tau_m \cap \sigma \neq \emptyset$ (since $I = [N]$). Since $\sigma_m \in \FP(G|_{\tau_m})$, we have $\sgn s_k^{\sigma_m} = -\idx(\sigma_m)$ and thus 
$$\sgn s_k^{\sigma}=\sgn s_k^{\sigma_m} \hspace{-.15in}\prod_{\ell\in [N]\setminus\{m\}}  \hspace{-.15in}\sgn s_k^{\sigma_\ell} = -\idx(\sigma_m)  \hspace{-.15in}\prod_{\ell\in [N]\setminus\{m\}}  \hspace{-.15in}\idx(\sigma_\ell) = -\prod_{\ell\in [N]} \idx(\sigma_\ell) = -\sgn s_j^\sigma.$$ 
Thus sign conditions are satisfied, and so $\sigma \in \FP(G)$.   
\end{proof}

\end{document}